\documentclass[prx,twocolumn,showpacs]{revtex4-1}
\usepackage{multirow}
\usepackage{graphicx}
\usepackage{dcolumn}
\usepackage{bm}
\usepackage[latin1]{inputenc}
\usepackage{amssymb, amsmath}
\usepackage{bbm}
\usepackage{nicefrac}
\usepackage{esvect}
\usepackage{sidecap}
\sidecaptionvpos{figure}{t}
\newcommand*{\citen}[1]{%
  \begingroup
    \romannumeral-`\x 
    \setcitestyle{numbers}%
    \cite{#1}%
  \endgroup
} 

\begin{document}
\title{Probing alloy formation using different excitonic species: \\ The particular case of InGaN}
\author{G. Callsen}
\email{gordon.callsen@epfl.ch}
\author{R. Butt\'{e}}
\author{N. Grandjean}
\affiliation{Institute of Physics, \'{E}cole Polytechnique F\'{e}d\'{e}rale de Lausanne (EPFL), CH-1015 Lausanne, Switzerland}
\date{\today}
\begin{abstract}
Since the early 1960s, alloys are commonly grouped into two classes, featuring bound states in the bandgap (I) or additional, non-discrete, band states (II). As a consequence, one can observe either a rich and informative zoo of excitons bound to isoelectronic impurities (I), or the typical bandedge emission of a semiconductor that shifts and broadens with rising isoelectronic doping concentration (II). Microscopic material parameters for class I alloys can directly be extracted from photoluminescence (PL) spectra, whereas any conclusions drawn for class II alloys usually remain indirect and limited to macroscopic assertions. Nonetheless, here, we present a comprehensive spectroscopic study on exciton localization in a so-called mixed crystal alloy (class II) that allows us to access microscopic alloy parameters. In order to exemplify our experimental approach we study bulk In$_x$Ga$_{1-x}$N epilayers at the onset of alloy formation (0\,$\leq$\,$x$\,$\leq$\,2.4\%) in order to understand the material's particular robustness to point and structural defects. Based on an in-depth PL analysis it is demonstrated how different excitonic complexes (free, bound, and complex bound excitons) can serve as a probe to monitor the dilute limit of class II alloys. From an $x$-dependent linewidth analysis we extract the length scales at which excitons become increasingly localized, meaning that they convert from a free to a bound particle upon alloy formation. Already at $x=2.4\%$ the average exciton diffusion length is reduced to $5.7 \pm 1.3\,\text{nm}$ at a temperature of 12\,K, thus, detrimental exciton transfer mechanisms towards non-radiative defects are suppressed. In addition, the associated low temperature luminescence data suggests that a single indium atom does not suffice in order to permanently capture an exciton. The low density of silicon impurities in our samples even allows studying their local, indium-enriched environment at the length scale of the exciton Bohr radius based on impurity bound excitons. The associated temperature-dependent PL data reveals an alloying dependence for the exciton-phonon coupling. Thus, the formation of the random alloy can not only directly be monitored by the emission of various excitonic complexes, but also more indirectly via the associated coupling(s) to the phonon bath. Micro-PL spectra even give access to a forthright probing of silicon bound excitons embedded in a particular environment of indium atoms, thanks to the emergence of a hierarchy of individual, energetically sharp emission lines (full width at half maximum\,$\approx$\,300\,$\mu$eV). Consequently, the present spectroscopic study allows us to extract first microscopic alloy properties formerly only accessible for class I alloys.
\end{abstract}
%
%
%
%
\maketitle
\section{Introduction}
\label{introduction}

Upon alloying, host atoms in insulators, semiconductors, or metals are replaced by atoms with an equivalent valence electron structure, giving rise to isoelectronic impurities. Despite the matching valency, which does not lead to $n$- or $p$-type doping in case of a semiconductor crystal, the total number of electrons is altered, inducing fundamental changes in the electronic band structure and the associated optical signatures of alloying \cite{Physics}. Already in 1966, Thomas pointed out \cite{Thomas1966a} that such isoelectronic impurities can be divided into two classes: For the first class (I), discrete electronic levels are formed in the bandgap that can be studied by the related emission of bound excitons in, e.g., GaP:N \cite{Wilkinson1959,Zhang2000}, GaP:Bi \cite{Faulkner1970,Francoeur2008}, GaAs:N \cite{Wolford2014}, ZnTe:O \cite{Merz1968}, CdS:Te \cite{Parz}, and ZnO:Hg \cite{Cullen2013}. However, such highly localized bound excitons caused by isoelectronic centers strongly differ from the wide range of bound excitons known for shallow impurities in semiconductors and cannot be described by an effective mass approach \cite{Faulkner1968,Phillips1969,Allen1971}. The second class of isoelectronic centers (II) evokes the formation of mixed crystal alloys like, e.g., SiGe \cite{Braunstein1958}, GaAsP \cite{Tietjen1965}, InGaAs \cite{J.P.LaurentiP.RoentgenK.WolterK.SeibertH.Kurz1988}, AlGaAs \cite{Schubert1984}, InGaN \cite{Butte2018a}, AlGaN \cite{Wagener2004}, CdSSe \cite{Notes1978,Klochikhin1999}, ZnSeTe \cite{Klochikhin1999}, and MgZnO \cite{Grundmann2009}. In these cases, no new electronic levels are formed in the bandgap, but rather in the bands themselves, a process often described as hybridization \cite{Bellaiche1999}. Clearly, such a simple classification is not always straightforward \cite{Dean1970} and even a continuous, concentration-dependent transition between these two classes of isoelectronic centers has been reported for the unique case of silver halides \cite{Kanzaki1969}.

The consequences regarding the existence of these two types of isoelectronic centers in nature for any optical material characterization are pivotal: For class I alloys the apparent rich optical signature \cite{Hopfield1966} allows to directly extract microscopic material parameters from macroscopic photoluminescence (PL) spectra. Not only the number of isoelectronic impurities per binding center can be extracted \cite{Francoeur2008}, but even the distance in between these impurities or their constellation can be determined \cite{Thomas1966,Zhang2000,Muller2006}. In contrast, for class II alloys, the spectroscopic analysis is hampered. Commonly, only a continuous shift of the bandedge luminescence is observable, which is additionally plagued by pronounced linewidth broadening \cite{Schubert1984,J.P.LaurentiP.RoentgenK.WolterK.SeibertH.Kurz1988}, preventing any detailed information to be extracted from PL spectra. Consequently, in contrast to class I alloys, only a rather indirect analysis of such mixed crystal alloys is feasible by optical techniques. What further worsens the situation is the fact that isoelectronic impurities related to class II are more abundant in nature in comparison to their type I counterparts \cite{Physics}. Furthermore, most technologically relevant alloys (e.g., InGaAs, AlGaAs \cite{Manfra2014}, AlInGaP \cite{GeraldB.Stringfellow1998}, InGaN, InAlN \cite{Chichibu2006a}, and AlGaN \cite{Kneissl2011}) belong to class II. Here, especially III-nitride alloys have recently proven their importance for a wide range of applications, rendering a spectroscopic analysis beyond the limits of their classification as mixed crystal alloys an auspicious task.

Over the past decade, III-nitride-based semiconductors have reached a high level of dissemination only second to silicon based on a wide range of applications covering both, electronics and optoelectronics \cite{Gil2014b}. Many aspects of our daily life are already impacted by III-nitride-based light-emitting and laser diodes (LEDs and LDs) \cite{Nakamura1991,S.Nakamura1997}, while high-power electronics is also on the rise \cite{Mishra2008,Rajan2013}. Here, ternary alloys like InGaN and InAlN play a crucial role, as essential properties like, e.g., their emission wavelength can be tuned over wide ranges \cite{Wachter2011} in order to suit the particular application of choice. Such alloys containing indium are often described as "special alloys", because many devices like LEDs based on InGaN/GaN quantum wells (QWs) perform astonishingly well, with internal quantum efficiencies $\eta_{int}$ well beyond 80\% \cite{Nippert2016d}, even though the threading dislocation density is high with values in the $10^{8}-10^{9}\,\text{cm}^{-2}$ range \cite{Ponce1997a}. Generally, the material proves particularly robust against structural and point defects in comparison to other III-V alloys (e.g. InGaAs, AlGaAs, AlInGaP, etc.) \cite{GeraldB.Stringfellow1998} as outlined in the seminal paper of Chichibu \textit{et al.} \cite{Chichibu2006a}. However, the physical origin behind this behavior is not well understood yet and several possible causes have been discussed in the literature. 

Early explanations accounting for the particular case of InGaN range from large structural defects like V-pits \cite{Hangleiter2005} and indium clusters \cite{Narukawa1997}, over indium-zig-zag chains \cite{Kent2001a,Chichibu2006a}, to the particular electronic environment of a single indium atom \cite{Bellaiche1999}. All explanations share the common idea that localization of carriers occurs in the alloy, ultimately leading to higher $\eta_{int}$ values. As soon as carriers are localized, their diffusion to non-radiative centers is suppressed, which in turn enhances $\eta_{int}$. While larger structural defects can nowadays be excluded for modern III-nitride materials based on a combination of structural [scanning transmission electron microscopy, nanoscale secondary ion mass spectroscopy, etc.] and optical [cathodoluminescence (CL), micro-photoluminescence ($\mu$-PL), etc.] techniques, the material analysis at the few to even sub-nanometer scale remains challenging. Even though recent technical progress made by atom probe tomography (APT) indicates that InGaN is a random alloy \cite{Galtrey2007,Cerezo2007,Rigutti2018}, the impact of assemblies of indium atoms like pairs, triplets, and larger sets on the material's optical signature remains unclear.



%
%
%
\begin{SCfigure*}[][h]
\includegraphics[width=9.5cm]{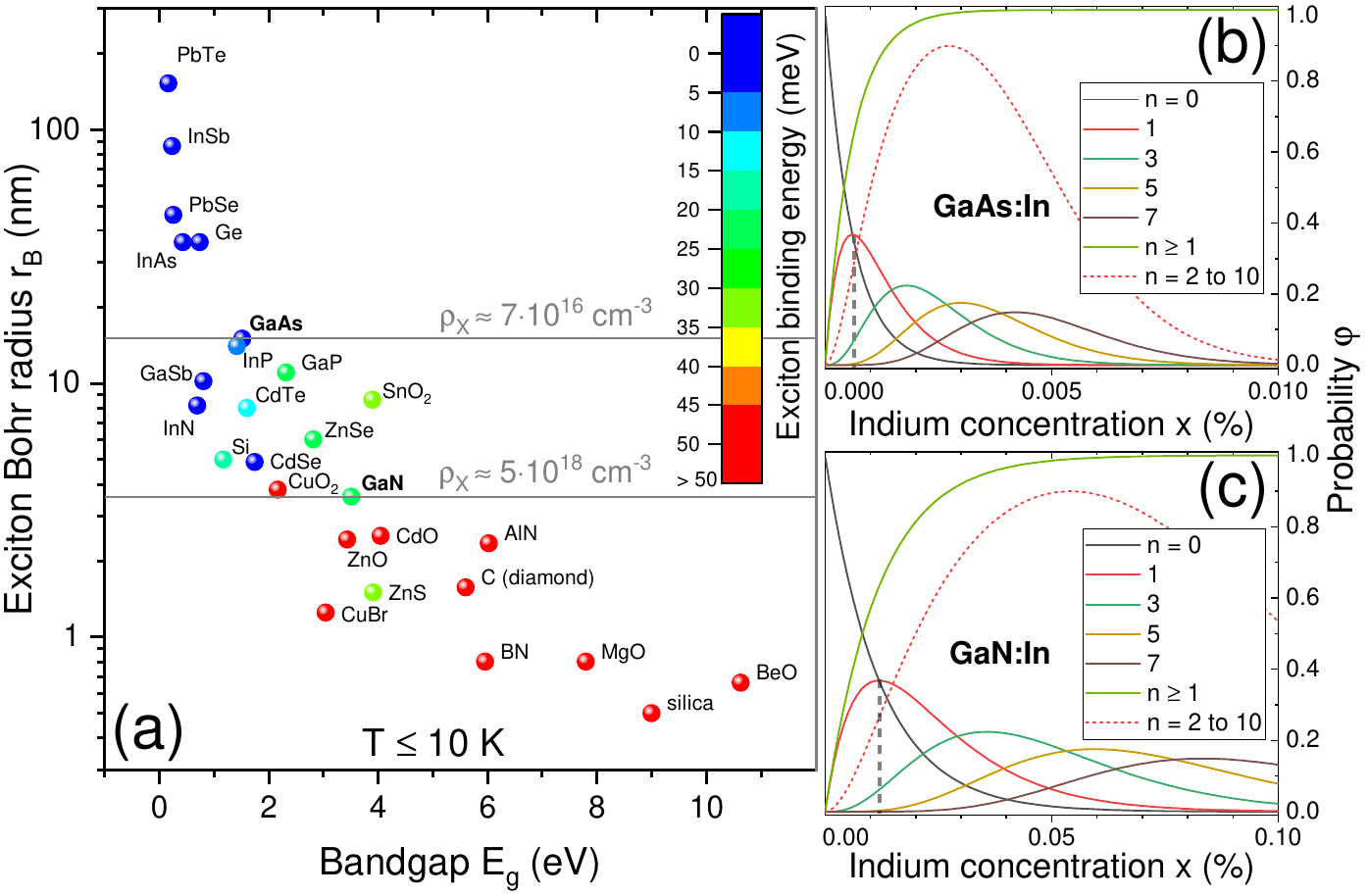} 
\caption{(color online) (a) Hydrogen-like exciton Bohr radius $r_B$ (excitonic ground state) for various material systems comprising semiconductors and insulators that exhibit a bandgap energy $E_{g}$ in between $\approx$\,0.2\,-\,10\,eV. With diminishing $r_B$ and increasing $E_{g}$ the exciton binding energy $E_{bind}$ rises (color encoded) - a trend which is also valid for all related alloys. The inverse of the exciton volume $V_{X}(r_{B})$ motivates a density $\rho_{X}$, which allows for a first estimate of the excitons' sensitivity to point and structural defects. To illustrate this matter, we compare isoelectronically doped (b) GaAs:In and (c) GaN:In, whose $r_B$ values differ by a factor of $\sim 4$. The indium concentration $x$ required to find a certain number of indium atoms $n$ in $V_X$ with the probability $\varphi$ is given by the Bernoulli distribution and varies by one order of magnitude in this comparison. While bound excitons average over $\approx V_X$, free excitons monitor the alloy over a larger averaging volume $V_a$ due to their eponymous movement.}
\label{Fig1}
\end{SCfigure*}

In this work, we show that even for class II alloys it is possible to extract microscopic material parameters from PL and $\mu$-PL spectra. A detailed PL analysis of bulk In$_{x}$Ga$_{1-x}$N epilayers (0\,$\leq$\,$x$\,$\leq$\,2.4\%) grown on freestanding (FS) GaN substrates (dislocation density $\sim10^{6}\,\text{cm}^{-2}$) allows us to extract the effective exciton diffusion length ($r_a$) and its dependence on indium content ($x$). By analyzing the linewidth broadening of free and bound excitons, one obtains probes that monitor the alloy formation on different length scales determined by $r_a$ and the exciton Bohr radius $r_B$, hence, in our case encompassing length scales ranging from tens to a few nanometers. With rising $x$ we measure a decrease of $r_a$ towards $r_B$ using the free exciton ($\mathrm{X_A}$) as a probe for the formation of a random alloy in which the excitonic center of mass (COM) movement becomes increasingly negligible at cryogenic temperatures. In addition, the exciton-phonon coupling is studied, e.g., for silicon-bound excitons ($\mathrm{Si^{0}X_{A}}$), representing an alternative tool to monitor the onset of alloy formation. The spectroscopy of excitons captured at individual indium-related centers proves challenging due to an apparent high density of emitters. Hence, we focus on an analysis of $\mu$-PL spectra of individual silicon-bound excitons embedded in particular configurations of indium atoms in their direct vicinity. As a result, we observe a hierarchy of energetically well-defined emission lines [full width at half maximum (FWHM) $\approx$\,300\,$\mu$eV] that originate from $\mathrm{Si^{0}X_{A}-In^{\textit{n}}}$ complexes in the InGaN alloy. Hence, our study opens a pathway towards a spectroscopic analysis of the microscopic properties of a class II alloy at the nanometer scale by probing distinct configurations of indium atoms at the onset of alloy formation. 

The paper is structured as follows: In Sec.\,\ref{context} we first compare the fundamental excitonic properties of a wide range of materials, before focusing the motivation to III-V semiconductors represented by indium doped GaAs and GaN as representatives of class II alloys. We present our results in Sec.\,\ref{results}. This section is subdivided into four parts labeled Secs.\,\ref{macro}-\ref{indirect}. In Sec.\,\ref{macro} we show how ensembles of free and bound excitons can serve as probes in the InGaN alloy using PL spectra. Temperature-dependent PL data allows us to analyze the local distribution of indium atoms around particular impurity bound excitons as demonstrated in Sec.\,\ref{local}. Subsequently, Sec.\,\ref{micro} introduces the $\mu$-PL traces of individual excitonic complexes bearing on an impurity embedded in a distinct environment of indium atoms. The spectra of such individual bound excitonic complexes are further analyzed in Sec.\,\ref{indirect}, motivating the existence of spatially direct and indirect excitonic complexes in the InGaN alloy. In a nutshell, Secs.\,\ref{macro}-\ref{indirect} describe the following transition: first, ensembles of excitonic complexes are probed in the alloy by PL. Second, the presented alloy probes become increasingly localized, culminating in the $\mu$-PL observation of individual impurity bound excitonic complexes embedded in a distinct configuration of indium atoms. Sec.\,\ref{discussion} finally relates our PL and $\mu$-PL results before Sec.\,\ref{conclusions} summarizes all findings. Experimental details regarding the spectroscopic and growth techniques can be found in the Supplementary Information (SI) in Sec.\,I.

\section{Context and motivation}
\label{context}
\begin{SCfigure*}[][h]
\includegraphics[width=11cm]{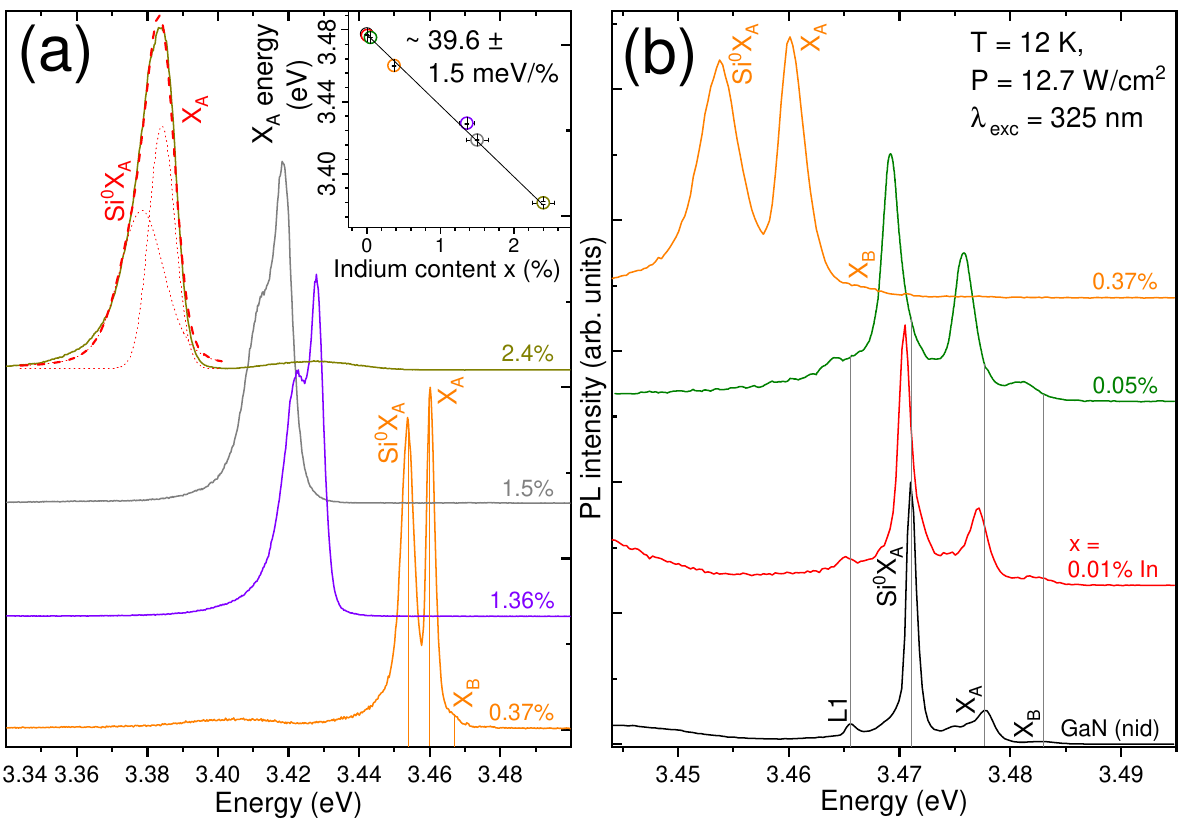} 
\caption{(color online) Overview PL spectra of the In$_x$Ga$_{1-x}$N sample series (0\,$\leq$\,$x$\,$\leq$\,2.4\%). The energetic shift and the linewidth of the two main emission lines related to the silicon bound exciton ($\mathrm{Si^{0}X_{A}}$) and the A-exciton ($\mathrm{X_{A}}$) can be followed up to an indium content of $x=\text{2.4}\%$ in (a) and (b). For the $\mathrm{X_{A}}$ transition a shift rate of 39.6\,$\pm$\,1.5\,meV/\% is extracted (inset) for 0\,$\leq$\,$x$\,$\leq$\,2.4\% at the given temperature of 12\,K. Additionally, a faint bound excitonic emission L1 and the emission of the B-exciton ($\mathrm{X_{B}}$) can be observed at the very onset of the composition range. At $x=\text{2.4}\%$ a simple fit routine based on two Voigt profiles highlights the two main emission peaks (illustrative purpose only). An inversion of the FWHM values related to $\mathrm{Si^{0}X_{A}}$ and $\mathrm{X_{A}}$ is directly visible by, e.g., comparing the spectra for $x=0.01\%$ and 0.37\% in (b).}
\label{Fig2}
\end{SCfigure*}

In order to obtain a general understanding of the sensitivity of excitons in alloys to crystal defects, Fig.\,\ref{Fig1}(a) introduces the reduction of $r_B$ (excitonic ground state) with rising bandgap energy $E_g$ for a large variety of materials (direct as well as indirect semiconductors and insulators). The higher the ionicity of the crystal - following the transition from IV-IV, over III-V, to II-VI semiconductors - the larger are the effective masses of the electrons ($m_e$) and holes ($m_h$) that move in the periodic potential of the lattice. In turn, the excitonic effective mass $\mu$ increases accordingly with $E_g$ like $1/\mu=1/m_{e}+1/m_{h}$, which is inversely proportional to $r_B$ \cite{Mathieu1992}. Such "heavy" excitons with small $r_B$ values ultimately lead to the transition from Wannier-Mott type excitons found in most semiconductors to self-trapped excitons frequently observed in silica \cite{Kalceff1995} and halides \cite{Tanimura1981}. As soon as $r_B$ is reduced, the exciton binding energy $E_{bind}$ rises as the Coulomb attraction between the electron and the hole is enhanced. Hence, $E_{bind}$ is proportional to $\mu$ as encoded in the data-point colors of Fig.\,\ref{Fig1}(a), leading to the experimental stability and the herewith interconnected accessibility of excitons beyond just cryogenic temperatures for large bandgap materials like, e.g., GaN \cite{Viswanath1998} and AlN \cite{Onuma2009}.

Already these basic considerations facilitate some conclusions that should be valid for any alloy of class I or II formed by the materials summarized in Fig.\,\ref{Fig1}(a). In alloys with low $E_g$ values an exciton averages at least over comparably large exciton volumes $V_{X} \propto r_{B}^{3}$. Hence, correspondingly low densities of, e.g., impurities and structural defects $\rho_X \propto V_{X}^{-1}$ suffice in order to affect or even trap an exciton if the associated perturbation is sufficiently large. In this sense, $\rho_X$ represents a tentative upper limit for the concentration of such centers affecting all excitons.  In contrast, high $E_g$ materials with lower $V_{X}$ values should be less sensitive to higher defect concentrations based on this simplistic comparison. In order to illustrate this matter, we highlight the corresponding $\rho_X$ values for GaAs and GaN in Fig.\,\ref{Fig1}(a), providing a first sensitivity estimate for excitons occurring in related alloys. As the $r_B$ value of both materials varies by a factor of $\sim 4$, the corresponding approximations of $\rho_X$ differ by almost two orders of magnitude. However, this comparison is complicated by the occurrence of free, bound (isovalent), and impurity bound (non-isovalent) excitons constituting two- or three-particle complexes. One can either observe free excitons that move and hence monitor the alloy over an averaging volume $V_{a} \geq V_{X}$ leading to lower critical $\rho_X$ values, or localized, shallow impurity bound excitons sensing the material within $\approx V_{X}$ (see Sec.\,\ref{macro} for a detailed analysis). Nevertheless, the lower $r_B$ for large $E_g$ materials, the less excitons  average over a random alloy and the more they just monitor the immediate neighborhood of a distribution of, e.g., (non-) isovalent centers in the lattice. This simple picture is further supported by the common neglect of any COM movement for excitons in an alloy \cite{Zimmermann1990} - a simplification that will be discussed in detail in Sec.\,\ref{macro} for InGaN.

We further motivate the present study by focussing our previous considerations to indium-doped GaAs:In (e.g. Ref.\,\citen{J.P.LaurentiP.RoentgenK.WolterK.SeibertH.Kurz1988}) and GaN:In (this work) as representatives for the extremal cases of relatively low and high $r_B$ values in III-V semiconductor alloys of class II. Based on this choice, the associated $E_{bind}$ values would always remain sufficiently high to ensure a straightforward PL analysis. The reasoning that further motivates this approach is threefold:

A) High quality material is available for InGaN and InGaAs alloys for studying the onset of alloy formation for class II alloys due to the availability of high quality substrates for the epitaxial growth of thick, ternary layers ($\geq 100\,\text{nm}$) as required for spectroscopy. 

B) It is known that InGaAs- and AlGaAs-based LEDs are rather sensitive to the density of point and structural defects, whereas InGaN proves to be much more robust \cite{GeraldB.Stringfellow1998,Chichibu2006a} against them. 

C) It was pointed out theoretically that individual indium atoms in GaAs and GaN do not lead to any bound states in the bandgap \cite{Bellaiche1999}, a matter that would rather require several indium atoms to form larger indium configurations \cite{Kent2001a}. 

In other words, both materials are unambiguous representatives of class II alloys with relatively simple spectra that facilitate a detailed tracking of the alloy formation. For class I alloys known bound states are mostly associated to single atoms or (extended) pairs and can be observed for a wide range of materials as summarized in Sec.\,\ref{introduction}. The absence of such optical signatures will allow us to study the very onset of alloy formation in In$_x$Ga$_{1-x}$N (0\,$\leq$\,$x$\,$\leq$\,2.4\%) by monitoring the undisturbed averaging process over the random alloy from the perspective of either free, bound, or impurity-(complex)-bound excitons.

Generally, the probability $\varphi$ to find $n$ indium atoms in $V_{X}$ for a certain indium concentration $x$ is given by the Bernoulli distribution in a random alloy:

\begin{equation}
\label{eq:Bernoulli}
\varphi(n) = \binom{K V_{X}}{n} x^{n}(1-x)^{K V_{X} - n}.
\end{equation}

Here, $K V_{X}$ is the cation number in the excitonic volume with $K \approx$ 0.0438\,{\AA}$^{-3}$ for wurtzite GaN in the indium composition range analyzed in this paper \cite{Vurgaftman2003,Butte2018a}. An illustration for the probability $\varphi$ to find, e.g., $0 \leq n \leq 7$ indium atoms in $V_{X}$ is given for GaN and GaAs ($K \approx$ 0.0221\,{\AA}$^{-3}$) \cite{Schubert1984} in Figs.\,\ref{Fig1}(b) and \ref{Fig1}(c), respectively. The indium concentration $x$ for which one can find, e.g., one indium atom in $V_{X}$ with $\varphi=0.5$ deviates by more than one order of magnitude, due to the about 4$\times$ larger $r_{B}$ value of GaAs in comparison to GaN. This fact will be crucial for the following PL linewidths analysis of the InGaN alloy, providing insight into the material's particular robustness against point and structural defects \cite{Chichibu2006a}.

\section{Results}
\label{results}

Figure\,\ref{Fig2} introduces the low temperature (12\,K) PL spectra of the bulk In$_x$Ga$_{1-x}$N sample series at hand with an indium content ranging from 0 to 2.4\% (see Sec.\,I in the SI for experimental details). The spectrum of the non-intentionally doped (nid) GaN reference sample shows the common optical traces of the A- and B-exciton ($\mathrm{X_{A}}$ and $\mathrm{X_{B}}$) along with a dominant neutral donor bound exciton line ($\mathrm{Si^{0}X_{A}}$). The energetic splitting in between $\mathrm{X_{A}}$ and $\mathrm{Si^{0}X_{A}}$ is known as the localization energy $E_{loc} \approx 7\,\text{meV}$ \cite{Callsen2018}. Generally, this terminology is not only used for excitons binding to non-isovalent impurities, but also for the localization of excitons in an alloy due to isoelectronic impurities that induce the formation of a potential landscape \cite{Chichibu2006a}. In both cases, the emission energy of the excitons exhibits a redshift with respect to $\mathrm{X_{A}}$, which will always be described by $E_{loc}$ in the following. Previous results have shown that the dominant impurity giving rise to the main bound excitonic emission is the neutral silicon center $\mathrm{Si^{0}}$, while other typical trace impurities in GaN like oxygen are negligible in our samples \cite{Callsen2018}. In addition, towards lower energies one observes the L1 emission line in Fig. \ref{Fig2}(b), whose origin is still under debate in the literature, ranging from a deep, over an ionized donor-bound exciton, to a neutral, shallow acceptor bound exciton \cite{Santic1998,Monemar2001a,Wysmoek2003,Fischer1997a}. 

Upon increasing indium content, the entire set of emission lines shifts continuously towards lower energies, while the level of detail in the spectra diminishes due to linewidth broadening as commonly observed for a class II alloy. The spectral shift of the $\mathrm{X_{A}}$ transition as a function of the change in bandgap energy $\delta E_{g}$ with rising $x$ is linear in the given indium concentration range and amounts to $\delta E_{g}/\delta x = 39.6\,\pm\,1.5\,\text{meV}/\%$ at a temperature of 12\,K as shown in the inset of Fig.\,\ref{Fig2}(a). Clear evidence for the emission line L1 is lost at indium concentrations exceeding 0.05\%, before even the spectral separation between $\mathrm{X_{A}}$ and $\mathrm{Si^{0}X_{A}}$ vanishes. Nevertheless, even at indium concentrations of 1.5\% and 2.4\% we can still reveal the presence of the $\mathrm{Si^{0}X_{A}}$ centers in the corresponding spectra by a simple lineshape fitting procedure employing two Voigt profiles as exemplified for $x=2.4\%$ in Fig.\,\ref{Fig2}(a). We find the Voigt profiles to be dominated by their Gaussian contribution due to the comparable small influence of homogeneous broadening at a temperature of 12\,K. Note that the Voigt profiles from Fig.\,\ref{Fig2}(a) only have an illustrative purpose, highlighting the two different spectral components. In the following all reported FWHM values $\Delta E$ originate from a manual data reading in order to avoid a troublesome and overparametrized lineshape fitting. We will subsequently analyze the temperature dependence of the linewidths in the context of Fig.\,\ref{Fig4} in Sec.\,\ref{local}.


At first glance, the present alloying series of GaN:In seems to match the case of GaAs:In as presented by Laurenti \textit{et al.} \cite{J.P.LaurentiP.RoentgenK.WolterK.SeibertH.Kurz1988} for indium concentrations down to 0.03\% in GaAs. A continuous shift and a continuous broadening of all optical transitions are observed in GaN:In down to $x=0.01\%$ as expected for a class II alloy. Not even at the very onset of the alloying range one can observe any additional emission lines related to excitons bound to single isoelectronic indium centers in agreement with the aforementioned theoretical predictions \cite{Bellaiche1999}. Note that an indium concentration of $x=0.01\%$ yields a probability of $\varphi \approx 0.35$ for a single indium atom to be present in the excitonic volume $V_X$ as shown in Fig.\,\ref{Fig1}(c). Hence, the indium concentration should be sufficiently low to observe the occurrence of any additional bound states induced by single indium atoms.
 
However, a more detailed inspection of PL and $\mu$-PL spectra will result in a manifold of interesting and so far unforeseen observations for the InGaN alloy as shown in Secs.\,\ref{macro}-\ref{indirect}. The latter mark the transition from an analysis dealing with ensembles of excitons down to individual ones giving access to the microscopic properties of a class II alloy.


%
\subsection{Probing alloying with free and bound excitons based on macro-photoluminescence}
\label{macro}

An in-depth analysis of the emission linewidth of $\mathrm{X_{A}}$ and $\mathrm{Si^{0}X_{A}}$ reveals an intriguing feature for GaN:In that is fostered by its large $E_{bind}$ values and, in comparison to GaAs:In, its smaller $r_B$ values. At a temperature of 12\,K and $x=0.01\%$ the $\mathrm{Si^{0}X_{A}}$ transition exhibits a FWHM value of ${\Delta E_{\mathrm{Si^{0}X_{A}}}=1.30\pm 0.06\,\text{meV}}$ while the FWHM of $\mathrm{X_{A}}$ is larger with ${\Delta E_{\mathrm{X_{A}}}=1.74\pm 0.19\,\text{meV}}$ due to the distribution of free excitons in momentum ($\bm{k}$) space. Interestingly, with rising indium content this FWHM ratio is reversed as seen, e.g., in the PL spectrum recorded for $x\,=\,0.37\%$, cf. Fig.\,\ref{Fig2}(b). At this indium concentration we find ${\Delta E_{\mathrm{X_{A}}}=2.63\,\pm\,0.04\,\text{meV}}$, while $\Delta E_{\mathrm{Si^{0}X_{A}}}$ has more than tripled to 4.47\,$\pm$\,0.08\,meV.

\begin{figure}[]
\includegraphics[width=8cm]{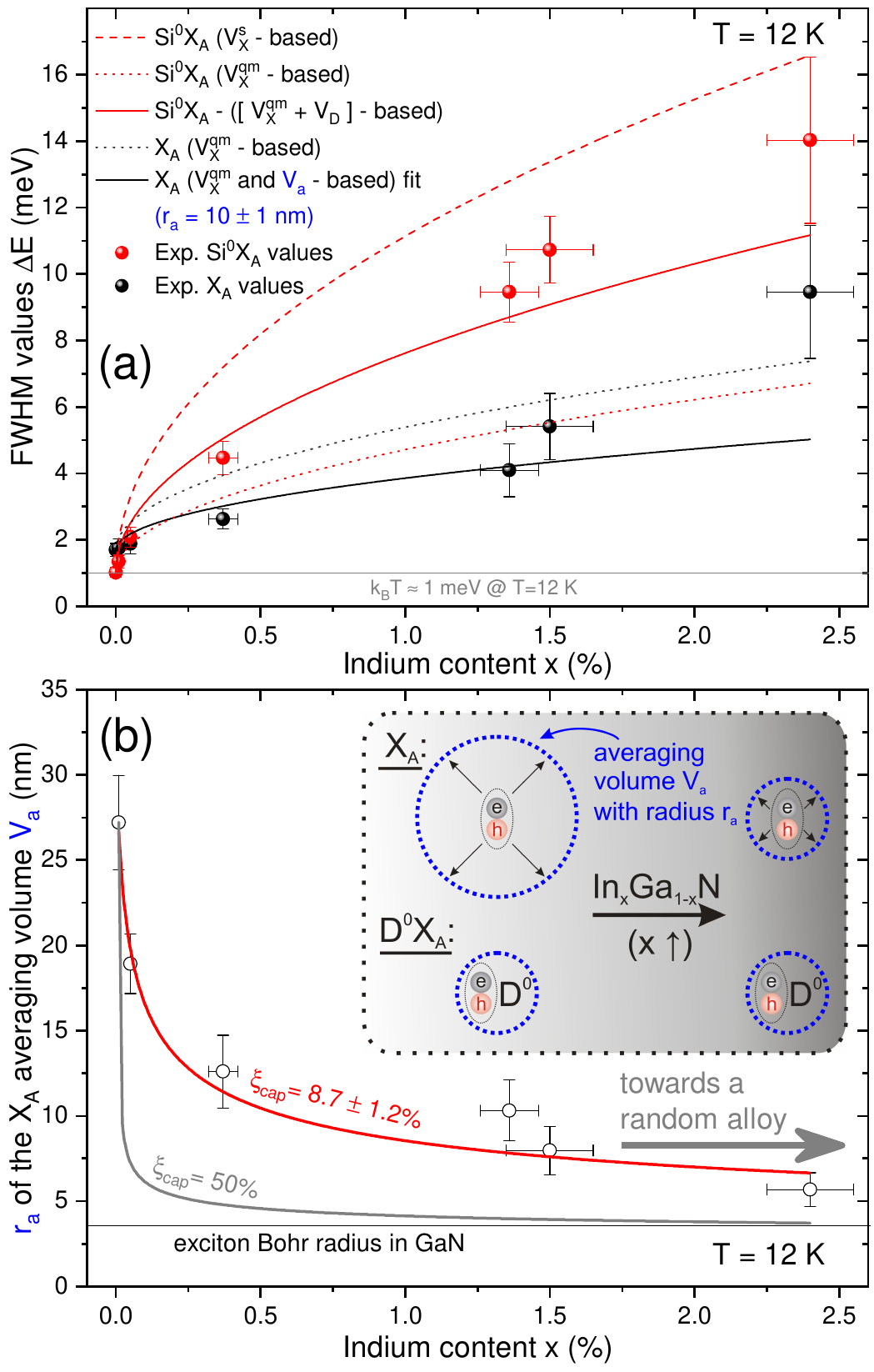} 
\caption{(color online) (a) Experimental FWHM values $\Delta E$ of the $\mathrm{Si^{0}X_{A}}$ (red symbols) and $\mathrm{X_{A}}$ (black symbols ) transitions versus indium content $x$ extracted from PL spectra. The red lines introduce step-by-step the modeling of the $\mathrm{Si^{0}X_{A}}$ linewidth depending on the particular excitonic volume $V_X$. Here, $V^{s}_X$ denotes a spherical and $V^{qm}_X$ the quantum-mechanical exciton volume, while $V_{D}$ is the volume ascribed to the donor electron. An exciton averaging volume ($V_{a}$) has to be introduced to model the $\Delta E$ values of the $\mathrm{X_{A}}$ complex (solid black line) in order to take into account its movement. All graphs are offset by the corresponding experimental FWHM value at $x=0$. (b) The values of $r_a$ are extracted from $V_{a}$ and converge towards the exciton Bohr radius because $\mathrm{X_{A}}$ gets increasingly localized with rising $x$ in contrast to $\mathrm{Si^{0}X_{A}}$ as illustrated in the inset. Hence, a random alloy is formed in which the excitonic center of mass movement is increasingly suppressed. See the main text for details.}
\label{Fig3}
\end{figure}

The evolution of the experimentally determined $\Delta E$ values for the $\mathrm{X_{A}}$ and $\mathrm{Si^{0}X_{A}}$ excitonic complexes is summarized in Fig.\,\ref{Fig3}(a) (black and red symbols). Both sets of FWHM values show a different evolution as averaging over the alloy occurs for different effective volumes. All $\Delta E$ values are commonly related to the peak width $w$ by $\Delta E=2\sqrt{2\text{ln}2}\,w$. Generally, such $w$ values are composed of an inhomogeneous component due to alloy broadening, $\sigma(x)$, and a homogeneous, temperature-dependent component $\Gamma(T)$, caused by phonon scattering, leading in first approximation to $w=\sigma+\Gamma$ \cite{Grundmann2009,ClausKlingshirn}. At low temperatures (12\,K) $\sigma$ dominates any phononic effects giving rise to the experimental trends shown in Fig.\,\ref{Fig3}(a) for the directly measured $\Delta E$ values. Additional broadening by $\Gamma(T)$ will subsequently be discussed for $\mathrm{Si^{0}X_{A}}$ in Sec.\,\ref{local} in the context of Fig.\,\ref{Fig4}, while the corresponding analysis for $\mathrm{X_{A}}$ is given in SI Sec.\,III. Such predominance of inhomogeneous broadening due to alloy disorder at cryogenic temperatures was also observed for other ternary alloys of class II such as MgZnO \cite{Grundmann2009}, AlGaAs \cite{Schubert1984}, CdZnTe \cite{G.NeuA.A.Mbaye1984}, and CdSSe \cite{Notes1978}.

Based on Eq.\,\ref{eq:Bernoulli}, the standard deviation of the bandgap energy $\sigma_{E_{g}}$ can be expressed by the binomial distribution \cite{Notes1978,Schubert1984}:

\begin{equation}
\label{eq:Broadening}
\sigma_{E_{g}} = \delta E_{g}/\delta x \sqrt{\left(\frac{x(1-x)}{K V_{X}}\right)}.
\end{equation}

Hence, for a two-particle system like $\mathrm{X_{A}}$, $\sigma_{E_{g}}$ should equal the peak width $w$ determined via $\Delta E$, while $\delta E_{g}/\delta x$ is deduced from Fig.\,\ref{Fig2}(a). As a first approach, $V_{X}$ can be replaced by the standard spherical exciton volume ($V_{X}^{s}=4/3 \pi r_{B}^{3}$) - a method that commonly yields a good agreement with experimental data \cite{Grundmann2009}. However, the quantum mechanical nature of the exciton is completely neglected by this approach, as one assumes a constant occupation probability density for the exciton over $V_{X}$. Instead, the quantum-mechanical (qm) exciton volume $V_{X}^{qm}=10 \pi r_{B}^{3}$ should rather be considered  in order to take the wavefunction of the exciton into account as outlined in Refs.\,\citen{Notes1978} and \citen{Zimmermann1990}.

Figure\,\ref{Fig3}(a) shows that both volumes ($V_{X}^{s}$ and $V_{X}^{qm}$) either lead to a complete over- (dashed red line) or underestimation (dotted red line) of the experimentally observed trend for $\mathrm{Si^{0}X_{A}}$ as the effect of the donor electron in this three-particle complex is not yet considered. The $\mathrm{X_{A}}$ complex as well as the donor electron sense the alloy formation, leading to energy fluctuations that are normally distributed. Hence, the square of the entire peak width of $\mathrm{Si^{0}X_{A}}$ amounts to: $w_{\mathrm{Si^{0}X_{A}}}^{2} = \lbrack w_{\mathrm{Si^{0}}}(V_{D}) \rbrack ^{2} +  \lbrack w_{\mathrm{X_{A}}}(V^{qm}_{\mathrm{X_A}}) \rbrack ^{2}$. Here, one obtains $V_{D}=10 \pi r_{D}^{3}$ based on Ref.\,\citen{G.NeuA.A.Mbaye1984} with the donor Bohr radius $r_{D}$ that is derived from the silicon donor binding energy of 28.5\,meV \cite{Callsen2018} calculated within the framework of the hydrogenic model \cite{Gil2007}. This consideration of $\mathrm{Si^{0}X_{A}}$ as a three-particle complex leads to a reasonable agreement between experiment and theory [Fig.\,\ref{Fig3}(a), solid red line].

In contrast, for the two-particle complex $\mathrm{X_{A}}$ the consideration of $V_{X}^{qm}$ (dotted black line) still yields an overestimation for most of the experimentally observed $\Delta E_{X_{A}}$ values as the COM movement of the exciton would need to be considered at the onset of the alloy formation. However, even the sophisticated theoretical treatment for alloys of Zimmermann given in Ref. \citen{Zimmermann1990} deems the COM movement of the exciton a task for future work, which has - to the best of our knowledge - never been accomplished. As a direct consequence of the excitons' COM motion, the entire set of $\mathrm{X_{A}}$ complexes averages over a larger fraction of the alloy, hence, we suggest to consider an effective, spherical averaging volume $V_{a}=a\,V_{X}^{qm}$ with scaling parameter $a$ as a measure for the $\mathrm{X_{A}}$ COM movement. A fit to the experimental data from Fig.\,\ref{Fig3}(a) for $\Delta E_{X_{A}}$ yields $a \approx 3$ (solid black line), while the offset between experimental data points and the fit increases with $x$. This indicates that $V_{a}$ scales with $x$ as described in the following.

For $\mathrm{X_{A}}$, Eq.\,\ref{eq:Broadening} can be solved for $r_{a}$ using $V_{X}=V_{X}^{qm}=V_{a}(r_{a})/a$, yielding the indium content dependent averaging radius $r_{a}(x)$ of the corresponding volume as shown in Fig.\,\ref{Fig3}(b) (black circles). The value of $r_a$ is continuously decreasing with rising indium content as $\mathrm{X_{A}}$ complexes get increasingly localized by indium-related trapping sites. Ultimately, $r_a$ approaches $r_{B}$ as the free exciton transforms into a bound two-particle complex as shown in Fig.\,\ref{Fig3}(b) and illustrated in the corresponding inset. In contrast, a $\mathrm{D^{0}X_{A}}$ complex like $\mathrm{Si^{0}X_{A}}$ always remains a bound complex at the onset of alloy formation and should, as a first approximation, maintain a constant averaging volume. At the present onset of alloy formation, we do not expect any strong influence of $x$ on $r_{B}$, which is taken in Fig.\,\ref{Fig3}(b) as that of pure GaN. Singlets, doublets, and with rising $x$ even more extended complexes of indium atoms [see Fig.\,\ref{Fig1}(c)] should lead to such increasing localization of $\mathrm{X_{A}}$ with rising $x$. In this respect, the reported $r_{a}(x)$ values represent a measure of the $\mathrm{X_{A}}$ diffusion length. According to Bellaiche \textit{et al.} \cite{Bellaiche1999} the hole as a building block of the $\mathrm{X_{A}}$ complex can get localized even in the case of a single indium atom. However, the corresponding energy level is still hybridized with the valence band states of GaN, meaning that no additional states appear in the bandgap at the onset of alloy formation in agreement with our observations from Fig.\,\ref{Fig2} (no new emission lines appear with rising $x$) and our expectations for a class II alloy. This particular case is fundamentally different from that of conventional excitons bound to isoelectronic centers (cf. examples given in Sec.\,\ref{context}) mostly situated on anionic lattice sites \cite{Hopfield1966}, whose strong localization leads to the appearance of new energy levels in the bandgap.

Based on a simple, static approximation, the probability to capture $\mathrm{X_{A}}$ at such indium-related sites can be approximated by $\xi_{cap} \propto \tilde{r}/r_{a}$ with the average distance in between the indium atoms in the dilute alloy $\tilde{r} = (K x)^{-1/3}$. This approximation relies on the common assumption made for most III-V \cite{Callsen2012a} and II-VI \cite{Wagner2011} semiconductors that the exciton capture time $\tau_{cap}$ is short compared to the radiative decay time, $\tau_{cap} < \tau_{rad}$. Consequently, based on the spherical averaging volume $V_{a}(r_{a})$ one straightforwardly finds the following relation to fit the data from Fig.\,\ref{Fig3}(b): $r_{a} \propto \xi_{cap}^{-1} (K x)^{-1/3}$. Clearly, only for $\xi_{cap}\ll1$ it is possible to obtain a reasonable fit to the experimental data as shown in Fig.\,\ref{Fig3}(b) by the solid red line. Hence, indeed on average a single indium atom per $V_{a}$ does not seem to provide a sufficiently deep potential in order to permanently capture the $\mathrm{X_{A}}$ complex at the given temperature \cite{Bellaiche1999}. This observation represents the main result of the present simple fitting model. For instance, already for $\xi_{cap} = 50\%$ the values for $r_{a}$ would rapidly diminish with rising $x$. This evolution is shown in Fig.\,\ref{Fig3}(b) (solid grey line) and constitutes the reason behind the common neglect of the excitons' COM movement in alloys \cite{Zimmermann1990}, which is not highly diluted, or in other words, just doped. 

The fit from Fig.\,\ref{Fig3}(b) yields $\xi_{cap} \approx 8.7 \pm 1.2\%$ (solid red line), suggesting that indium-related complexes comprising $\approx$\,10 indium atoms predominantly contribute to the capture of $\mathrm{X_{A}}$ in the InGaN alloy at 12\,K. Clearly, the occurrence of such complexes becomes more likely with rising $x$, explaining the reduction of $r_{a}$ towards the $r_{B}$ value of GaN in Fig.\,\ref{Fig3}(b), while $V_{a}$ also approaches $V_{X}^{qm}$. Hence, around 2.4\% of indium are required to form an InGaN alloy for which the neglect of the excitons' COM movement is justified at cryogenic temperatures. Therefore, the difference in between the data points for $\Delta E_{X_{A}}$ and the corresponding fit in Fig.\,\ref{Fig3}(a) (solid black line) starting from $x \approx 1.36\%$ is likely caused by the increasing localization of $\mathrm{X_{A}}$, i.e., the transition from a free to a bound two-particle complex in a class II alloy.

This transitional regime of increasing excitonic localization at 12\,K by indium-related complexes is challenging from a theoretical point of view as exciton diffusion needs to be considered. In this context, the present simplistic derivation of $\xi_{cap}$ just represents a pragmatic approach that seems well suited, given the error and the scatter of the underlying data points, cf. Fig.\,\ref{Fig3}(b). Furthermore, we plausibly assume that our findings are related to the indium zig-zag chains \cite{Kent2001a} described by Chichibu \textit{et al.} \cite{Chichibu2006a}, however, the precise sub-structure at the atomic scale still remains of speculative nature. In this regard, Sec. \ref{micro} and \ref{indirect} will describe a pathway towards an analysis of the particular configuration of indium atoms by spectroscopic means at the onset of alloy formation. In the following we will always refer to such dilute indium assemblies, because larger, non-dilute indium aggregates should lead to the formation of electronic states in the bandgap \cite{Kent2001a} [i.e., quantum dot (QD) like states]. 

We suggest that the apparent localization of $\mathrm{X_{A}}$ complexes with rising $x$ enhances the robustness of the InGaN alloy against point and structural defects in contrast to other III-V alloys as outlined by Chichibu \textit{et al.} in Ref.\,\citen{Chichibu2006}. Such a picture also accounts for the intensity increase of $\mathrm{X_{A}}$ relative to $\mathrm{Si^{0}X_{A}}$ as shown in Fig.\,\ref{Fig2}. As exciton diffusion is inhibited with rising $x$, the probability to reach non-radiative exciton trapping sites is reduced. In this respect, we wish also to note that the overall intensity of the bandedge luminescence shown in Fig.\,\ref{Fig2} continuously increases with rising indium content at the onset of alloy formation (not shown) as frequently reported in the literature \cite{Shu1998a,Shen1999,Shen1999a,Kumano2001}. Naturally, with rising $x$ the trapping potential becomes deeper as larger indium assemblies are formed that ultimately even govern the $\bm{k}$-distribution of $\mathrm{X_{A}}$ at a temperature of 300\,K with localization energies already in excess of $k_{B}T$ at $x \approx 3\,\%$ \cite{Butte2018a}. In addition, larger indium assemblies can not only lead to exciton localization but also to confinement, which commonly boosts excitonic decay rates.  Such an inhibited exciton diffusion should also be relevant for the impact of threading dislocations, which could directly be analyzed by means of, e.g., CL \cite{Liu2016a} in bulk In$_x$Ga$_{1-x}$N epilayers with rising $x$. However, the data presented in Fig.\,\ref{Fig3}(b) is of a more general nature as the trapping of excitons is monitored over a large PL excitation area (excitation spot diameter of $\approx 100\,\mu \text{m}$) and does not depend on the specific strain-driven impurity distribution around a structural defect \cite{Hayama2017}. Obviously, such increasing exciton localization with rising $x$ is less beneficial for alloys like In$_x$Ga$_{1-x}$As as already much larger $r_B$ values prevail in the related binary compounds as indicated in Fig.\,\ref{Fig1}(a).

\subsection{Analysis of the local indium distribution around impurity bound excitons}
\label{local}
\begin{figure}[]
\includegraphics[width=8cm]{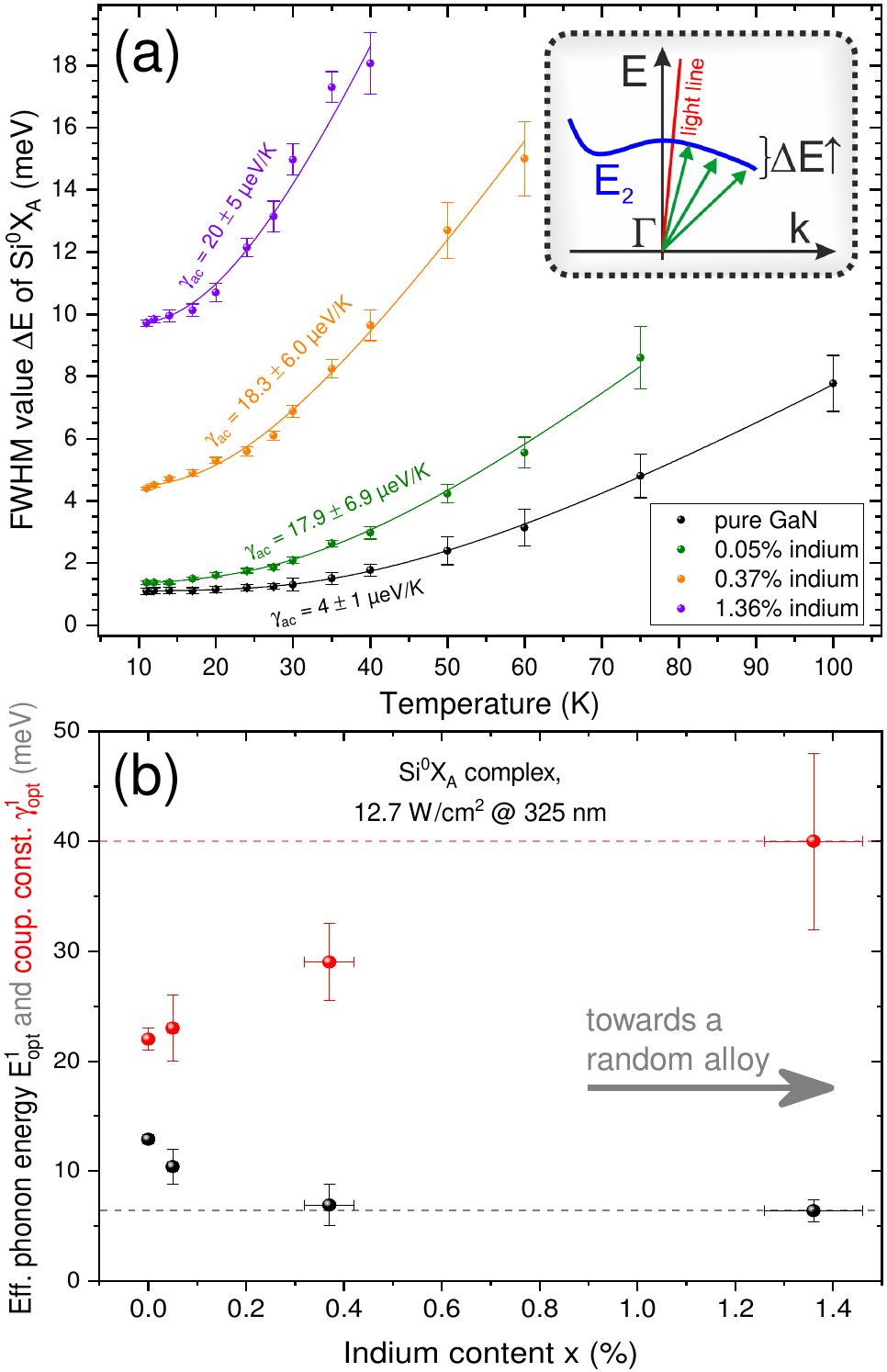}
\caption{(color online) (a) Temperature dependence of the linewidth broadening for the $\mathrm{Si^{0}X_{A}}$ complex extracted from PL spectra. Solid lines show fits to the data based on Eq.\,\ref{eq:T}. The value of the fitting parameter $\gamma_{ac}$ is inscribed, whereas the particular scaling behavior of the fitting parameters $E_{opt}^{1}$ and $\gamma^{1}_{opt}$ is shown in (b). Here, the values for $E_{opt}^{1}$ and $\gamma^{1}_{opt}$ seem to converge with rising $x$ (dashed lines). The temperature dependence of the linewidth broadening for $\mathrm{Si^{0}X_{A}}$ is strongly affected by the indium content in the alloy $x$, in contrast to the case of $\mathrm{X_{A}}$ (cf. Fig.\,S3). An optical phonon population with an effective energy $E_{opt}^1$ (weighted by the corresponding density of states) is responsible for the particular broadening of $\mathrm{Si^{0}X_{A}}$. The maximum energy of phonons within this phonon population corresponds to the energy of the $E_{2}^{low}$ phonon in GaN ($17.85 \pm 0.05\,\text{meV}$). The inset of (a) sketches the $E_2$ phonon dispersion relation in the first Brillouin zone near the $\Gamma$-point, cf. Sec\,\ref{local}.}
\label{Fig4}
\end{figure}

The transition from a free to a bound excitonic two-particle complex summarized in Fig.\,\ref{Fig3}(b) represents a spectroscopic probe, whose sensitivity to the local environment in the alloy increases with rising indium content as $r_a$ converges to $r_B$. In contrast, an impurity bound three-particle complex always remains localized, giving rise to a permanently local alloy probe as illustrated in the inset of Fig.\,\ref{Fig3}(b).

In the following, we will show that the impurity bound exciton complex $\mathrm{Si^{0}X_{A}}$ also provides an extremely versatile tool to study the formation of dilute indium assemblies in GaN:In. While the temperature-dependent linewidth of $\mathrm{X_{A}}$ does not show any clear indium content dependent trend (see Fig.\,S3), the broadening of $\mathrm{Si^{0}X_{A}}$ upon rising temperature increases with $x$ as shown in Fig.\,\ref{Fig4}(a) - a quite unexpected finding in the light of previous alloy studies \cite{Notes1978}. In addition to the data shown in Fig.\,\ref{Fig4}, SI Secs.\,II and III provide an overview about the underlying, temperature-dependent PL spectra and the corresponding analysis of peak positions and linewidths.

Generally, the phonon-induced linewidth broadening $\Gamma(T)$ is given by

\begin{equation}
\label{eq:T}
\Gamma(T) = \gamma_{ac} T + \sum_{i=1}^2 \frac{\gamma_{opt}^{i}}{\text{exp}(E_{opt}^{i}/k_{B}T)-1},
\end{equation}

with the acoustical ($\gamma_{ac}$) and the optical ($\gamma_{opt}^{i}$) phonon coupling constants as well as the corresponding effective optical phonon energies ($E_{opt}^{i}$) numbered by $i$. The best fit to the evolution of $\Delta E(T)$ shown in Fig.\,\ref{Fig4}(a) for $\mathrm{Si^{0}X_{A}}$ at $x=0$ (solid black line) is obtained for $\gamma_{ac} = 4 \pm 1 \, \mu \text{eV/K}$ and one set of $\gamma_{opt}^{1}$, $E_{opt}^{1}$ values as summarized in Fig.\,\ref{Fig4}(b). Here, $E_{opt}^{1}=12.9 \pm 0.4\,\text{meV}$ is smaller than the energy of the $E_{2}^{low}$ phonon mode in GaN of $17.85 \pm 0.05\,\text{meV}$ \cite{Callsen2011}. This difference can be explained by the curvature of the corresponding phonon dispersion relation $E(k)$ around $k_{0} \approx 0$ with $\partial^2 E / \partial^2 k|_{k=k_0}<0$ as illustrated in the inset of Fig.\,\ref{Fig4}(a). The exciton-phonon coupling related to $\mathrm{Si^{0}X_{A}}$ probes an extended phonon energy interval due to the localization of the bound exciton in real space, providing access to a larger fraction of the first Brillouin zone near the $\Gamma$-point. Hence, the temperature-dependent PL data yields an effective phonon energy $E_{opt}^{1}$ that is weighted by the corresponding one-phonon density of states \cite{Davydov1998a,Passler2001,Song2006}.

For $\mathrm{Si^{0}X_{A}}$ the effective phonon energy $E_{opt}^{1}$ gradually decreases from $12.9 \pm 0.4\,\text{meV}$ to $6.4 \pm 1.0\,\text{meV}$ for ${0 \leq x \leq 1.36\%}$ as depicted in Fig.\,\ref{Fig4}(b). The corresponding $\gamma_{ac}$ fitting parameters denoted in Fig.\,\ref{Fig4}(a) approach values equal to the ones observed for $\mathrm{X_{A}}$ within the given error bars with rising $x$ (see Fig.\,S3 and the corresponding discussion in SI Sec.\,III). The observed trend for $E_{opt}^{1}$ is likely originating from an alloying-induced local variation of the $\mathrm{Si^{0}}$ environment. Callsen \textit{et al.} have found for single excitons trapped in GaN/AlN QDs that, e.g., the exciton-LO-phonon interaction averages over a volume with a radius on the order of $r_{B}$ \cite{Callsen2015d}. Hence, alloying in thick InGaN epilayers can not only \textit{directly} be monitored by emission energy shifts (see Fig.\,\ref{Fig2}) caused by the immediate indium atom configuration in the vicinity of $\mathrm{Si^{0}}$ impurities, which in total leads to the formation of bound excitons. More \textit{indirectly}, the interaction of bound excitons with $E_{2}^{low}$ phonons can also be used as an alloy probe as shown in Fig.\,\ref{Fig4}. 


As $\Gamma(T)$ for $\mathrm{X_{A}}$ does not show any pronounced indium content dependence (see Fig.\,S3), we suggest that the environment of $\mathrm{Si^{0}}$ centers is richer in indium atoms, a feature only noticeable at the very onset of alloy formation. Thus, here it can be expected that the phonon energies and coupling constants (e.g., $E_{opt}^{1}$ and $\gamma_{opt}^{1}$) will rapidly converge with rising $x$. Any local variation in the indium content caused by the particular distribution of $\mathrm{Si^{0}}$ atoms will get increasingly masked by random alloy fluctuations with rising $x$. As a signature of this masking, $E_{opt}^{1}(x)$ converges towards $\approx\,6\,\text{meV}$ as shown in Fig.\,\ref{Fig4}(b). The optical phonon coupling constant $\gamma_{opt}^{1}$ continuously rises with $x$, but exhibits an opposite convergence behavior in Fig.\,\ref{Fig4}(b) at the onset of alloy formation. Indium atoms that surround the $\mathrm{Si^{0}}$ center within $r_B$ can act as strongly localized isoelectronic centers, whose broad combined distribution in $\bm{k}$-space seems to foster the interaction with phonons that deviate from the Brillouin zone center. Therefore, the aforementioned bowing of the associated phonon dispersion relation ($\partial^2 E / \partial^2 k|_{k=k_0}<0$) leads to an effective reduction of the measured $E_{opt}^{1}$ values that is accompanied by a rise in $\gamma_{opt}^{1}$. Clearly, the total reduction in the effective phonon energy $E_{opt}^{1}$ cannot exclusively be explained by a local rise in indium content as even pure InN still exhibits an $E_{2}^{low}$ phonon energy of around $10.9\pm 0.1\,\text{meV}$ \cite{Davydov1999}. In addition, non-dilute indium assemblies matching the size of $V_X$ could have been detected by APT in state-of-the-art InGaN/GaN QW samples \cite{Cerezo2007,Galtrey2007} commonly comprising silicon concentrations $\approx 1 \times 10^{16}\,\text{cm}^{-3}$ (cf. SI Sec.\,I). Thus, as a possible reason for the local indium enrichment in the vicinity of $\mathrm{Si^{0}}$ centers, we suggest tensile strain that is commonly introduced upon silicon doping of GaN \cite{Nenstiel2015a}. Hence, it is energetically more favorable for a rather large atom like indium to incorporate close to a silicon atom (distance on the order of $r_B$), in order to approach the lattice's strain equilibrium at the onset of the alloy formation. Bezyazychnaya \textit{et al.} have theoretically predicted the impact of point defects (vacancies) in InGaN and InGaAs on the indium distribution in these alloys \cite{Bezyazychnaya2015}. Here, we experimentally find - to a certain degree - a similar situation for the $\mathrm{Si^{0}}$ impurity. Future theoretical work is needed to validate this picture of a point defect embedded in a dilute assembly of indium atoms. This image differs from the common concept of direct complex formation \cite{Liu2016} that often just considers nearest- or second-nearest-neighbor sites on the cationic sub-lattice.

\subsection{Individual bound excitonic complexes analyzed by micro-photoluminescence}
\label{micro}

\begin{figure}[h!]
\includegraphics[width=8cm]{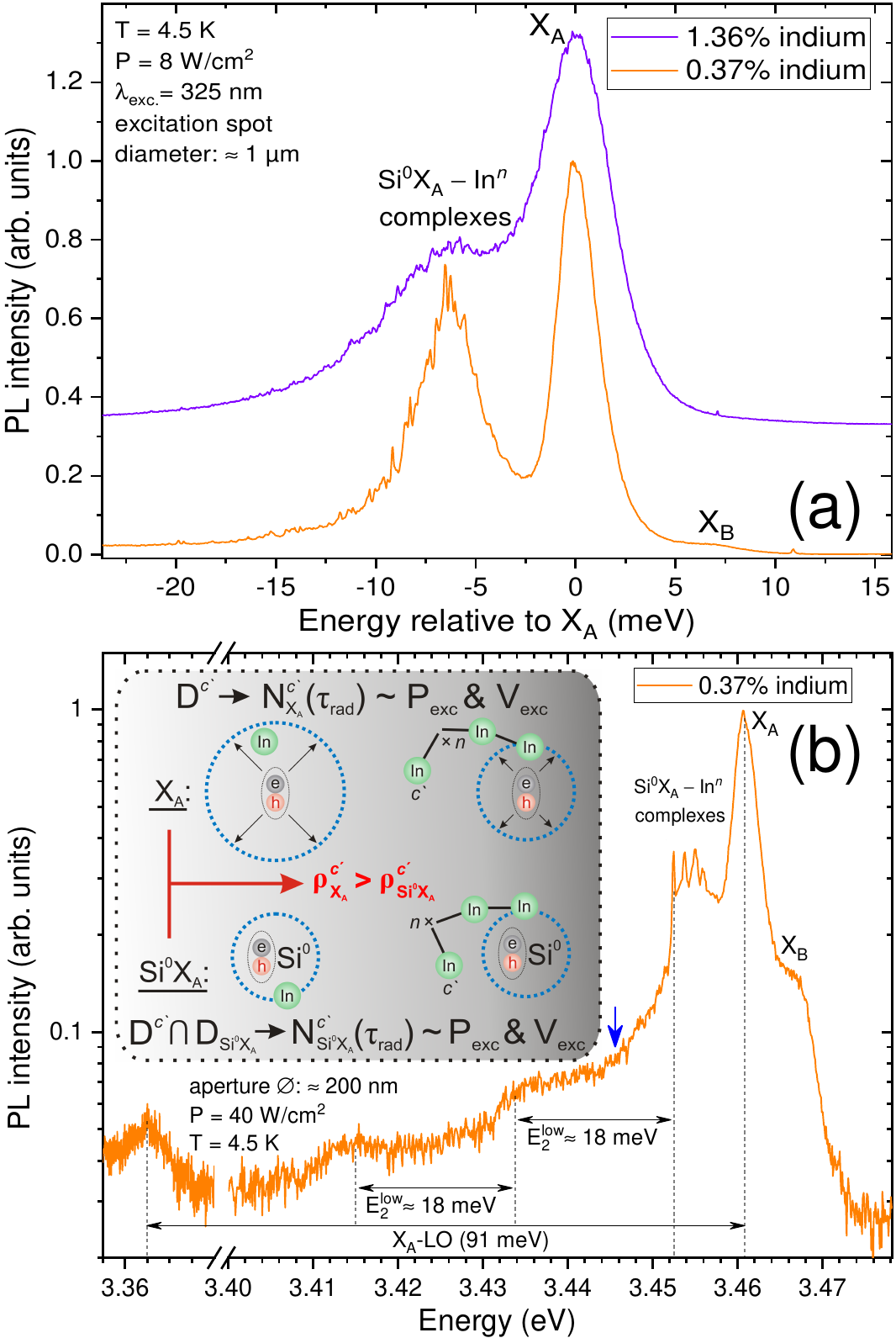} 
\caption{(color online) (a) $\mu$-PL spectra of two samples containing 0.37\% and 1.36\% of indium, respectively, show a structured spectrum  in the energy range of $\mathrm{Si^{0}X_{A}}$, while the emission band related to $\mathrm{X_{A}}$ remains unstructured (spectra are shown on a relative energy scale for better comparison). Both spectra are typical for the entire, bare surface area of our samples. The density of sharp emission lines noticeable around $\mathrm{Si^{0}X_{A}}$ increases with rising $x$. (b) At an indium concentration of ${x = 0.37\%}$ and a temperature of 4.5\,K the $\mathrm{X_{A}}$ and $\mathrm{X_{B}}$ transitions still appear as unstructured bands even if measured under a metal aperture with a diameter of 200\,nm by means of $\mu$-PL. In contrast, the $\mathrm{Si^{0}X_{A}}$ emission band consists of sharp emission lines with FWHM values down to $\approx$\,300\,$\mu$eV. At the given low excitation density ($\sim 40\,\text{W/cm}^2$), several phonon replicas are noticeable [$E^{low}_2$ and longitudinal-optical (LO) phonon replica]. Here, the blue arrow indicates a particular phonon energy $E_{opt}^1 =\,6.9\,\text{meV}$ as extracted in Fig.\,\ref{Fig4}(b) for $x$\,=\,0.37\%. The inset illustrates why $\mathrm{X_{A}}$ exhibits an unstructured emission band, whereas the luminescence around $\mathrm{Si^{0}X_{A}}$ [cf. Fig.\,\ref{Fig2}(b)] shows several sharp emission lines related to individual complexes comprising the $\mathrm{Si^{0}}$ center and specific indium assemblies ($\mathrm{In^{\textit{n}}}$) with a certain atomic configuration $c'$ comprising $n$ indium atoms.}
\label{Fig5}
\end{figure}

Fortunately, $\mu$-PL measurements with an excitation spot diameter $\approx 1\,\mu \text{m}$ can  provide more detailed information regarding the particular configurations of indium atoms close to a $\mathrm{Si^{0}}$ donor at the onset of alloy formation. Even on completely unprocessed samples one can resolve individual emission lines around the $\mathrm{Si^{0}X_{A}}$ transition for $x=0.37\%$ and partially for 1.36\% as show in Fig.\,\ref{Fig5}(a), while the $\mathrm{X_{A}}$ emission remains a rather unstructured emission band due to the given probe volume and excitation power density ($P=40\,\text{W/cm}^2$). This observation is of high importance as the subsequently described processing of metal apertures on the sample could lead to exciton localization by any damage to the sample as discussed in SI Sec.\,IV.

\begin{SCfigure*}[][h!]
\includegraphics[width=11cm]{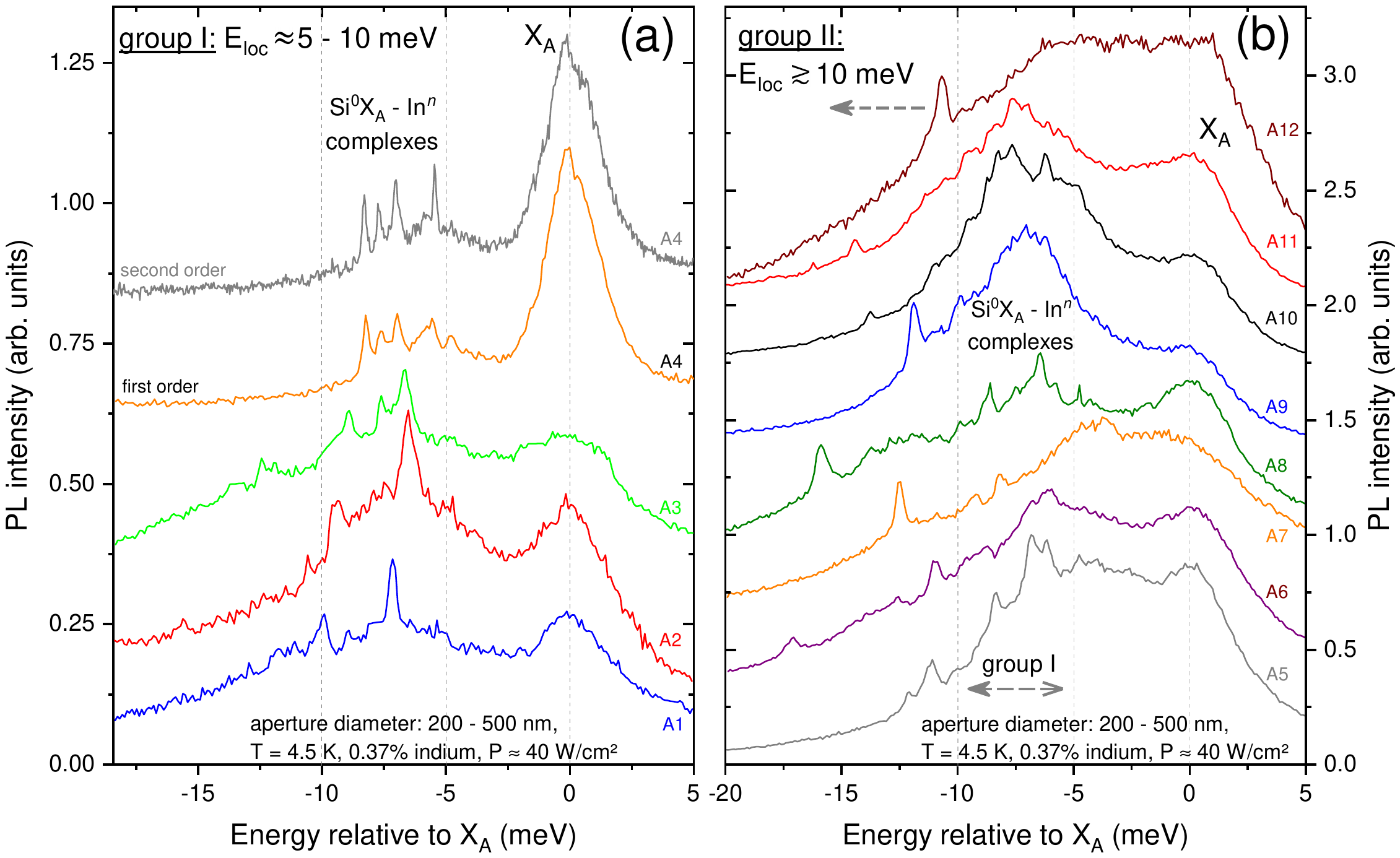}
\caption{(color online) The sharp emission lines introduced in Fig.\,\ref{Fig5} appear due to the formation of dilute $\mathrm{Si^{0}X_{A}-In^{\textit{n}}}$ assemblies and can be divided in two major groups I and II based on $\mu$-PL spectra. (a) Emission lines of group I exhibit localization energies $E_{loc}$ of $\approx$\,5\,-\,10\,meV and show small energetic shifts upon increasing excitation power density. (b) In contrast, emission lines of group II typically show $E_{loc}$ values in excess of $10\,\text{meV}$ and exhibit pronounced spectral shifts for any variation of the excitation power density. This difference in the peak shift rates for group I and II emission lines is highlighted in Fig.\,\ref{Fig7}. Generally, the emission lines of group I appear more frequently than their more deeply localized group II counterparts. A1 - A12 denote the particular metal apertures selected from more than 50.}
\label{Fig6}
\end{SCfigure*}

We further reduced the size of the probe volume by processing apertures into an aluminum film with diameters down to $200\,\text{nm}$ on the sample with $x=0.37\%$. As a result, a set of spectrally well separated emission lines can be resolved around $\mathrm{Si^{0}X_{A}}$ along with particular phonon sidebands as shown in Fig.\,\ref{Fig5}(b). An equally spaced ($\approx$\,18\,meV) double step appears on the low energy side of the set of sharp emission lines due to coupling with $E_{2}^{low}$ phonons. Interestingly, the optical signature of the exciton-phonon coupling does not appear as a commonly observed \cite{Callsen2012a} isolated peak [see the LO-replica of $\mathrm{X_{A}}$ in Fig.\,\ref{Fig5}(b)], but as a step due to the energetically broad range of contributing phonons with $E_{2}$ symmetry. About 7\,meV below the first step related to the zero phonon lines in the spectrum of Fig.\,\ref{Fig5}(b) (marked by a blue arrow) one even observes an additional, step-like increase in intensity and hence phonon coupling strength in accordance with the findings of Fig.\,\ref{Fig4}(b) ($E_{opt}^{1}=6.9\,\text{meV}$ for $x=0.37\%$). Hence, the $\mu$-PL spectrum shown in Fig.\,\ref{Fig5}(b) represents a coarse probe of the phonon density of states that stands in direct relation to the homogeneous broadening $\Gamma(T)$ of the $\mathrm{Si^{0}X_{A}}$ complex introduced in Fig.\,\ref{Fig4}(a).

The sharp emission lines from Fig.\,\ref{Fig5}(b)  present the most direct evidence for an alloying-induced perturbation of the immediate $\mathrm{Si^{0}}$ environment at $x=0.37\%$. Here, emission lines with a linewidth of just 320\,$\pm$\,30\,$\mu$eV can be resolved, opening the perspective for a spectroscopic study of the $\mathrm{Si^{0}}$ environment at the nanometer scale based on the emission of such dilute $\mathrm{Si^{0}X_{A}-In^{\textit{n}}}$ assemblies. In contrast to $\mathrm{Si^{0}X_{A}}$, the $\mathrm{X_{A}}$ emission remains an unstructured band in our $\mu$-PL spectra as the density $\rho^{\,c'}_{\mathrm{X_A}}$ of $\mathrm{X_{A}}$ that attach to a specific indium atom configuration $c'$ with $n$ indium atoms is comparably larger than the corresponding density of $\mathrm{Si^{0}X_{A}}$ centers $\rho^{\,c'}_{\mathrm{Si^{0}X_{A}}}$ with an identical configuration of indium atoms in their vicinity. 

Generally, the number of \textit{any} (transition from ${c'\,\rightarrow\,c}$) per time interval decaying, indium-related complex $N_{\mathrm{X_{A} , Si^{0}X_{A}}}^{\,c}$ is proportional to the $\mu$-PL excitation volume as well as pump power ($V_{exc}$ and $P_{exc}$) and depends on the corresponding radiative decay time $\tau_{rad}$. However, for the case of impurity bound excitons, as illustrated in the inset of Fig.\,\ref{Fig5}(b), the intersecting set of a certain distribution of exemplary indium configurations (e.g., $D^{c'}$) and $\mathrm{Si^{0}X_{A}}$ centers ($D_\mathrm{Si^{0}X_{A}}$) defines the total number of experimentally accessible complexes ($N$) via $D^{c'} \cap D_\mathrm{Si^{0}X_{A}} \rightarrow N^{c'}_\mathrm{Si^{0}X_{A}}$. However, in this simplified description no varying distances are considered in between these sites. Here, $r_B$ could represent a meaningful upper limit for the distance in between a certain indium configuration and the $\mathrm{Si^{0}}$ center. Our experimental findings from Sec.\,\ref{indirect} will further elucidate this point.

Hence, only the $\mathrm{Si^{0}X_{A}}$ emission band falls apart into individual emission lines in the $\mu$-PL spectra of Fig.\,\ref{Fig5}, while a similar observation for $\mathrm{X_{A}}$ would require an even lower excitation density or a smaller probe volume. For instance, the impact of individual indium atoms on the emission of $\mathrm{Si^{0}X_{A}}$ could possibly be resolved for smaller indium concentrations (e.g., $x=0.01\%$). However, the temperature-induced emission line shifts shown in Fig.\,S1 and summarized in Fig.\,S2 (comparison in between $x=0,\,0.37\%,\,\text{and}\, 1.36\%$) suggest corresponding $E_{loc}$ values $\lesssim 0.5\,\text{meV}$. Hence, the emission energy difference in between, e.g., $\mathrm{Si^{0}X_{A}}$ and $\mathrm{Si^{0}X_{A}-In^{\textit{n}}}$ centers at $x=0.01\%$ cannot straightforwardly be resolved with the present emission linewidths in our samples ($\approx300\,\mu$eV). So far the impact of individual indium atoms can only be monitored by the localization onset of $\mathrm{X_A}$ as demonstrated in Figs.\,\ref{Fig3}(a) and \ref{Fig3}(b) based on PL measurements.


%
\subsection{Statistical analysis of individual bound excitonic complexes}
\label{indirect}

The observation of sharp emission lines in Fig.\,\ref{Fig5} caused by dilute $\mathrm{Si^{0}X_{A}-In^{\textit{n}}}$ assemblies directly evokes the need for a statistical analysis of the underlying emitters. Figure\,\ref{Fig6}(a) shows selected $\mu$-PL spectra recorded for four different metal apertures labeled A1-A4. Here, the two spectra for aperture A4 (measured using the first and second order of the optical grating) provide a detailed view of the $\mu$-PL spectrum shown in Fig.\,\ref{Fig5}(b). All spectra in Fig.\,\ref{Fig6}(a) show a common optical signature of sharp emission lines related to dilute $\mathrm{Si^{0}X_{A}-In^{\textit{n}}}$ assemblies in addition to a rather unstructured emission band related to $\mathrm{X_{A}}$. For the latter case, a smaller spectroscopic probe would be required in order to resolve any spectral details. The localization energy $E_{loc}$ (energetic spacing between $\mathrm{X_{A}}$ and the $\mathrm{Si^{0}X_{A}-In^{\textit{n}}}$ complexes) commonly ranges in between $\approx 5-10\,\text{meV}$. Figure\,\ref{Fig6}(b) introduces eight additional $\mu$-PL spectra (A5-A12) that also show emission lines with $E_{loc} \gtrsim 10\,\text{meV}$ that we assign to group II. Generally, the emission lines of group I occur more frequently than their group II counterparts. While almost all spectra in Fig.\,\ref{Fig6} show emission lines $\approx 5-10\,\text{meV}$ below $\mathrm{X_{A}}$ (partially overlapping likely due to a high density of the associated excitonic complexes), only a few apertures show a clear signature of the emission lines of group II. Here, apertures A5-A12 illustrate the rare cases of these group II emission lines that we extracted from $\mu$-PL measurements on more than $50$ apertures.

\begin{figure}[]
\includegraphics[width=8cm]{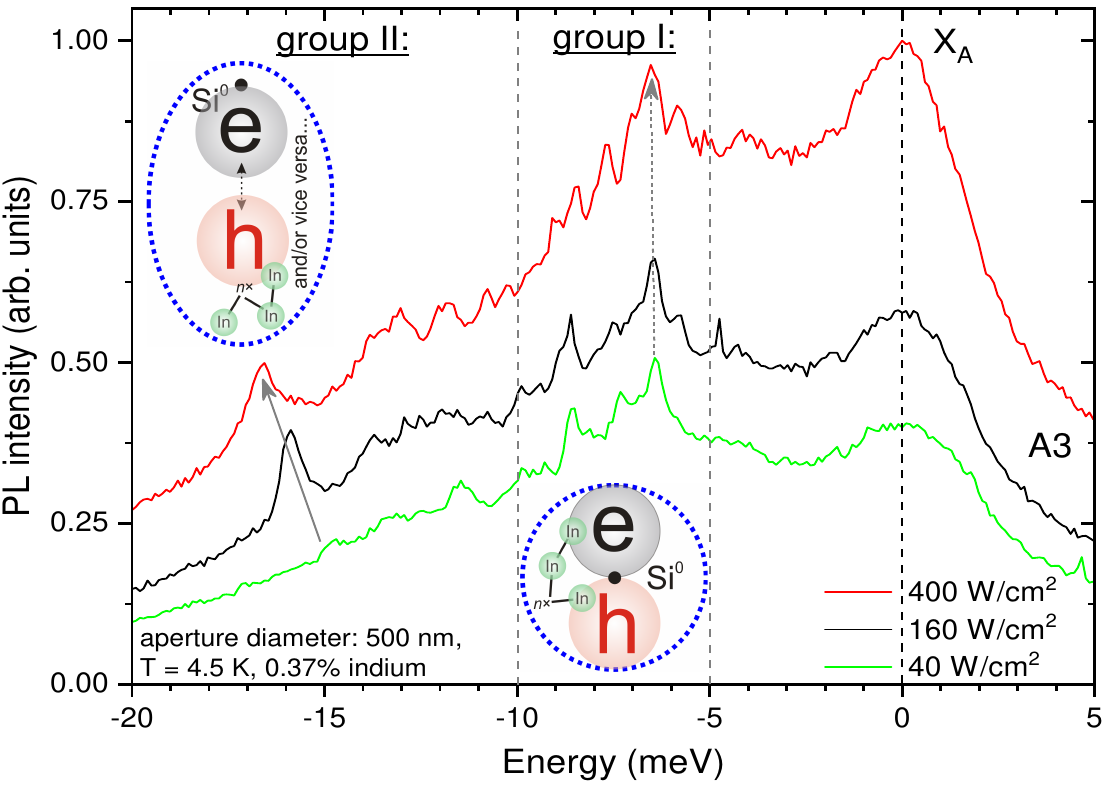}
\caption{(color online) Excitation power dependent $\mu$-PL measurements for the third aperture (A3) from Fig.\,\ref{Fig6}(a). Emission lines of group I only exhibit shifts on the order of $100\,\mu \text{eV}$ upon a tenfold increase in excitation power (see dashed gray arrow), while the distinct emission line of group II shifts by $\approx 2\,\text{meV}$ (see solid gray arrow). Similar observations hold for all apertures analyzed in Fig.\,\ref{Fig6}. The drawings depict possible configuration for the excitonic complexes belonging to group I or II.}
\label{Fig7}
\end{figure}

The classification of the sharp emission lines into two groups, I and II, becomes clearer based on Fig.\,\ref{Fig7}, showing excitation power dependent $\mu$-PL spectra for aperture A3. This excitation power series shows a sequence of spectra that is typical for all apertures showing emission lines of groups I and II. When varying the excitation power density by a factor of ten, group I emission lines only exhibit minor energetic shifts on the order of $\approx 100\,\mu\text{eV}$. In contrast, the emission line of group II from Fig.\,\ref{Fig7} shifts by $\approx\,2\,\text{meV}$ towards lower energies.

The emission lines of group I seem to behave like common bound excitons in GaN upon rising excitation power that do not exhibit any pronounced energetic shifts \cite{Santic1998}. The effects of bandgap renormalization and band filling almost perfectly compensate each other over a wide excitation range in GaN, hence, only negligible shifts of bound excitons are noticeable \cite{Reynolds2000a}. At the given low indium concentration of 0.37\% (cf. Figs.\,\ref{Fig6} and \ref{Fig7}) we do not expect any strong deviation from this common behavior of bulk GaN. Hence, we assume that the emission lines of group I belong to $\mathrm{Si^{0}X_{A}-In^{\textit{n}}}$ complexes comprising a unique configuration of indium atoms within the Bohr radius of the exciton $r_B$. In this respect, the two electrons and the single hole of this complex occupy a well-defined region and can therefore be referred to as spatially direct, neutral, complex bound excitons as sketched in Fig.\,\ref{Fig7} for group I. Here in Fig.\,\ref{Fig7} an exemplary dilute assembly of $n$ indium atoms is sketched, along with the $\mathrm{Si^{0}}$ center and the most relevant charge carriers. 



The pronounced redshift of the emission lines belonging to group II points towards a gradual change in the effective Coulomb interaction in between the charge particles that contribute to the overall $\mathrm{Si^{0}X_{A}-In^{\textit{n}}}$ emission. A spatially indirect transition as drawn in Fig.\,\ref{Fig7} for group II could explain such a particular sensitivity to the excitation power as well as the larger $E_{loc}$ values with respect to the emission lines of group I. Clearly, with rising $x$ such spatially indirect transitions involving $\mathrm{In^n}$ complexes with increasing size and density become more probable. At first glance, the observation of group II emitters resembles the common picture of excitons trapped in $c$-plane (In)GaN QDs that are embedded in, e.g., Al$_{x}$Ga$_{1-x}$N (0\,$\leq$\,$x$\,$\leq$\,1) as the matrix material \cite{Kindel2010}. Charge fluctuations commonly occur in the matrix material due to point defects, which in-turn leads to a pronounced linewidth broadening known to be particularly strong in III-nitrides \cite{Kindel2014}. Upon changing the laser excitation power, the occupation probability of charges in the vicinity of such QDs is altered. As a result of this statistical process, a pronounced red-shift of single QD emission lines can be observed with rising excitation power \cite{Kindel2010,Holmes2015}, which resembles the emission line shift of the group II emitter from Fig.\,\ref{Fig7}. Hence, the pronounced energetic shift of group II emission lines upon varying excitation power could originate from a spectral diffusion phenomenon, which scales with the size of the excitonic dipole moment \cite{Kindel2014}. Clearly, indirect excitonic transitions as sketched in Fig.\,\ref{Fig7} exhibit a larger excitonic dipole moment and are consequently more sensitive to charge fluctuations. Future work should analyze if any preferential orientation exists for this excitonic dipole. However, currently, a more detailed analysis of these group II emitters is hindered by the comparably long integration times for spectra on the order of tens of minutes and the spectral overlap with other emitters, preventing a detailed linewidths and emission line jitter analysis.



%
\section{Discussion}
\label{discussion}
More detailed $\mu$-PL measurements need to be performed based on further optimized samples in order to reach a conclusive picture regarding the sharp emission lines shown in Figs.\,\ref{Fig5}\,-\,\ref{Fig7}. In addition, future temperature-dependent $\mu$-PL measurements should provide further insights. So far we only observed that the emission lines summarized in Fig.\,\ref{Fig6} rapidly broaden and thermalize (traceable up to $\sim\,30\,\text{K}$) due to the associated exciton-phonon coupling involving the $E_{2}^{low}$ mode and acoustic phonons as described for Figs.\,\ref{Fig4} and \ref{Fig5} (not shown). The distinction made between emission lines of group I and II is mainly motivated by experiments and helps to access the mechanisms of alloying. Generally, a continuous transition can be expected in between these two groups of emission lines, e.g., with rising $x$ the probability for spatially indirect transitions should first increase in the dilute alloy limit.

The appearance of spatially direct and indirect $\mathrm{Si^{0}X_{A}-In^{\textit{n}}}$ recombinations is likely interlinked with the $x$-dependence of the homogeneous linewidth broadening $\Gamma(T,x)$ (see Eq.\,\ref{eq:T}) shown in Fig.\,\ref{Fig4}(a). Spatially more indirect transitions exhibit larger excitonic dipole moments that would enable a strong coupling to, e.g., polar phonons like LO-phonons via the Fr\"ohlich interaction \cite{Callsen2015}. However, the limited $E_{loc}$ values of $\mathrm{Si^{0}X_{A}-In^{\textit{n}}}$ render the contribution of the exciton-LO-phonon coupling negligible in the present temperature range (due to the large LO-phonon energies in nitrides \cite{Davydov1999,Callsen2011}) and the coupling to acoustic and non-polar, e.g., $E_{2}^{low}$ phonons becomes most relevant as introduced in the context of Fig.\,\ref{Fig4}. The associated exciton-phonon interaction is dominated by the deformation potential coupling, which does not depend on the excitonic dipole moment. In addition, a less prominent piezoelectric coupling will occur \cite{Ostapenko2012}. Hence, we can expect $\mathrm{Si^{0}X_{A}-In^{\textit{n}}}$ complexes with large $E_{loc}$ values (group II) to be more temperature-stable \cite{Meyer2010a} than their group I counterparts as long as a dominant contribution of LO-phonons can be excluded. Therefore, with rising temperature the overall emission band related to $\mathrm{Si^{0}X_{A}-In^{\textit{n}}}$ should first become increasingly influenced by complexes with effectively larger $E_{loc}$ values (group II), while the more spatially direct recombination channels (group I) should preferentially dissociate. However, already at 12\,K the emitters of group II seem to increasingly contribute to the linewidth broadening of $\mathrm{Si^{0}X_{A}}$ with rising $x$. Thus, the offset in between the associated experimental $\Delta E$ values and the model (solid red line) increases with $x$ as shown in Fig.\,\ref{Fig3}(a). Finally, the evolution of the weighting in between spatially direct and indirect $\mathrm{Si^{0}X_{A}-In^{\textit{n}}}$ centers is likely involved in the indium content dependence of $\Gamma(T,x)$ at the onset of the alloy formation. Finally, we suggest a twofold reasoning for the particular evolution of $\Gamma(T,x)$: At the onset of alloy formation one can observe a local indium enrichment in the vicinity of $\mathrm{Si^{0}}$ centers that affects the exciton-phonon coupling. In addition, spatially direct and indirect $\mathrm{Si^{0}X_{A}-In^{\textit{n}}}$ transitions exhibit different thermalization behaviors, contributing to the particular evolution of $\Gamma(T,x)$.


%
\section{Conclusions}
\label{conclusions}

In summary, we have demonstrated the detailed spectroscopic analysis of a III-V mixed crystal alloy by PL and $\mu$-PL constituting a macro- and even microscopic material characterization. Thus, the present work on a class II alloy approaches the high level of spectroscopic sophistication known for class I alloys that relies on the emission of excitons bound to isoelectronic centers. As no such strongly localized excitons appear in the investigated In$_x$Ga$_{1-x}$N epilayers (0\,$\leq$\,$x$\,$\leq$\,2.4\%), we utilized shallow impurities forming bound states as a tool to study the particular distribution of isoelectronic centers at the onset of alloy formation. By means of $\mu$-PL we directly observed a hierarchy of bound excitons related to dilute silicon-indium assemblies as individual, energetically sharp (FWHM\,$\approx$\,300\,$\mu$eV) emission lines appear. Consequently, we introduced a classification of the underlying emitters into spatially direct and indirect bound excitonic complexes, whose balance is weighted by the indium-induced localization of charge carriers at the very onset of alloy formation. However, not only such $\mu$-PL data, but even conventional macro PL spectra allowed us to extract crucial material parameters for the mixed crystal alloy at hand. Based on ensembles of impurity bound excitons (three-particle complexes) we studied the indium-enriched environment of neutral silicon donors in InGaN at the length scale of the exciton Bohr radius. The analysis of the related exciton phonon coupling revealed a reduction of the average optical phonon energy that governs the temperature-dependent emission line broadening from $12.9\,\pm\,0.4\,\text{meV}$ to $6.4\,\pm\,1.0\,\text{meV}$ upon rising indium content from pure GaN up to $x=1.36\%$. Interestingly, for $\mathrm{Si^{0}X_{A}}$ we found an alloying dependence for the homogeneous emission line broadening caused by local indium enrichment and an indium-induced delocalization of the bound exciton yielding a transition from spatially direct to indirect transitions excitonic complexes. Based on the luminescence traces of free excitons that become increasingly trapped upon alloy formation, we extracted microscopic material properties from conventional PL data. We motivated that upon increasing indium content $x$ the radius of the excitonic averaging volume $r_a$ reduces gradually from $27.2\,\pm\, 3.7\,\text{nm}$ at $x=0.01\%$ down to $5.7\,\pm\,1.3\,\text{nm}$ at $x=2.4\%$. This transition illustrates the evolution from a free exciton in a doped semiconductor to a bound exciton  (two-particle complex) in a class II alloy. As a result, the exciton capture by point- and structural defects will be diminished, supporting the high $\eta_{int}$ values of the InGaN alloy even at defect densities that are detrimental for other III-V binary semiconductors and alloys like, e.g, (In)GaAs \cite{Takano1991,Chichibu2006a}, AlInGaP \cite{GeraldB.Stringfellow1998}, and AlGaAs \cite{Schubert1984,Chichibu2006a}. Based on our detailed linewidth analysis we estimated that $\approx$\,10 indium atoms are required to form assemblies capable of effectively capturing an exciton at a temperature of 12\,K. In contrast, a single indium atom does not introduce an electronic state in the bandgap of InGaN as confirmed by our PL experiments and first theoretically predicted by Bellaiche \textit{et al.} \cite{Bellaiche1999}. Clearly, this justifies the categorization of InGaN as a class II alloy. The present results open the perspective to utilize ensembles, but also individual excitonic complexes as a beneficial tools for \textit{any} class II alloy analysis at the few nanometer scale. In this regard, our results are not limited to the InGaN alloy. However, the constraints regarding a suitable doping interval are strict and explain - to the best of our knowledge - the absence of corresponding data in the literature for any other class II alloys.
\\
\section*{Acknowledgments}
\label{Acknowledgments}

This work is supported by the Marie Sk\l{}odowska-Curie action "PhotoHeatEffect" (Grant No. 749565) within the European Union's Horizon 2020 research and innovation program. The authors wish to thank I. M. Rousseau for his support with the experimental setup. Furthermore, we highly acknowledge D. Martin for growing the samples and K. Shojiki for processing the metal apertures.

\bibliography{Alloying1}

\begin{thebibliography}{87}%
\makeatletter
\providecommand \@ifxundefined [1]{%
 \@ifx{#1\undefined}
}%
\providecommand \@ifnum [1]{%
 \ifnum #1\expandafter \@firstoftwo
 \else \expandafter \@secondoftwo
 \fi
}%
\providecommand \@ifx [1]{%
 \ifx #1\expandafter \@firstoftwo
 \else \expandafter \@secondoftwo
 \fi
}%
\providecommand \natexlab [1]{#1}%
\providecommand \enquote  [1]{``#1''}%
\providecommand \bibnamefont  [1]{#1}%
\providecommand \bibfnamefont [1]{#1}%
\providecommand \citenamefont [1]{#1}%
\providecommand \href@noop [0]{\@secondoftwo}%
\providecommand \href [0]{\begingroup \@sanitize@url \@href}%
\providecommand \@href[1]{\@@startlink{#1}\@@href}%
\providecommand \@@href[1]{\endgroup#1\@@endlink}%
\providecommand \@sanitize@url [0]{\catcode `\\12\catcode `\$12\catcode
  `\&12\catcode `\#12\catcode `\^12\catcode `\_12\catcode `\%12\relax}%
\providecommand \@@startlink[1]{}%
\providecommand \@@endlink[0]{}%
\providecommand \url  [0]{\begingroup\@sanitize@url \@url }%
\providecommand \@url [1]{\endgroup\@href {#1}{\urlprefix }}%
\providecommand \urlprefix  [0]{URL }%
\providecommand \Eprint [0]{\href }%
\providecommand \doibase [0]{http://dx.doi.org/}%
\providecommand \selectlanguage [0]{\@gobble}%
\providecommand \bibinfo  [0]{\@secondoftwo}%
\providecommand \bibfield  [0]{\@secondoftwo}%
\providecommand \translation [1]{[#1]}%
\providecommand \BibitemOpen [0]{}%
\providecommand \bibitemStop [0]{}%
\providecommand \bibitemNoStop [0]{.\EOS\space}%
\providecommand \EOS [0]{\spacefactor3000\relax}%
\providecommand \BibitemShut  [1]{\csname bibitem#1\endcsname}%
\let\auto@bib@innerbib\@empty
\bibitem [{\citenamefont {Czaja}(1971)}]{Physics}%
  \BibitemOpen
  \bibfield  {author} {\bibinfo {author} {\bibfnamefont {W.}~\bibnamefont
  {Czaja}},\ }\href@noop {} {\emph {\bibinfo {title} {{Festk{\"{o}}rperprobleme
  XI, Advances in Solid State Physics}}}},\ edited by\ \bibinfo {editor}
  {\bibfnamefont {O.}~\bibnamefont {Madelung}}\ (\bibinfo  {publisher}
  {Pergamon, Vieweg},\ \bibinfo {address} {Marburg, Braunschweig},\ \bibinfo
  {year} {1971})\ pp.\ \bibinfo {pages} {65--85}\BibitemShut {NoStop}%
\bibitem [{\citenamefont {Thomas}(1966)}]{Thomas1966a}%
  \BibitemOpen
  \bibfield  {author} {\bibinfo {author} {\bibfnamefont {D.~G.}\ \bibnamefont
  {Thomas}},\ }\href@noop {} {\bibfield  {journal} {\bibinfo  {journal} {J.
  Phys. Soc. Japan}\ }\textbf {\bibinfo {volume} {21}},\ \bibinfo {pages} {265}
  (\bibinfo {year} {1966})}\BibitemShut {NoStop}%
\bibitem [{\citenamefont {Thomas}\ \emph {et~al.}(1965)\citenamefont {Thomas},
  \citenamefont {Hopfield},\ and\ \citenamefont {Frosch}}]{Wilkinson1959}%
  \BibitemOpen
  \bibfield  {author} {\bibinfo {author} {\bibfnamefont {D.~G.}\ \bibnamefont
  {Thomas}}, \bibinfo {author} {\bibfnamefont {J.~J.}\ \bibnamefont
  {Hopfield}}, \ and\ \bibinfo {author} {\bibfnamefont {C.~J.}\ \bibnamefont
  {Frosch}},\ }\href {\doibase 10.1103/PhysRev.150.680} {\bibfield  {journal}
  {\bibinfo  {journal} {Physical Review Letters}\ }\textbf {\bibinfo {volume}
  {15}},\ \bibinfo {pages} {857} (\bibinfo {year} {1965})}\BibitemShut
  {NoStop}%
\bibitem [{\citenamefont {Zhang}\ \emph {et~al.}(2000)\citenamefont {Zhang},
  \citenamefont {Fluegel}, \citenamefont {Mascarenhas}, \citenamefont {Xin},\
  and\ \citenamefont {Tu}}]{Zhang2000}%
  \BibitemOpen
  \bibfield  {author} {\bibinfo {author} {\bibfnamefont {Y.}~\bibnamefont
  {Zhang}}, \bibinfo {author} {\bibfnamefont {B.}~\bibnamefont {Fluegel}},
  \bibinfo {author} {\bibfnamefont {A.}~\bibnamefont {Mascarenhas}}, \bibinfo
  {author} {\bibfnamefont {H.}~\bibnamefont {Xin}}, \ and\ \bibinfo {author}
  {\bibfnamefont {C.}~\bibnamefont {Tu}},\ }\href {\doibase
  10.1103/PhysRevB.62.4493} {\bibfield  {journal} {\bibinfo  {journal}
  {Physical Review B}\ }\textbf {\bibinfo {volume} {62}},\ \bibinfo {pages}
  {4493} (\bibinfo {year} {2000})}\BibitemShut {NoStop}%
\bibitem [{\citenamefont {Faulkner}\ and\ \citenamefont
  {Dean}(1970)}]{Faulkner1970}%
  \BibitemOpen
  \bibfield  {author} {\bibinfo {author} {\bibfnamefont {R.~A.}\ \bibnamefont
  {Faulkner}}\ and\ \bibinfo {author} {\bibfnamefont {P.~J.}\ \bibnamefont
  {Dean}},\ }\href@noop {} {\bibfield  {journal} {\bibinfo  {journal} {Journal
  of Luminescence}\ }\textbf {\bibinfo {volume} {1, 2}},\ \bibinfo {pages}
  {552} (\bibinfo {year} {1970})}\BibitemShut {NoStop}%
\bibitem [{\citenamefont {Francoeur}\ \emph {et~al.}(2008)\citenamefont
  {Francoeur}, \citenamefont {Tixier}, \citenamefont {Young}, \citenamefont
  {Tiedje},\ and\ \citenamefont {Mascarenhas}}]{Francoeur2008}%
  \BibitemOpen
  \bibfield  {author} {\bibinfo {author} {\bibfnamefont {S.}~\bibnamefont
  {Francoeur}}, \bibinfo {author} {\bibfnamefont {S.}~\bibnamefont {Tixier}},
  \bibinfo {author} {\bibfnamefont {E.}~\bibnamefont {Young}}, \bibinfo
  {author} {\bibfnamefont {T.}~\bibnamefont {Tiedje}}, \ and\ \bibinfo {author}
  {\bibfnamefont {A.}~\bibnamefont {Mascarenhas}},\ }\href {\doibase
  10.1103/PhysRevB.77.085209} {\bibfield  {journal} {\bibinfo  {journal}
  {Physical Review B}\ }\textbf {\bibinfo {volume} {77}},\ \bibinfo {pages}
  {085209} (\bibinfo {year} {2008})}\BibitemShut {NoStop}%
\bibitem [{\citenamefont {Wolford}(1979)}]{Wolford2014}%
  \BibitemOpen
  \bibfield  {author} {\bibinfo {author} {\bibfnamefont {D.}~\bibnamefont
  {Wolford}},\ }\href {\doibase 10.1016/0022-2313(79)90252-7} {\bibfield
  {journal} {\bibinfo  {journal} {Journal of Luminescence}\ }\textbf {\bibinfo
  {volume} {18/19}},\ \bibinfo {pages} {863} (\bibinfo {year}
  {1979})}\BibitemShut {NoStop}%
\bibitem [{\citenamefont {Merz}(1968)}]{Merz1968}%
  \BibitemOpen
  \bibfield  {author} {\bibinfo {author} {\bibfnamefont {J.~L.}\ \bibnamefont
  {Merz}},\ }\href {\doibase 10.1103/PhysRev.176.961} {\bibfield  {journal}
  {\bibinfo  {journal} {Physical Review}\ }\textbf {\bibinfo {volume} {176}},\
  \bibinfo {pages} {961} (\bibinfo {year} {1968})}\BibitemShut {NoStop}%
\bibitem [{\citenamefont {Fukushima}\ and\ \citenamefont
  {Shionoya}(1976)}]{Parz}%
  \BibitemOpen
  \bibfield  {author} {\bibinfo {author} {\bibfnamefont {T.}~\bibnamefont
  {Fukushima}}\ and\ \bibinfo {author} {\bibfnamefont {S.}~\bibnamefont
  {Shionoya}},\ }\href@noop {} {\bibfield  {journal} {\bibinfo  {journal}
  {Japanese Journal of Applied Physics}\ }\textbf {\bibinfo {volume} {15}},\
  \bibinfo {pages} {813} (\bibinfo {year} {1976})}\BibitemShut {NoStop}%
\bibitem [{\citenamefont {Cullen}\ \emph {et~al.}(2013)\citenamefont {Cullen},
  \citenamefont {Johnston}, \citenamefont {Dunker}, \citenamefont {McGlynn},
  \citenamefont {Yakovlev}, \citenamefont {Bayer},\ and\ \citenamefont
  {Henry}}]{Cullen2013}%
  \BibitemOpen
  \bibfield  {author} {\bibinfo {author} {\bibfnamefont {J.}~\bibnamefont
  {Cullen}}, \bibinfo {author} {\bibfnamefont {K.}~\bibnamefont {Johnston}},
  \bibinfo {author} {\bibfnamefont {D.}~\bibnamefont {Dunker}}, \bibinfo
  {author} {\bibfnamefont {E.}~\bibnamefont {McGlynn}}, \bibinfo {author}
  {\bibfnamefont {D.~R.}\ \bibnamefont {Yakovlev}}, \bibinfo {author}
  {\bibfnamefont {M.}~\bibnamefont {Bayer}}, \ and\ \bibinfo {author}
  {\bibfnamefont {M.~O.}\ \bibnamefont {Henry}},\ }\href@noop {} {\bibfield
  {journal} {\bibinfo  {journal} {Journal of Applied Physics}\ }\textbf
  {\bibinfo {volume} {114}},\ \bibinfo {pages} {193515} (\bibinfo {year}
  {2013})}\BibitemShut {NoStop}%
\bibitem [{\citenamefont {Faulkner}(1968)}]{Faulkner1968}%
  \BibitemOpen
  \bibfield  {author} {\bibinfo {author} {\bibfnamefont {R.~A.}\ \bibnamefont
  {Faulkner}},\ }\href {\doibase 10.1103/PhysRev.175.991} {\bibfield  {journal}
  {\bibinfo  {journal} {Physical Review}\ }\textbf {\bibinfo {volume} {175}},\
  \bibinfo {pages} {991} (\bibinfo {year} {1968})}\BibitemShut {NoStop}%
\bibitem [{\citenamefont {Phillips}(1969)}]{Phillips1969}%
  \BibitemOpen
  \bibfield  {author} {\bibinfo {author} {\bibfnamefont {J.~C.}\ \bibnamefont
  {Phillips}},\ }\href@noop {} {\bibfield  {journal} {\bibinfo  {journal}
  {Physical Review Letters}\ }\textbf {\bibinfo {volume} {22}},\ \bibinfo
  {pages} {285} (\bibinfo {year} {1969})}\BibitemShut {NoStop}%
\bibitem [{\citenamefont {Allen}(1971)}]{Allen1971}%
  \BibitemOpen
  \bibfield  {author} {\bibinfo {author} {\bibfnamefont {J.~W.}\ \bibnamefont
  {Allen}},\ }\href {\doibase 10.1088/0022-3719/4/14/008} {\bibfield  {journal}
  {\bibinfo  {journal} {Journal of Physics C: Solid State Physics}\ }\textbf
  {\bibinfo {volume} {4}},\ \bibinfo {pages} {1936} (\bibinfo {year}
  {1971})}\BibitemShut {NoStop}%
\bibitem [{\citenamefont {Braunstein}\ \emph {et~al.}(1958)\citenamefont
  {Braunstein}, \citenamefont {Moore},\ and\ \citenamefont
  {Herman}}]{Braunstein1958}%
  \BibitemOpen
  \bibfield  {author} {\bibinfo {author} {\bibfnamefont {R.}~\bibnamefont
  {Braunstein}}, \bibinfo {author} {\bibfnamefont {A.~R.}\ \bibnamefont
  {Moore}}, \ and\ \bibinfo {author} {\bibfnamefont {F.}~\bibnamefont
  {Herman}},\ }\href {\doibase 10.1103/PhysRev.109.695} {\bibfield  {journal}
  {\bibinfo  {journal} {Physical Review}\ }\textbf {\bibinfo {volume} {109}},\
  \bibinfo {pages} {695} (\bibinfo {year} {1958})}\BibitemShut {NoStop}%
\bibitem [{\citenamefont {Tietjen}\ and\ \citenamefont
  {Weisberg}(1965)}]{Tietjen1965}%
  \BibitemOpen
  \bibfield  {author} {\bibinfo {author} {\bibfnamefont {J.~J.}\ \bibnamefont
  {Tietjen}}\ and\ \bibinfo {author} {\bibfnamefont {L.~R.}\ \bibnamefont
  {Weisberg}},\ }\href {\doibase 10.1063/1.1754248} {\bibfield  {journal}
  {\bibinfo  {journal} {Applied Physics Letters}\ }\textbf {\bibinfo {volume}
  {7}},\ \bibinfo {pages} {261} (\bibinfo {year} {1965})}\BibitemShut {NoStop}%
\bibitem [{\citenamefont {Laurenti}\ \emph {et~al.}(1988)\citenamefont
  {Laurenti}, \citenamefont {Roentgen}, \citenamefont {Wolter}, \citenamefont
  {Seibert}, \citenamefont {Kurz},\ and\ \citenamefont
  {Camassel}}]{J.P.LaurentiP.RoentgenK.WolterK.SeibertH.Kurz1988}%
  \BibitemOpen
  \bibfield  {author} {\bibinfo {author} {\bibfnamefont {J.~P.}\ \bibnamefont
  {Laurenti}}, \bibinfo {author} {\bibfnamefont {P.}~\bibnamefont {Roentgen}},
  \bibinfo {author} {\bibfnamefont {K.}~\bibnamefont {Wolter}}, \bibinfo
  {author} {\bibfnamefont {K.}~\bibnamefont {Seibert}}, \bibinfo {author}
  {\bibfnamefont {H.}~\bibnamefont {Kurz}}, \ and\ \bibinfo {author}
  {\bibfnamefont {J.}~\bibnamefont {Camassel}},\ }\href@noop {} {\bibfield
  {journal} {\bibinfo  {journal} {Physical Review B}\ }\textbf {\bibinfo
  {volume} {37}},\ \bibinfo {pages} {4155} (\bibinfo {year}
  {1988})}\BibitemShut {NoStop}%
\bibitem [{\citenamefont {Schubert}\ \emph {et~al.}(1984)\citenamefont
  {Schubert}, \citenamefont {G{\"{o}}bel}, \citenamefont {Horikoshi},
  \citenamefont {Ploog},\ and\ \citenamefont {Queisser}}]{Schubert1984}%
  \BibitemOpen
  \bibfield  {author} {\bibinfo {author} {\bibfnamefont {E.~F.}\ \bibnamefont
  {Schubert}}, \bibinfo {author} {\bibfnamefont {E.~O.}\ \bibnamefont
  {G{\"{o}}bel}}, \bibinfo {author} {\bibfnamefont {Y.}~\bibnamefont
  {Horikoshi}}, \bibinfo {author} {\bibfnamefont {K.}~\bibnamefont {Ploog}}, \
  and\ \bibinfo {author} {\bibfnamefont {H.~J.}\ \bibnamefont {Queisser}},\
  }\href {\doibase 10.1103/PhysRevB.30.813} {\bibfield  {journal} {\bibinfo
  {journal} {Physical Review B}\ }\textbf {\bibinfo {volume} {30}},\ \bibinfo
  {pages} {813} (\bibinfo {year} {1984})}\BibitemShut {NoStop}%
\bibitem [{\citenamefont {Butt{\'{e}}}\ \emph {et~al.}(2018)\citenamefont
  {Butt{\'{e}}}, \citenamefont {Lahourcade}, \citenamefont {U{\v{z}}davinys},
  \citenamefont {Callsen}, \citenamefont {Mensi}, \citenamefont {Glauser},
  \citenamefont {Rossbach}, \citenamefont {Martin}, \citenamefont {Carlin},
  \citenamefont {Marcinkevi{\v{c}}ius},\ and\ \citenamefont
  {Grandjean}}]{Butte2018a}%
  \BibitemOpen
  \bibfield  {author} {\bibinfo {author} {\bibfnamefont {R.}~\bibnamefont
  {Butt{\'{e}}}}, \bibinfo {author} {\bibfnamefont {L.}~\bibnamefont
  {Lahourcade}}, \bibinfo {author} {\bibfnamefont {T.~K.}\ \bibnamefont
  {U{\v{z}}davinys}}, \bibinfo {author} {\bibfnamefont {G.}~\bibnamefont
  {Callsen}}, \bibinfo {author} {\bibfnamefont {M.}~\bibnamefont {Mensi}},
  \bibinfo {author} {\bibfnamefont {M.}~\bibnamefont {Glauser}}, \bibinfo
  {author} {\bibfnamefont {G.}~\bibnamefont {Rossbach}}, \bibinfo {author}
  {\bibfnamefont {D.}~\bibnamefont {Martin}}, \bibinfo {author} {\bibfnamefont
  {J.-F.}\ \bibnamefont {Carlin}}, \bibinfo {author} {\bibfnamefont
  {S.}~\bibnamefont {Marcinkevi{\v{c}}ius}}, \ and\ \bibinfo {author}
  {\bibfnamefont {N.}~\bibnamefont {Grandjean}},\ }\href@noop {} {\bibfield
  {journal} {\bibinfo  {journal} {Applied Physics Letters}\ }\textbf {\bibinfo
  {volume} {112}},\ \bibinfo {pages} {032106} (\bibinfo {year}
  {2018})}\BibitemShut {NoStop}%
\bibitem [{\citenamefont {Wagener}\ \emph {et~al.}(2004)\citenamefont
  {Wagener}, \citenamefont {James}, \citenamefont {Leitch},\ and\ \citenamefont
  {Omn{\`{e}}s}}]{Wagener2004}%
  \BibitemOpen
  \bibfield  {author} {\bibinfo {author} {\bibfnamefont {M.~C.}\ \bibnamefont
  {Wagener}}, \bibinfo {author} {\bibfnamefont {G.~R.}\ \bibnamefont {James}},
  \bibinfo {author} {\bibfnamefont {A.~W.~R.}\ \bibnamefont {Leitch}}, \ and\
  \bibinfo {author} {\bibfnamefont {F.}~\bibnamefont {Omn{\`{e}}s}},\ }\href
  {\doibase 10.1002/pssc.200404838} {\bibfield  {journal} {\bibinfo  {journal}
  {Phys. Stat. Sol. (c)}\ }\textbf {\bibinfo {volume} {1}},\ \bibinfo {pages}
  {2322} (\bibinfo {year} {2004})}\BibitemShut {NoStop}%
\bibitem [{\citenamefont {Goede}\ \emph {et~al.}(1978)\citenamefont {Goede},
  \citenamefont {John},\ and\ \citenamefont {Hennig}}]{Notes1978}%
  \BibitemOpen
  \bibfield  {author} {\bibinfo {author} {\bibfnamefont {O.}~\bibnamefont
  {Goede}}, \bibinfo {author} {\bibfnamefont {L.}~\bibnamefont {John}}, \ and\
  \bibinfo {author} {\bibfnamefont {D.}~\bibnamefont {Hennig}},\ }\href@noop {}
  {\bibfield  {journal} {\bibinfo  {journal} {Physica Status Solidi B}\
  }\textbf {\bibinfo {volume} {89}},\ \bibinfo {pages} {K183} (\bibinfo {year}
  {1978})}\BibitemShut {NoStop}%
\bibitem [{\citenamefont {Klochikhin}\ \emph {et~al.}(1999)\citenamefont
  {Klochikhin}, \citenamefont {Reznitsky}, \citenamefont {Permogorov},
  \citenamefont {Breitkopf}, \citenamefont {Gr{\"{u}}n}, \citenamefont
  {Hetterich}, \citenamefont {Klingshirn}, \citenamefont {Lyssenko},
  \citenamefont {Langbein},\ and\ \citenamefont {Hvam}}]{Klochikhin1999}%
  \BibitemOpen
  \bibfield  {author} {\bibinfo {author} {\bibfnamefont {A.}~\bibnamefont
  {Klochikhin}}, \bibinfo {author} {\bibfnamefont {A.}~\bibnamefont
  {Reznitsky}}, \bibinfo {author} {\bibfnamefont {S.}~\bibnamefont
  {Permogorov}}, \bibinfo {author} {\bibfnamefont {T.}~\bibnamefont
  {Breitkopf}}, \bibinfo {author} {\bibfnamefont {M.}~\bibnamefont
  {Gr{\"{u}}n}}, \bibinfo {author} {\bibfnamefont {M.}~\bibnamefont
  {Hetterich}}, \bibinfo {author} {\bibfnamefont {C.}~\bibnamefont
  {Klingshirn}}, \bibinfo {author} {\bibfnamefont {V.}~\bibnamefont
  {Lyssenko}}, \bibinfo {author} {\bibfnamefont {W.}~\bibnamefont {Langbein}},
  \ and\ \bibinfo {author} {\bibfnamefont {J.~M.}\ \bibnamefont {Hvam}},\
  }\href@noop {} {\bibfield  {journal} {\bibinfo  {journal} {Physical Review
  B}\ }\textbf {\bibinfo {volume} {59}},\ \bibinfo {pages} {947} (\bibinfo
  {year} {1999})}\BibitemShut {NoStop}%
\bibitem [{\citenamefont {Grundmann}\ and\ \citenamefont
  {Dietrich}(2009)}]{Grundmann2009}%
  \BibitemOpen
  \bibfield  {author} {\bibinfo {author} {\bibfnamefont {M.}~\bibnamefont
  {Grundmann}}\ and\ \bibinfo {author} {\bibfnamefont {C.~P.}\ \bibnamefont
  {Dietrich}},\ }\href@noop {} {\bibfield  {journal} {\bibinfo  {journal}
  {Journal of Applied Physics}\ }\textbf {\bibinfo {volume} {106}},\ \bibinfo
  {pages} {123521} (\bibinfo {year} {2009})}\BibitemShut {NoStop}%
\bibitem [{\citenamefont {Bellaiche}\ \emph {et~al.}(1999)\citenamefont
  {Bellaiche}, \citenamefont {Mattila}, \citenamefont {Wang}, \citenamefont
  {Wei},\ and\ \citenamefont {Zunger}}]{Bellaiche1999}%
  \BibitemOpen
  \bibfield  {author} {\bibinfo {author} {\bibfnamefont {L.}~\bibnamefont
  {Bellaiche}}, \bibinfo {author} {\bibfnamefont {T.}~\bibnamefont {Mattila}},
  \bibinfo {author} {\bibfnamefont {L.~W.}\ \bibnamefont {Wang}}, \bibinfo
  {author} {\bibfnamefont {S.~H.}\ \bibnamefont {Wei}}, \ and\ \bibinfo
  {author} {\bibfnamefont {A.}~\bibnamefont {Zunger}},\ }\href {\doibase
  10.1063/1.123687} {\bibfield  {journal} {\bibinfo  {journal} {Applied Physics
  Letters}\ }\textbf {\bibinfo {volume} {74}},\ \bibinfo {pages} {1842}
  (\bibinfo {year} {1999})}\BibitemShut {NoStop}%
\bibitem [{\citenamefont {Dean}(1970)}]{Dean1970}%
  \BibitemOpen
  \bibfield  {author} {\bibinfo {author} {\bibfnamefont {P.~J.}\ \bibnamefont
  {Dean}},\ }\href@noop {} {\bibfield  {journal} {\bibinfo  {journal} {Journal
  of Luminescence}\ }\textbf {\bibinfo {volume} {1}},\ \bibinfo {pages} {398}
  (\bibinfo {year} {1970})}\BibitemShut {NoStop}%
\bibitem [{\citenamefont {Kanzaki}\ and\ \citenamefont
  {Sakuragi}(1969)}]{Kanzaki1969}%
  \BibitemOpen
  \bibfield  {author} {\bibinfo {author} {\bibfnamefont {H.}~\bibnamefont
  {Kanzaki}}\ and\ \bibinfo {author} {\bibfnamefont {S.}~\bibnamefont
  {Sakuragi}},\ }\href {\doibase 10.1143/JPSJ.61.3267} {\bibfield  {journal}
  {\bibinfo  {journal} {Journal of the Physical Society of Japan}\ }\textbf
  {\bibinfo {volume} {27}},\ \bibinfo {pages} {109} (\bibinfo {year}
  {1969})}\BibitemShut {NoStop}%
\bibitem [{\citenamefont {Hopfield}\ \emph {et~al.}(1966)\citenamefont
  {Hopfield}, \citenamefont {Thomas},\ and\ \citenamefont
  {Lynch}}]{Hopfield1966}%
  \BibitemOpen
  \bibfield  {author} {\bibinfo {author} {\bibfnamefont {J.~J.}\ \bibnamefont
  {Hopfield}}, \bibinfo {author} {\bibfnamefont {D.~G.}\ \bibnamefont
  {Thomas}}, \ and\ \bibinfo {author} {\bibfnamefont {R.~T.}\ \bibnamefont
  {Lynch}},\ }\href@noop {} {\bibfield  {journal} {\bibinfo  {journal}
  {Physical Review}\ }\textbf {\bibinfo {volume} {17}},\ \bibinfo {pages} {312}
  (\bibinfo {year} {1966})}\BibitemShut {NoStop}%
\bibitem [{\citenamefont {Thomas}\ and\ \citenamefont
  {Hopfield}(1966)}]{Thomas1966}%
  \BibitemOpen
  \bibfield  {author} {\bibinfo {author} {\bibfnamefont {D.~G.}\ \bibnamefont
  {Thomas}}\ and\ \bibinfo {author} {\bibfnamefont {J.~J.}\ \bibnamefont
  {Hopfield}},\ }\href {\doibase 10.1103/PhysRev.150.680} {\bibfield  {journal}
  {\bibinfo  {journal} {Physical Review}\ }\textbf {\bibinfo {volume} {150}},\
  \bibinfo {pages} {680} (\bibinfo {year} {1966})}\BibitemShut {NoStop}%
\bibitem [{\citenamefont {Muller}\ \emph {et~al.}(2006)\citenamefont {Muller},
  \citenamefont {Bianucci}, \citenamefont {Piermarocchi}, \citenamefont
  {Fornari}, \citenamefont {Robin}, \citenamefont {Andr{\'{e}}},\ and\
  \citenamefont {Shih}}]{Muller2006}%
  \BibitemOpen
  \bibfield  {author} {\bibinfo {author} {\bibfnamefont {A.}~\bibnamefont
  {Muller}}, \bibinfo {author} {\bibfnamefont {P.}~\bibnamefont {Bianucci}},
  \bibinfo {author} {\bibfnamefont {C.}~\bibnamefont {Piermarocchi}}, \bibinfo
  {author} {\bibfnamefont {M.}~\bibnamefont {Fornari}}, \bibinfo {author}
  {\bibfnamefont {I.~C.}\ \bibnamefont {Robin}}, \bibinfo {author}
  {\bibfnamefont {R.}~\bibnamefont {Andr{\'{e}}}}, \ and\ \bibinfo {author}
  {\bibfnamefont {C.~K.}\ \bibnamefont {Shih}},\ }\href {\doibase
  10.1103/PhysRevB.73.081306} {\bibfield  {journal} {\bibinfo  {journal}
  {Physical Review B}\ }\textbf {\bibinfo {volume} {73}},\ \bibinfo {pages}
  {081306(R)} (\bibinfo {year} {2006})}\BibitemShut {NoStop}%
\bibitem [{\citenamefont {Manfra}(2014)}]{Manfra2014}%
  \BibitemOpen
  \bibfield  {author} {\bibinfo {author} {\bibfnamefont {M.~J.}\ \bibnamefont
  {Manfra}},\ }\href {\doibase 10.1146/annurev-conmatphys-031113-133905}
  {\bibfield  {journal} {\bibinfo  {journal} {Annual Review of Condensed Matter
  Physics}\ }\textbf {\bibinfo {volume} {5}},\ \bibinfo {pages} {347} (\bibinfo
  {year} {2014})}\BibitemShut {NoStop}%
\bibitem [{\citenamefont {Stringfellow}\ and\ \citenamefont
  {Craford}(1998)}]{GeraldB.Stringfellow1998}%
  \BibitemOpen
  \bibfield  {author} {\bibinfo {author} {\bibfnamefont {G.~B.}\ \bibnamefont
  {Stringfellow}}\ and\ \bibinfo {author} {\bibfnamefont {M.~G.}\ \bibnamefont
  {Craford}},\ }\href@noop {} {\emph {\bibinfo {title} {{High Brightness Light
  Emitting Diodes: Vol 48 (Semiconductors and Semimetals)}}}}\ (\bibinfo
  {publisher} {Academic Press Inc.},\ \bibinfo {address} {San Diego},\ \bibinfo
  {year} {1998})\BibitemShut {NoStop}%
\bibitem [{\citenamefont {Chichibu}\ \emph
  {et~al.}(2006{\natexlab{a}})\citenamefont {Chichibu}, \citenamefont {Uedono},
  \citenamefont {Onuma}, \citenamefont {Haskell}, \citenamefont {Chakraborty},
  \citenamefont {Koyama}, \citenamefont {Fini}, \citenamefont {Keller},
  \citenamefont {DenBaars}, \citenamefont {Speck}, \citenamefont {Mishra},
  \citenamefont {Nakamura}, \citenamefont {Yamaguchi}, \citenamefont
  {Kamiyama}, \citenamefont {Amano}, \citenamefont {Akasaki}, \citenamefont
  {Han},\ and\ \citenamefont {Sota}}]{Chichibu2006a}%
  \BibitemOpen
  \bibfield  {author} {\bibinfo {author} {\bibfnamefont {S.~F.}\ \bibnamefont
  {Chichibu}}, \bibinfo {author} {\bibfnamefont {A.}~\bibnamefont {Uedono}},
  \bibinfo {author} {\bibfnamefont {T.}~\bibnamefont {Onuma}}, \bibinfo
  {author} {\bibfnamefont {B.~A.}\ \bibnamefont {Haskell}}, \bibinfo {author}
  {\bibfnamefont {A.}~\bibnamefont {Chakraborty}}, \bibinfo {author}
  {\bibfnamefont {T.}~\bibnamefont {Koyama}}, \bibinfo {author} {\bibfnamefont
  {P.~T.}\ \bibnamefont {Fini}}, \bibinfo {author} {\bibfnamefont
  {S.}~\bibnamefont {Keller}}, \bibinfo {author} {\bibfnamefont {S.~P.}\
  \bibnamefont {DenBaars}}, \bibinfo {author} {\bibfnamefont {J.~S.}\
  \bibnamefont {Speck}}, \bibinfo {author} {\bibfnamefont {U.~K.}\ \bibnamefont
  {Mishra}}, \bibinfo {author} {\bibfnamefont {S.}~\bibnamefont {Nakamura}},
  \bibinfo {author} {\bibfnamefont {S.}~\bibnamefont {Yamaguchi}}, \bibinfo
  {author} {\bibfnamefont {S.}~\bibnamefont {Kamiyama}}, \bibinfo {author}
  {\bibfnamefont {H.}~\bibnamefont {Amano}}, \bibinfo {author} {\bibfnamefont
  {I.}~\bibnamefont {Akasaki}}, \bibinfo {author} {\bibfnamefont
  {J.}~\bibnamefont {Han}}, \ and\ \bibinfo {author} {\bibfnamefont
  {T.}~\bibnamefont {Sota}},\ }\href {\doibase 10.1038/nmat1726} {\bibfield
  {journal} {\bibinfo  {journal} {Nature Materials}\ }\textbf {\bibinfo
  {volume} {5}},\ \bibinfo {pages} {810} (\bibinfo {year}
  {2006}{\natexlab{a}})}\BibitemShut {NoStop}%
\bibitem [{\citenamefont {Kneissl}\ \emph {et~al.}(2011)\citenamefont
  {Kneissl}, \citenamefont {Kolbe}, \citenamefont {Chua}, \citenamefont
  {Kueller}, \citenamefont {Lobo}, \citenamefont {Stellmach}, \citenamefont
  {Knauer}, \citenamefont {Rodriguez}, \citenamefont {Einfeldt}, \citenamefont
  {Yang}, \citenamefont {Johnson},\ and\ \citenamefont {Weyers}}]{Kneissl2011}%
  \BibitemOpen
  \bibfield  {author} {\bibinfo {author} {\bibfnamefont {M.}~\bibnamefont
  {Kneissl}}, \bibinfo {author} {\bibfnamefont {T.}~\bibnamefont {Kolbe}},
  \bibinfo {author} {\bibfnamefont {C.}~\bibnamefont {Chua}}, \bibinfo {author}
  {\bibfnamefont {V.}~\bibnamefont {Kueller}}, \bibinfo {author} {\bibfnamefont
  {N.}~\bibnamefont {Lobo}}, \bibinfo {author} {\bibfnamefont {J.}~\bibnamefont
  {Stellmach}}, \bibinfo {author} {\bibfnamefont {A.}~\bibnamefont {Knauer}},
  \bibinfo {author} {\bibfnamefont {H.}~\bibnamefont {Rodriguez}}, \bibinfo
  {author} {\bibfnamefont {S.}~\bibnamefont {Einfeldt}}, \bibinfo {author}
  {\bibfnamefont {Z.}~\bibnamefont {Yang}}, \bibinfo {author} {\bibfnamefont
  {N.~M.}\ \bibnamefont {Johnson}}, \ and\ \bibinfo {author} {\bibfnamefont
  {M.}~\bibnamefont {Weyers}},\ }\href@noop {} {\bibfield  {journal} {\bibinfo
  {journal} {Semiconductor Science and Technology}\ }\textbf {\bibinfo {volume}
  {26}},\ \bibinfo {pages} {014036} (\bibinfo {year} {2011})}\BibitemShut
  {NoStop}%
\bibitem [{\citenamefont {Gil}(2014)}]{Gil2014b}%
  \BibitemOpen
  \bibfield  {author} {\bibinfo {author} {\bibfnamefont {B.}~\bibnamefont
  {Gil}},\ }\href@noop {} {\emph {\bibinfo {title} {{Physics of Wurtzite
  Nitrides and Oxides - Passport to Devices}}}}\ (\bibinfo  {publisher}
  {Springer},\ \bibinfo {address} {Heidelberg},\ \bibinfo {year}
  {2014})\BibitemShut {NoStop}%
\bibitem [{\citenamefont {Nakamura}\ \emph {et~al.}(1991)\citenamefont
  {Nakamura}, \citenamefont {Mukai},\ and\ \citenamefont
  {Senoh}}]{Nakamura1991}%
  \BibitemOpen
  \bibfield  {author} {\bibinfo {author} {\bibfnamefont {S.}~\bibnamefont
  {Nakamura}}, \bibinfo {author} {\bibfnamefont {T.}~\bibnamefont {Mukai}}, \
  and\ \bibinfo {author} {\bibfnamefont {M.}~\bibnamefont {Senoh}},\ }\href
  {\doibase 10.1143/JJAP.30.L1998} {\bibfield  {journal} {\bibinfo  {journal}
  {Japanese Journal of Applied Physics}\ }\textbf {\bibinfo {volume} {30}},\
  \bibinfo {pages} {L1998} (\bibinfo {year} {1991})}\BibitemShut {NoStop}%
\bibitem [{\citenamefont {Nakamura}\ \emph {et~al.}(1997)\citenamefont
  {Nakamura}, \citenamefont {Pearton},\ and\ \citenamefont
  {Fasol}}]{S.Nakamura1997}%
  \BibitemOpen
  \bibfield  {author} {\bibinfo {author} {\bibfnamefont {S.}~\bibnamefont
  {Nakamura}}, \bibinfo {author} {\bibfnamefont {S.}~\bibnamefont {Pearton}}, \
  and\ \bibinfo {author} {\bibfnamefont {G.}~\bibnamefont {Fasol}},\
  }\href@noop {} {\emph {\bibinfo {title} {{The Blue Laser Diode}}}}\ (\bibinfo
   {publisher} {Springer},\ \bibinfo {address} {Berlin},\ \bibinfo {year}
  {1997})\BibitemShut {NoStop}%
\bibitem [{\citenamefont {Mishra}\ \emph {et~al.}(2008)\citenamefont {Mishra},
  \citenamefont {Shen}, \citenamefont {Kazior},\ and\ \citenamefont
  {Wu}}]{Mishra2008}%
  \BibitemOpen
  \bibfield  {author} {\bibinfo {author} {\bibfnamefont {U.~K.}\ \bibnamefont
  {Mishra}}, \bibinfo {author} {\bibfnamefont {L.}~\bibnamefont {Shen}},
  \bibinfo {author} {\bibfnamefont {T.~E.}\ \bibnamefont {Kazior}}, \ and\
  \bibinfo {author} {\bibfnamefont {Y.-f.}\ \bibnamefont {Wu}},\ }\href
  {\doibase 10.1109/JPROC.2007.911060} {\bibfield  {journal} {\bibinfo
  {journal} {Proceedings of the IEEE}\ }\textbf {\bibinfo {volume} {96}},\
  \bibinfo {pages} {287} (\bibinfo {year} {2008})}\BibitemShut {NoStop}%
\bibitem [{\citenamefont {Rajan}\ and\ \citenamefont {Jena}(2013)}]{Rajan2013}%
  \BibitemOpen
  \bibfield  {author} {\bibinfo {author} {\bibfnamefont {S.}~\bibnamefont
  {Rajan}}\ and\ \bibinfo {author} {\bibfnamefont {D.}~\bibnamefont {Jena}},\
  }\href {\doibase 10.1088/0268-1242/28/7/070301} {\bibfield  {journal}
  {\bibinfo  {journal} {Semiconductor Science and Technology}\ }\textbf
  {\bibinfo {volume} {28}},\ \bibinfo {pages} {070301} (\bibinfo {year}
  {2013})}\BibitemShut {NoStop}%
\bibitem [{\citenamefont {W{\"{a}}chter}\ \emph {et~al.}(2011)\citenamefont
  {W{\"{a}}chter}, \citenamefont {Meyer}, \citenamefont {Metzner},
  \citenamefont {Jetter}, \citenamefont {Bertram}, \citenamefont {Christen},\
  and\ \citenamefont {Michler}}]{Wachter2011}%
  \BibitemOpen
  \bibfield  {author} {\bibinfo {author} {\bibfnamefont {C.}~\bibnamefont
  {W{\"{a}}chter}}, \bibinfo {author} {\bibfnamefont {A.}~\bibnamefont
  {Meyer}}, \bibinfo {author} {\bibfnamefont {S.}~\bibnamefont {Metzner}},
  \bibinfo {author} {\bibfnamefont {M.}~\bibnamefont {Jetter}}, \bibinfo
  {author} {\bibfnamefont {F.}~\bibnamefont {Bertram}}, \bibinfo {author}
  {\bibfnamefont {J.}~\bibnamefont {Christen}}, \ and\ \bibinfo {author}
  {\bibfnamefont {P.}~\bibnamefont {Michler}},\ }\href {\doibase
  10.1002/pssb.201046369} {\bibfield  {journal} {\bibinfo  {journal} {Physica
  Status Solidi (B)}\ }\textbf {\bibinfo {volume} {248}},\ \bibinfo {pages}
  {605} (\bibinfo {year} {2011})}\BibitemShut {NoStop}%
\bibitem [{\citenamefont {Nippert}\ \emph {et~al.}(2016)\citenamefont
  {Nippert}, \citenamefont {Karpov}, \citenamefont {Callsen}, \citenamefont
  {Galler}, \citenamefont {Kure}, \citenamefont {Nenstiel}, \citenamefont
  {Wagner}, \citenamefont {Stra{\ss}burg}, \citenamefont {Lugauer},\ and\
  \citenamefont {Hoffmann}}]{Nippert2016d}%
  \BibitemOpen
  \bibfield  {author} {\bibinfo {author} {\bibfnamefont {F.}~\bibnamefont
  {Nippert}}, \bibinfo {author} {\bibfnamefont {S.~Y.}\ \bibnamefont {Karpov}},
  \bibinfo {author} {\bibfnamefont {G.}~\bibnamefont {Callsen}}, \bibinfo
  {author} {\bibfnamefont {B.}~\bibnamefont {Galler}}, \bibinfo {author}
  {\bibfnamefont {T.}~\bibnamefont {Kure}}, \bibinfo {author} {\bibfnamefont
  {C.}~\bibnamefont {Nenstiel}}, \bibinfo {author} {\bibfnamefont {M.~R.}\
  \bibnamefont {Wagner}}, \bibinfo {author} {\bibfnamefont {M.}~\bibnamefont
  {Stra{\ss}burg}}, \bibinfo {author} {\bibfnamefont {H.~J.}\ \bibnamefont
  {Lugauer}}, \ and\ \bibinfo {author} {\bibfnamefont {A.}~\bibnamefont
  {Hoffmann}},\ }\href {http://dx.doi.org/10.1063/1.4965298} {\bibfield
  {journal} {\bibinfo  {journal} {Applied Physics Letters}\ }\textbf {\bibinfo
  {volume} {109}},\ \bibinfo {pages} {161103} (\bibinfo {year}
  {2016})}\BibitemShut {NoStop}%
\bibitem [{\citenamefont {Ponce}\ and\ \citenamefont
  {Bour}(1997)}]{Ponce1997a}%
  \BibitemOpen
  \bibfield  {author} {\bibinfo {author} {\bibfnamefont {F.~A.}\ \bibnamefont
  {Ponce}}\ and\ \bibinfo {author} {\bibfnamefont {D.~P.}\ \bibnamefont
  {Bour}},\ }\href {\doibase 10.1038/386351a0} {\bibfield  {journal} {\bibinfo
  {journal} {Nature}\ }\textbf {\bibinfo {volume} {386}},\ \bibinfo {pages}
  {351} (\bibinfo {year} {1997})}\BibitemShut {NoStop}%
\bibitem [{\citenamefont {Hangleiter}\ \emph {et~al.}(2005)\citenamefont
  {Hangleiter}, \citenamefont {Hitzel}, \citenamefont {Netzel}, \citenamefont
  {Fuhrmann}, \citenamefont {Rossow}, \citenamefont {Ade},\ and\ \citenamefont
  {Hinze}}]{Hangleiter2005}%
  \BibitemOpen
  \bibfield  {author} {\bibinfo {author} {\bibfnamefont {A.}~\bibnamefont
  {Hangleiter}}, \bibinfo {author} {\bibfnamefont {F.}~\bibnamefont {Hitzel}},
  \bibinfo {author} {\bibfnamefont {C.}~\bibnamefont {Netzel}}, \bibinfo
  {author} {\bibfnamefont {D.}~\bibnamefont {Fuhrmann}}, \bibinfo {author}
  {\bibfnamefont {U.}~\bibnamefont {Rossow}}, \bibinfo {author} {\bibfnamefont
  {G.}~\bibnamefont {Ade}}, \ and\ \bibinfo {author} {\bibfnamefont
  {P.}~\bibnamefont {Hinze}},\ }\href {\doibase 10.1103/PhysRevLett.95.127402}
  {\bibfield  {journal} {\bibinfo  {journal} {Physical Review Letters}\
  }\textbf {\bibinfo {volume} {95}},\ \bibinfo {pages} {127402} (\bibinfo
  {year} {2005})}\BibitemShut {NoStop}%
\bibitem [{\citenamefont {Narukawa}\ \emph {et~al.}(1997)\citenamefont
  {Narukawa}, \citenamefont {Kawakami}, \citenamefont {Funato}, \citenamefont
  {Fujita}, \citenamefont {Fujita},\ and\ \citenamefont
  {Nakamura}}]{Narukawa1997}%
  \BibitemOpen
  \bibfield  {author} {\bibinfo {author} {\bibfnamefont {Y.}~\bibnamefont
  {Narukawa}}, \bibinfo {author} {\bibfnamefont {Y.}~\bibnamefont {Kawakami}},
  \bibinfo {author} {\bibfnamefont {M.}~\bibnamefont {Funato}}, \bibinfo
  {author} {\bibfnamefont {S.}~\bibnamefont {Fujita}}, \bibinfo {author}
  {\bibfnamefont {S.}~\bibnamefont {Fujita}}, \ and\ \bibinfo {author}
  {\bibfnamefont {S.}~\bibnamefont {Nakamura}},\ }\href {\doibase
  10.1063/1.118455} {\bibfield  {journal} {\bibinfo  {journal} {Applied Physics
  Letters}\ }\textbf {\bibinfo {volume} {70}},\ \bibinfo {pages} {981}
  (\bibinfo {year} {1997})}\BibitemShut {NoStop}%
\bibitem [{\citenamefont {Kent}\ and\ \citenamefont
  {Zunger}(2001)}]{Kent2001a}%
  \BibitemOpen
  \bibfield  {author} {\bibinfo {author} {\bibfnamefont {P.~R.~C.}\
  \bibnamefont {Kent}}\ and\ \bibinfo {author} {\bibfnamefont {A.}~\bibnamefont
  {Zunger}},\ }\href {\doibase 10.1063/1.1405003} {\bibfield  {journal}
  {\bibinfo  {journal} {Applied Physics Letters}\ }\textbf {\bibinfo {volume}
  {79}},\ \bibinfo {pages} {1977} (\bibinfo {year} {2001})}\BibitemShut
  {NoStop}%
\bibitem [{\citenamefont {Galtrey}\ \emph {et~al.}(2007)\citenamefont
  {Galtrey}, \citenamefont {Oliver}, \citenamefont {Kappers}, \citenamefont
  {Humphreys}, \citenamefont {Stokes}, \citenamefont {Clifton},\ and\
  \citenamefont {Cerezo}}]{Galtrey2007}%
  \BibitemOpen
  \bibfield  {author} {\bibinfo {author} {\bibfnamefont {M.~J.}\ \bibnamefont
  {Galtrey}}, \bibinfo {author} {\bibfnamefont {R.~A.}\ \bibnamefont {Oliver}},
  \bibinfo {author} {\bibfnamefont {M.~J.}\ \bibnamefont {Kappers}}, \bibinfo
  {author} {\bibfnamefont {C.~J.}\ \bibnamefont {Humphreys}}, \bibinfo {author}
  {\bibfnamefont {D.~J.}\ \bibnamefont {Stokes}}, \bibinfo {author}
  {\bibfnamefont {P.~H.}\ \bibnamefont {Clifton}}, \ and\ \bibinfo {author}
  {\bibfnamefont {A.}~\bibnamefont {Cerezo}},\ }\href {\doibase
  10.1063/1.2431573} {\bibfield  {journal} {\bibinfo  {journal} {Applied
  Physics Letters}\ }\textbf {\bibinfo {volume} {90}},\ \bibinfo {pages}
  {061903} (\bibinfo {year} {2007})}\BibitemShut {NoStop}%
\bibitem [{\citenamefont {Cerezo}\ \emph {et~al.}(2007)\citenamefont {Cerezo},
  \citenamefont {Clifton}, \citenamefont {Galtrey}, \citenamefont {Humphreys},
  \citenamefont {Kelly}, \citenamefont {Larson}, \citenamefont {Lozano-Perez},
  \citenamefont {Marquis}, \citenamefont {Oliver}, \citenamefont {Sha},
  \citenamefont {Thompson}, \citenamefont {Zandbergen},\ and\ \citenamefont
  {Alvis}}]{Cerezo2007}%
  \BibitemOpen
  \bibfield  {author} {\bibinfo {author} {\bibfnamefont {A.}~\bibnamefont
  {Cerezo}}, \bibinfo {author} {\bibfnamefont {P.~H.}\ \bibnamefont {Clifton}},
  \bibinfo {author} {\bibfnamefont {M.~J.}\ \bibnamefont {Galtrey}}, \bibinfo
  {author} {\bibfnamefont {C.~J.}\ \bibnamefont {Humphreys}}, \bibinfo {author}
  {\bibfnamefont {T.~F.}\ \bibnamefont {Kelly}}, \bibinfo {author}
  {\bibfnamefont {D.~J.}\ \bibnamefont {Larson}}, \bibinfo {author}
  {\bibfnamefont {S.}~\bibnamefont {Lozano-Perez}}, \bibinfo {author}
  {\bibfnamefont {E.~A.}\ \bibnamefont {Marquis}}, \bibinfo {author}
  {\bibfnamefont {R.~A.}\ \bibnamefont {Oliver}}, \bibinfo {author}
  {\bibfnamefont {G.}~\bibnamefont {Sha}}, \bibinfo {author} {\bibfnamefont
  {K.}~\bibnamefont {Thompson}}, \bibinfo {author} {\bibfnamefont
  {M.}~\bibnamefont {Zandbergen}}, \ and\ \bibinfo {author} {\bibfnamefont
  {R.~L.}\ \bibnamefont {Alvis}},\ }\href {\doibase
  10.1016/S1369-7021(07)70306-1} {\bibfield  {journal} {\bibinfo  {journal}
  {Materials Today}\ }\textbf {\bibinfo {volume} {10}},\ \bibinfo {pages} {36}
  (\bibinfo {year} {2007})}\BibitemShut {NoStop}%
\bibitem [{\citenamefont {Rigutti}\ \emph {et~al.}(2018)\citenamefont
  {Rigutti}, \citenamefont {Bonef}, \citenamefont {Speck}, \citenamefont
  {Tang},\ and\ \citenamefont {Oliver}}]{Rigutti2018}%
  \BibitemOpen
  \bibfield  {author} {\bibinfo {author} {\bibfnamefont {L.}~\bibnamefont
  {Rigutti}}, \bibinfo {author} {\bibfnamefont {B.}~\bibnamefont {Bonef}},
  \bibinfo {author} {\bibfnamefont {J.}~\bibnamefont {Speck}}, \bibinfo
  {author} {\bibfnamefont {F.}~\bibnamefont {Tang}}, \ and\ \bibinfo {author}
  {\bibfnamefont {R.~A.}\ \bibnamefont {Oliver}},\ }\href {\doibase
  10.1016/j.scriptamat.2016.12.034} {\bibfield  {journal} {\bibinfo  {journal}
  {Scripta Materialia}\ }\textbf {\bibinfo {volume} {148}},\ \bibinfo {pages}
  {75} (\bibinfo {year} {2018})}\BibitemShut {NoStop}%
\bibitem [{\citenamefont {Mathieu}\ \emph {et~al.}(1992)\citenamefont
  {Mathieu}, \citenamefont {Lefebvre},\ and\ \citenamefont
  {Christol}}]{Mathieu1992}%
  \BibitemOpen
  \bibfield  {author} {\bibinfo {author} {\bibfnamefont {H.}~\bibnamefont
  {Mathieu}}, \bibinfo {author} {\bibfnamefont {P.}~\bibnamefont {Lefebvre}}, \
  and\ \bibinfo {author} {\bibfnamefont {P.}~\bibnamefont {Christol}},\
  }\href@noop {} {\bibfield  {journal} {\bibinfo  {journal} {Physical Review
  B}\ }\textbf {\bibinfo {volume} {46}},\ \bibinfo {pages} {4092} (\bibinfo
  {year} {1992})}\BibitemShut {NoStop}%
\bibitem [{\citenamefont {Kalceff}\ and\ \citenamefont
  {Phillips}(1995)}]{Kalceff1995}%
  \BibitemOpen
  \bibfield  {author} {\bibinfo {author} {\bibfnamefont {M.~A.~S.}\
  \bibnamefont {Kalceff}}\ and\ \bibinfo {author} {\bibfnamefont {M.~R.}\
  \bibnamefont {Phillips}},\ }\href {\doibase 10.1103/PhysRevB.52.3122}
  {\bibfield  {journal} {\bibinfo  {journal} {Physical Review B}\ }\textbf
  {\bibinfo {volume} {52}},\ \bibinfo {pages} {3122} (\bibinfo {year}
  {1995})}\BibitemShut {NoStop}%
\bibitem [{\citenamefont {Tanimura}\ and\ \citenamefont
  {Itoh}(1981)}]{Tanimura1981}%
  \BibitemOpen
  \bibfield  {author} {\bibinfo {author} {\bibfnamefont {K.}~\bibnamefont
  {Tanimura}}\ and\ \bibinfo {author} {\bibfnamefont {N.}~\bibnamefont
  {Itoh}},\ }\href {\doibase 10.1016/0022-3697(81)90016-0} {\bibfield
  {journal} {\bibinfo  {journal} {Journal of Physics and Chemistry of Solids}\
  }\textbf {\bibinfo {volume} {42}},\ \bibinfo {pages} {901} (\bibinfo {year}
  {1981})}\BibitemShut {NoStop}%
\bibitem [{\citenamefont {Viswanath}\ \emph {et~al.}(1998)\citenamefont
  {Viswanath}, \citenamefont {Lee}, \citenamefont {Kim}, \citenamefont {Lee},\
  and\ \citenamefont {Leem}}]{Viswanath1998}%
  \BibitemOpen
  \bibfield  {author} {\bibinfo {author} {\bibfnamefont {A.~K.}\ \bibnamefont
  {Viswanath}}, \bibinfo {author} {\bibfnamefont {J.~I.}\ \bibnamefont {Lee}},
  \bibinfo {author} {\bibfnamefont {D.}~\bibnamefont {Kim}}, \bibinfo {author}
  {\bibfnamefont {C.~R.}\ \bibnamefont {Lee}}, \ and\ \bibinfo {author}
  {\bibfnamefont {J.~Y.}\ \bibnamefont {Leem}},\ }\href {\doibase
  10.1103/PhysRevB.58.16333} {\bibfield  {journal} {\bibinfo  {journal}
  {Physical Review B}\ }\textbf {\bibinfo {volume} {58}},\ \bibinfo {pages}
  {16333 } (\bibinfo {year} {1998})}\BibitemShut {NoStop}%
\bibitem [{\citenamefont {Onuma}\ \emph {et~al.}(2009)\citenamefont {Onuma},
  \citenamefont {Shibata}, \citenamefont {Kosaka}, \citenamefont {Asai},
  \citenamefont {Sumiya}, \citenamefont {Tanaka}, \citenamefont {Sota},
  \citenamefont {Uedono},\ and\ \citenamefont {Chichibu}}]{Onuma2009}%
  \BibitemOpen
  \bibfield  {author} {\bibinfo {author} {\bibfnamefont {T.}~\bibnamefont
  {Onuma}}, \bibinfo {author} {\bibfnamefont {T.}~\bibnamefont {Shibata}},
  \bibinfo {author} {\bibfnamefont {K.}~\bibnamefont {Kosaka}}, \bibinfo
  {author} {\bibfnamefont {K.}~\bibnamefont {Asai}}, \bibinfo {author}
  {\bibfnamefont {S.}~\bibnamefont {Sumiya}}, \bibinfo {author} {\bibfnamefont
  {M.}~\bibnamefont {Tanaka}}, \bibinfo {author} {\bibfnamefont
  {T.}~\bibnamefont {Sota}}, \bibinfo {author} {\bibfnamefont {A.}~\bibnamefont
  {Uedono}}, \ and\ \bibinfo {author} {\bibfnamefont {S.~F.}\ \bibnamefont
  {Chichibu}},\ }\href {\doibase 10.1063/1.3068335} {\bibfield  {journal}
  {\bibinfo  {journal} {Journal of Applied Physics}\ }\textbf {\bibinfo
  {volume} {105}},\ \bibinfo {pages} {023529} (\bibinfo {year}
  {2009})}\BibitemShut {NoStop}%
\bibitem [{\citenamefont {Zimmermann}(1990)}]{Zimmermann1990}%
  \BibitemOpen
  \bibfield  {author} {\bibinfo {author} {\bibfnamefont {R.}~\bibnamefont
  {Zimmermann}},\ }\href {\doibase 10.1016/0022-0248(90)90993-U} {\bibfield
  {journal} {\bibinfo  {journal} {Journal of Crystal Growth}\ }\textbf
  {\bibinfo {volume} {101}},\ \bibinfo {pages} {346} (\bibinfo {year}
  {1990})}\BibitemShut {NoStop}%
\bibitem [{\citenamefont {Vurgaftman}\ and\ \citenamefont
  {Meyer}(2003)}]{Vurgaftman2003}%
  \BibitemOpen
  \bibfield  {author} {\bibinfo {author} {\bibfnamefont {I.}~\bibnamefont
  {Vurgaftman}}\ and\ \bibinfo {author} {\bibfnamefont {J.~R.}\ \bibnamefont
  {Meyer}},\ }\href
  {http://scitation.aip.org/content/aip/journal/jap/94/6/10.1063/1.1600519
  papers2://publication/doi/10.1063/1.1600519} {\bibfield  {journal} {\bibinfo
  {journal} {Journal of Applied Physics}\ }\textbf {\bibinfo {volume} {94}},\
  \bibinfo {pages} {3675} (\bibinfo {year} {2003})}\BibitemShut {NoStop}%
\bibitem [{\citenamefont {Callsen}\ \emph {et~al.}(2018)\citenamefont
  {Callsen}, \citenamefont {Kure}, \citenamefont {Wagner}, \citenamefont
  {Butt{\'{e}}},\ and\ \citenamefont {Grandjean}}]{Callsen2018}%
  \BibitemOpen
  \bibfield  {author} {\bibinfo {author} {\bibfnamefont {G.}~\bibnamefont
  {Callsen}}, \bibinfo {author} {\bibfnamefont {T.}~\bibnamefont {Kure}},
  \bibinfo {author} {\bibfnamefont {M.~R.}\ \bibnamefont {Wagner}}, \bibinfo
  {author} {\bibfnamefont {R.}~\bibnamefont {Butt{\'{e}}}}, \ and\ \bibinfo
  {author} {\bibfnamefont {N.}~\bibnamefont {Grandjean}},\ }\href {\doibase
  10.1063/1.5028370} {\bibfield  {journal} {\bibinfo  {journal} {Journal of
  Applied Physics}\ }\textbf {\bibinfo {volume} {123}},\ \bibinfo {pages}
  {215702} (\bibinfo {year} {2018})}\BibitemShut {NoStop}%
\bibitem [{\citenamefont {{\v{S}}antic}\ \emph {et~al.}(1997)\citenamefont
  {{\v{S}}antic}, \citenamefont {Merz}, \citenamefont {Kaufmann}, \citenamefont
  {Niebuhr}, \citenamefont {Obloh},\ and\ \citenamefont {Bachem}}]{Santic1998}%
  \BibitemOpen
  \bibfield  {author} {\bibinfo {author} {\bibfnamefont {B.}~\bibnamefont
  {{\v{S}}antic}}, \bibinfo {author} {\bibfnamefont {C.}~\bibnamefont {Merz}},
  \bibinfo {author} {\bibfnamefont {U.}~\bibnamefont {Kaufmann}}, \bibinfo
  {author} {\bibfnamefont {R.}~\bibnamefont {Niebuhr}}, \bibinfo {author}
  {\bibfnamefont {H.}~\bibnamefont {Obloh}}, \ and\ \bibinfo {author}
  {\bibfnamefont {K.}~\bibnamefont {Bachem}},\ }\href {\doibase
  10.1063/1.119415} {\bibfield  {journal} {\bibinfo  {journal} {Applied Physics
  Letters}\ }\textbf {\bibinfo {volume} {71}},\ \bibinfo {pages} {1837}
  (\bibinfo {year} {1997})}\BibitemShut {NoStop}%
\bibitem [{\citenamefont {Monemar}\ \emph {et~al.}(2001)\citenamefont
  {Monemar}, \citenamefont {Chen}, \citenamefont {Paskov}, \citenamefont
  {Paskova}, \citenamefont {Pozina},\ and\ \citenamefont
  {Bergman}}]{Monemar2001a}%
  \BibitemOpen
  \bibfield  {author} {\bibinfo {author} {\bibfnamefont {B.}~\bibnamefont
  {Monemar}}, \bibinfo {author} {\bibfnamefont {W.~M.}\ \bibnamefont {Chen}},
  \bibinfo {author} {\bibfnamefont {P.~P.}\ \bibnamefont {Paskov}}, \bibinfo
  {author} {\bibfnamefont {T.}~\bibnamefont {Paskova}}, \bibinfo {author}
  {\bibfnamefont {G.}~\bibnamefont {Pozina}}, \ and\ \bibinfo {author}
  {\bibfnamefont {J.~P.}\ \bibnamefont {Bergman}},\ }\href {\doibase
  10.1002/1521-3951(200111)228:2<489::AID-PSSB489>3.0.CO;2-N} {\bibfield
  {journal} {\bibinfo  {journal} {Physica Status Solidi B}\ }\textbf {\bibinfo
  {volume} {228}},\ \bibinfo {pages} {489} (\bibinfo {year}
  {2001})}\BibitemShut {NoStop}%
\bibitem [{\citenamefont {Wysmo{\l}ek}\ \emph {et~al.}(2003)\citenamefont
  {Wysmo{\l}ek}, \citenamefont {Potemski}, \citenamefont {Paku{\l}a},
  \citenamefont {Baranowski}, \citenamefont {Grzegory}, \citenamefont
  {Porowski}, \citenamefont {Martinez},\ and\ \citenamefont
  {Wyder}}]{Wysmoek2003}%
  \BibitemOpen
  \bibfield  {author} {\bibinfo {author} {\bibfnamefont {A.}~\bibnamefont
  {Wysmo{\l}ek}}, \bibinfo {author} {\bibfnamefont {M.}~\bibnamefont
  {Potemski}}, \bibinfo {author} {\bibfnamefont {K.}~\bibnamefont {Paku{\l}a}},
  \bibinfo {author} {\bibfnamefont {J.~M.}\ \bibnamefont {Baranowski}},
  \bibinfo {author} {\bibfnamefont {I.}~\bibnamefont {Grzegory}}, \bibinfo
  {author} {\bibfnamefont {S.}~\bibnamefont {Porowski}}, \bibinfo {author}
  {\bibfnamefont {G.}~\bibnamefont {Martinez}}, \ and\ \bibinfo {author}
  {\bibfnamefont {P.}~\bibnamefont {Wyder}},\ }\href {\doibase
  10.1103/PhysRevLett.91.226404} {\bibfield  {journal} {\bibinfo  {journal}
  {Physical Review Letters}\ }\textbf {\bibinfo {volume} {91}},\ \bibinfo
  {pages} {226404} (\bibinfo {year} {2003})}\BibitemShut {NoStop}%
\bibitem [{\citenamefont {Fischer}\ \emph {et~al.}(1997)\citenamefont
  {Fischer}, \citenamefont {Volm}, \citenamefont {Kovalev}, \citenamefont
  {Averboukh}, \citenamefont {Graber}, \citenamefont {Alt},\ and\ \citenamefont
  {Meyer}}]{Fischer1997a}%
  \BibitemOpen
  \bibfield  {author} {\bibinfo {author} {\bibfnamefont {S.}~\bibnamefont
  {Fischer}}, \bibinfo {author} {\bibfnamefont {D.}~\bibnamefont {Volm}},
  \bibinfo {author} {\bibfnamefont {D.}~\bibnamefont {Kovalev}}, \bibinfo
  {author} {\bibfnamefont {B.}~\bibnamefont {Averboukh}}, \bibinfo {author}
  {\bibfnamefont {A.}~\bibnamefont {Graber}}, \bibinfo {author} {\bibfnamefont
  {H.~C.}\ \bibnamefont {Alt}}, \ and\ \bibinfo {author} {\bibfnamefont
  {B.~K.}\ \bibnamefont {Meyer}},\ }\href@noop {} {\bibfield  {journal}
  {\bibinfo  {journal} {Materials Science and Engineering B}\ }\textbf
  {\bibinfo {volume} {43}},\ \bibinfo {pages} {192} (\bibinfo {year}
  {1997})}\BibitemShut {NoStop}%
\bibitem [{\citenamefont {{Claus Klingshirn}}(2005)}]{ClausKlingshirn}%
  \BibitemOpen
  \bibfield  {author} {\bibinfo {author} {\bibnamefont {{Claus Klingshirn}}},\
  }\href {\doibase 10.1007/978-3-662-09855-4} {\emph {\bibinfo {title}
  {{Semiconductor Optics}}}},\ \bibinfo {edition} {2nd}\ ed.\ (\bibinfo
  {publisher} {Springer},\ \bibinfo {address} {Berlin Heidelberg New York},\
  \bibinfo {year} {2005})\BibitemShut {NoStop}%
\bibitem [{\citenamefont {Neu}\ \emph {et~al.}(1984)\citenamefont {Neu},
  \citenamefont {Mbaye},\ and\ \citenamefont {Triboulet}}]{G.NeuA.A.Mbaye1984}%
  \BibitemOpen
  \bibfield  {author} {\bibinfo {author} {\bibfnamefont {G.}~\bibnamefont
  {Neu}}, \bibinfo {author} {\bibfnamefont {A.~A.}\ \bibnamefont {Mbaye}}, \
  and\ \bibinfo {author} {\bibfnamefont {R.}~\bibnamefont {Triboulet}},\
  }\href@noop {} {\bibfield  {journal} {\bibinfo  {journal} {Proceedings of the
  17th International Conference on the Physics of Semiconductors}\ }\textbf
  {\bibinfo {volume} {1}},\ \bibinfo {pages} {1029} (\bibinfo {year}
  {1984})}\BibitemShut {NoStop}%
\bibitem [{\citenamefont {Gil}\ \emph {et~al.}(2007)\citenamefont {Gil},
  \citenamefont {Bigenwald}, \citenamefont {Leroux}, \citenamefont {Paskov},\
  and\ \citenamefont {Monemar}}]{Gil2007}%
  \BibitemOpen
  \bibfield  {author} {\bibinfo {author} {\bibfnamefont {B.}~\bibnamefont
  {Gil}}, \bibinfo {author} {\bibfnamefont {P.}~\bibnamefont {Bigenwald}},
  \bibinfo {author} {\bibfnamefont {M.}~\bibnamefont {Leroux}}, \bibinfo
  {author} {\bibfnamefont {P.~P.}\ \bibnamefont {Paskov}}, \ and\ \bibinfo
  {author} {\bibfnamefont {B.}~\bibnamefont {Monemar}},\ }\href {\doibase
  10.1103/PhysRevB.75.085204} {\bibfield  {journal} {\bibinfo  {journal}
  {Physical Review B}\ }\textbf {\bibinfo {volume} {75}},\ \bibinfo {pages}
  {085204} (\bibinfo {year} {2007})}\BibitemShut {NoStop}%
\bibitem [{\citenamefont {Callsen}\ \emph {et~al.}(2012)\citenamefont
  {Callsen}, \citenamefont {Wagner}, \citenamefont {Kure}, \citenamefont
  {Reparaz}, \citenamefont {B{\"{u}}gler}, \citenamefont {Brunnmeier},
  \citenamefont {Nenstiel}, \citenamefont {Hoffmann}, \citenamefont {Hoffmann},
  \citenamefont {Tweedie}, \citenamefont {Bryan}, \citenamefont {Aygun},
  \citenamefont {Kirste}, \citenamefont {Collazo},\ and\ \citenamefont
  {Sitar}}]{Callsen2012a}%
  \BibitemOpen
  \bibfield  {author} {\bibinfo {author} {\bibfnamefont {G.}~\bibnamefont
  {Callsen}}, \bibinfo {author} {\bibfnamefont {M.}~\bibnamefont {Wagner}},
  \bibinfo {author} {\bibfnamefont {T.}~\bibnamefont {Kure}}, \bibinfo {author}
  {\bibfnamefont {J.}~\bibnamefont {Reparaz}}, \bibinfo {author} {\bibfnamefont
  {M.}~\bibnamefont {B{\"{u}}gler}}, \bibinfo {author} {\bibfnamefont
  {J.}~\bibnamefont {Brunnmeier}}, \bibinfo {author} {\bibfnamefont
  {C.}~\bibnamefont {Nenstiel}}, \bibinfo {author} {\bibfnamefont
  {A.}~\bibnamefont {Hoffmann}}, \bibinfo {author} {\bibfnamefont
  {M.}~\bibnamefont {Hoffmann}}, \bibinfo {author} {\bibfnamefont
  {J.}~\bibnamefont {Tweedie}}, \bibinfo {author} {\bibfnamefont
  {Z.}~\bibnamefont {Bryan}}, \bibinfo {author} {\bibfnamefont
  {S.}~\bibnamefont {Aygun}}, \bibinfo {author} {\bibfnamefont
  {R.}~\bibnamefont {Kirste}}, \bibinfo {author} {\bibfnamefont
  {R.}~\bibnamefont {Collazo}}, \ and\ \bibinfo {author} {\bibfnamefont
  {Z.}~\bibnamefont {Sitar}},\ }\href
  {http://link.aps.org/doi/10.1103/PhysRevB.86.075207
  file:///Users/gordito/Documents/Papers2/Articles/2012/Callsen/Phys Rev B/Phys
  Rev B 2012 Callsen.pdf papers2://publication/doi/10.1103/PhysRevB.86.075207}
  {\bibfield  {journal} {\bibinfo  {journal} {Physical Review B}\ }\textbf
  {\bibinfo {volume} {86}},\ \bibinfo {pages} {075207} (\bibinfo {year}
  {2012})}\BibitemShut {NoStop}%
\bibitem [{\citenamefont {Wagner}\ \emph {et~al.}(2011)\citenamefont {Wagner},
  \citenamefont {Callsen}, \citenamefont {Reparaz}, \citenamefont {Schulze},
  \citenamefont {Kirste}, \citenamefont {Cobet}, \citenamefont {Ostapenko},
  \citenamefont {Rodt}, \citenamefont {Nenstiel}, \citenamefont {Kaiser},
  \citenamefont {Hoffmann}, \citenamefont {Rodina}, \citenamefont {Phillips},
  \citenamefont {Lautenschlaeger}, \citenamefont {Eisermann},\ and\
  \citenamefont {Meyer}}]{Wagner2011}%
  \BibitemOpen
  \bibfield  {author} {\bibinfo {author} {\bibfnamefont {M.~R.}\ \bibnamefont
  {Wagner}}, \bibinfo {author} {\bibfnamefont {G.}~\bibnamefont {Callsen}},
  \bibinfo {author} {\bibfnamefont {J.~S.}\ \bibnamefont {Reparaz}}, \bibinfo
  {author} {\bibfnamefont {J.~H.}\ \bibnamefont {Schulze}}, \bibinfo {author}
  {\bibfnamefont {R.}~\bibnamefont {Kirste}}, \bibinfo {author} {\bibfnamefont
  {M.}~\bibnamefont {Cobet}}, \bibinfo {author} {\bibfnamefont {I.~A.}\
  \bibnamefont {Ostapenko}}, \bibinfo {author} {\bibfnamefont {S.}~\bibnamefont
  {Rodt}}, \bibinfo {author} {\bibfnamefont {C.}~\bibnamefont {Nenstiel}},
  \bibinfo {author} {\bibfnamefont {M.}~\bibnamefont {Kaiser}}, \bibinfo
  {author} {\bibfnamefont {A.}~\bibnamefont {Hoffmann}}, \bibinfo {author}
  {\bibfnamefont {A.~V.}\ \bibnamefont {Rodina}}, \bibinfo {author}
  {\bibfnamefont {M.~R.}\ \bibnamefont {Phillips}}, \bibinfo {author}
  {\bibfnamefont {S.}~\bibnamefont {Lautenschlaeger}}, \bibinfo {author}
  {\bibfnamefont {S.}~\bibnamefont {Eisermann}}, \ and\ \bibinfo {author}
  {\bibfnamefont {B.~K.}\ \bibnamefont {Meyer}},\ }\href
  {http://gateway.webofknowledge.com/gateway/Gateway.cgi?GWVersion=2{\&}SrcAuth=mekentosj{\&}SrcApp=Papers{\&}DestLinkType=FullRecord{\&}DestApp=WOS{\&}KeyUT=000293129200009
  file:///Users/gordito/Documents/Papers2/Articles/2011/Wagner/Phys Rev B/Phys
  Rev B 2011 Wagner.pdf} {\bibfield  {journal} {\bibinfo  {journal} {Physical
  Review B}\ }\textbf {\bibinfo {volume} {84}},\ \bibinfo {pages} {035313}
  (\bibinfo {year} {2011})}\BibitemShut {NoStop}%
\bibitem [{\citenamefont {Chichibu}\ \emph
  {et~al.}(2006{\natexlab{b}})\citenamefont {Chichibu}, \citenamefont {Onuma},
  \citenamefont {Kubota}, \citenamefont {Uedono}, \citenamefont {Sota},
  \citenamefont {Tsukazaki}, \citenamefont {Ohtomo},\ and\ \citenamefont
  {Kawasaki}}]{Chichibu2006}%
  \BibitemOpen
  \bibfield  {author} {\bibinfo {author} {\bibfnamefont {S.~F.}\ \bibnamefont
  {Chichibu}}, \bibinfo {author} {\bibfnamefont {T.}~\bibnamefont {Onuma}},
  \bibinfo {author} {\bibfnamefont {M.}~\bibnamefont {Kubota}}, \bibinfo
  {author} {\bibfnamefont {A.}~\bibnamefont {Uedono}}, \bibinfo {author}
  {\bibfnamefont {T.}~\bibnamefont {Sota}}, \bibinfo {author} {\bibfnamefont
  {A.}~\bibnamefont {Tsukazaki}}, \bibinfo {author} {\bibfnamefont
  {A.}~\bibnamefont {Ohtomo}}, \ and\ \bibinfo {author} {\bibfnamefont
  {M.}~\bibnamefont {Kawasaki}},\ }\href
  {http://link.aip.org/link/?JAPIAU/99/093505/1
  papers2://publication/uuid/2DBDCD9F-F487-4A04-A67C-E744C636658D} {\bibfield
  {journal} {\bibinfo  {journal} {Journal of Applied Physics}\ }\textbf
  {\bibinfo {volume} {99}},\ \bibinfo {pages} {093505} (\bibinfo {year}
  {2006}{\natexlab{b}})}\BibitemShut {NoStop}%
\bibitem [{\citenamefont {Shu}\ \emph {et~al.}(1998)\citenamefont {Shu},
  \citenamefont {Ou}, \citenamefont {Lin}, \citenamefont {Chen},\ and\
  \citenamefont {Lee}}]{Shu1998a}%
  \BibitemOpen
  \bibfield  {author} {\bibinfo {author} {\bibfnamefont {C.~K.}\ \bibnamefont
  {Shu}}, \bibinfo {author} {\bibfnamefont {J.}~\bibnamefont {Ou}}, \bibinfo
  {author} {\bibfnamefont {H.~C.}\ \bibnamefont {Lin}}, \bibinfo {author}
  {\bibfnamefont {W.~K.}\ \bibnamefont {Chen}}, \ and\ \bibinfo {author}
  {\bibfnamefont {M.~C.}\ \bibnamefont {Lee}},\ }\href {\doibase
  10.1063/1.121933} {\bibfield  {journal} {\bibinfo  {journal} {Applied Physics
  Letters}\ }\textbf {\bibinfo {volume} {73}},\ \bibinfo {pages} {641}
  (\bibinfo {year} {1998})}\BibitemShut {NoStop}%
\bibitem [{\citenamefont {Shen}\ \emph {et~al.}(1999)\citenamefont {Shen},
  \citenamefont {Ramvall}, \citenamefont {Riblet},\ and\ \citenamefont
  {Aoyagi}}]{Shen1999}%
  \BibitemOpen
  \bibfield  {author} {\bibinfo {author} {\bibfnamefont {X.}~\bibnamefont
  {Shen}}, \bibinfo {author} {\bibfnamefont {P.}~\bibnamefont {Ramvall}},
  \bibinfo {author} {\bibfnamefont {P.}~\bibnamefont {Riblet}}, \ and\ \bibinfo
  {author} {\bibfnamefont {Y.}~\bibnamefont {Aoyagi}},\ }\href {\doibase
  10.1143/JJAP.38.L411} {\bibfield  {journal} {\bibinfo  {journal} {Japanese
  Journal of Applied Physics}\ }\textbf {\bibinfo {volume} {38}},\ \bibinfo
  {pages} {L411} (\bibinfo {year} {1999})}\BibitemShut {NoStop}%
\bibitem [{\citenamefont {Shen}\ and\ \citenamefont
  {Aoyagi}(1999)}]{Shen1999a}%
  \BibitemOpen
  \bibfield  {author} {\bibinfo {author} {\bibfnamefont {X.-Q.}\ \bibnamefont
  {Shen}}\ and\ \bibinfo {author} {\bibfnamefont {Y.}~\bibnamefont {Aoyagi}},\
  }\href {\doibase 10.1143/JJAP.38.L14} {\bibfield  {journal} {\bibinfo
  {journal} {Japanese Journal of Applied Physics}\ }\textbf {\bibinfo {volume}
  {38}},\ \bibinfo {pages} {L14} (\bibinfo {year} {1999})}\BibitemShut
  {NoStop}%
\bibitem [{\citenamefont {Kumano}\ \emph {et~al.}(2001)\citenamefont {Kumano},
  \citenamefont {Hoshi}, \citenamefont {Tanaka}, \citenamefont {Suemune},
  \citenamefont {Shen}, \citenamefont {Riblet}, \citenamefont {Ramvall},\ and\
  \citenamefont {Aoyagi}}]{Kumano2001}%
  \BibitemOpen
  \bibfield  {author} {\bibinfo {author} {\bibfnamefont {H.}~\bibnamefont
  {Kumano}}, \bibinfo {author} {\bibfnamefont {K.-i.}\ \bibnamefont {Hoshi}},
  \bibinfo {author} {\bibfnamefont {S.}~\bibnamefont {Tanaka}}, \bibinfo
  {author} {\bibfnamefont {I.}~\bibnamefont {Suemune}}, \bibinfo {author}
  {\bibfnamefont {X.-Q.}\ \bibnamefont {Shen}}, \bibinfo {author}
  {\bibfnamefont {P.}~\bibnamefont {Riblet}}, \bibinfo {author} {\bibfnamefont
  {P.}~\bibnamefont {Ramvall}}, \ and\ \bibinfo {author} {\bibfnamefont
  {Y.}~\bibnamefont {Aoyagi}},\ }\href {\doibase 10.1063/1.125178} {\bibfield
  {journal} {\bibinfo  {journal} {Applied Physics Letters}\ }\textbf {\bibinfo
  {volume} {75}},\ \bibinfo {pages} {2879} (\bibinfo {year}
  {2001})}\BibitemShut {NoStop}%
\bibitem [{\citenamefont {Liu}\ \emph {et~al.}(2016{\natexlab{a}})\citenamefont
  {Liu}, \citenamefont {Carlin}, \citenamefont {Grandjean}, \citenamefont
  {Deveaud},\ and\ \citenamefont {Jacopin}}]{Liu2016a}%
  \BibitemOpen
  \bibfield  {author} {\bibinfo {author} {\bibfnamefont {W.}~\bibnamefont
  {Liu}}, \bibinfo {author} {\bibfnamefont {J.-F.}\ \bibnamefont {Carlin}},
  \bibinfo {author} {\bibfnamefont {N.}~\bibnamefont {Grandjean}}, \bibinfo
  {author} {\bibfnamefont {B.}~\bibnamefont {Deveaud}}, \ and\ \bibinfo
  {author} {\bibfnamefont {G.}~\bibnamefont {Jacopin}},\ }\href {\doibase
  10.1063/1.4959832} {\bibfield  {journal} {\bibinfo  {journal} {Applied
  Physics Letters}\ }\textbf {\bibinfo {volume} {109}},\ \bibinfo {pages}
  {042101} (\bibinfo {year} {2016}{\natexlab{a}})}\BibitemShut {NoStop}%
\bibitem [{\citenamefont {Hayama}\ \emph {et~al.}(2017)\citenamefont {Hayama},
  \citenamefont {Takahashi},\ and\ \citenamefont {Usami}}]{Hayama2017}%
  \BibitemOpen
  \bibfield  {author} {\bibinfo {author} {\bibfnamefont {Y.}~\bibnamefont
  {Hayama}}, \bibinfo {author} {\bibfnamefont {I.}~\bibnamefont {Takahashi}}, \
  and\ \bibinfo {author} {\bibfnamefont {N.}~\bibnamefont {Usami}},\ }\href
  {\doibase 10.1016/j.egypro.2017.09.088} {\bibfield  {journal} {\bibinfo
  {journal} {Energy Procedia}\ }\textbf {\bibinfo {volume} {124}},\ \bibinfo
  {pages} {734} (\bibinfo {year} {2017})}\BibitemShut {NoStop}%
\bibitem [{\citenamefont {Callsen}\ \emph {et~al.}(2011)\citenamefont
  {Callsen}, \citenamefont {Reparaz}, \citenamefont {Wagner}, \citenamefont
  {Kirste}, \citenamefont {Nenstiel}, \citenamefont {Hoffmann},\ and\
  \citenamefont {Phillips}}]{Callsen2011}%
  \BibitemOpen
  \bibfield  {author} {\bibinfo {author} {\bibfnamefont {G.}~\bibnamefont
  {Callsen}}, \bibinfo {author} {\bibfnamefont {J.~S.}\ \bibnamefont
  {Reparaz}}, \bibinfo {author} {\bibfnamefont {M.~R.}\ \bibnamefont {Wagner}},
  \bibinfo {author} {\bibfnamefont {R.}~\bibnamefont {Kirste}}, \bibinfo
  {author} {\bibfnamefont {C.}~\bibnamefont {Nenstiel}}, \bibinfo {author}
  {\bibfnamefont {A.}~\bibnamefont {Hoffmann}}, \ and\ \bibinfo {author}
  {\bibfnamefont {M.~R.}\ \bibnamefont {Phillips}},\ }\href {\doibase
  10.1063/1.3554434} {\bibfield  {journal} {\bibinfo  {journal} {Applied
  Physics Letters}\ }\textbf {\bibinfo {volume} {98}},\ \bibinfo {pages}
  {061906} (\bibinfo {year} {2011})}\BibitemShut {NoStop}%
\bibitem [{\citenamefont {Davydov}\ \emph {et~al.}(1998)\citenamefont
  {Davydov}, \citenamefont {Kitaev}, \citenamefont {Goncharuk}, \citenamefont
  {Smirnov}, \citenamefont {Graul}, \citenamefont {Semchinova}, \citenamefont
  {Uffmann}, \citenamefont {Smirnov}, \citenamefont {Mirgorodsky},\ and\
  \citenamefont {Evarestov}}]{Davydov1998a}%
  \BibitemOpen
  \bibfield  {author} {\bibinfo {author} {\bibfnamefont {V.~Y.}\ \bibnamefont
  {Davydov}}, \bibinfo {author} {\bibfnamefont {Y.~E.}\ \bibnamefont {Kitaev}},
  \bibinfo {author} {\bibfnamefont {I.~N.}\ \bibnamefont {Goncharuk}}, \bibinfo
  {author} {\bibfnamefont {A.~N.}\ \bibnamefont {Smirnov}}, \bibinfo {author}
  {\bibfnamefont {J.}~\bibnamefont {Graul}}, \bibinfo {author} {\bibfnamefont
  {O.}~\bibnamefont {Semchinova}}, \bibinfo {author} {\bibfnamefont
  {D.}~\bibnamefont {Uffmann}}, \bibinfo {author} {\bibfnamefont {M.~B.}\
  \bibnamefont {Smirnov}}, \bibinfo {author} {\bibfnamefont {A.~P.}\
  \bibnamefont {Mirgorodsky}}, \ and\ \bibinfo {author} {\bibfnamefont {R.~A.}\
  \bibnamefont {Evarestov}},\ }\href
  {papers2://publication/uuid/CEFAA132-96B6-40A0-9CFF-ABA66FB56581} {\bibfield
  {journal} {\bibinfo  {journal} {Physical Review B}\ }\textbf {\bibinfo
  {volume} {58}},\ \bibinfo {pages} {12899} (\bibinfo {year}
  {1998})}\BibitemShut {NoStop}%
\bibitem [{\citenamefont {P{\"{a}}ssler}(2001)}]{Passler2001}%
  \BibitemOpen
  \bibfield  {author} {\bibinfo {author} {\bibfnamefont {R.}~\bibnamefont
  {P{\"{a}}ssler}},\ }\href {\doibase 10.1063/1.1369407} {\bibfield  {journal}
  {\bibinfo  {journal} {Journal of Applied Physics}\ }\textbf {\bibinfo
  {volume} {89}},\ \bibinfo {pages} {6235} (\bibinfo {year}
  {2001})}\BibitemShut {NoStop}%
\bibitem [{\citenamefont {Song}\ \emph {et~al.}(2006)\citenamefont {Song},
  \citenamefont {Basavaraj}, \citenamefont {Nikishin}, \citenamefont {Holtz},
  \citenamefont {Soukhoveev}, \citenamefont {Usikov},\ and\ \citenamefont
  {Dmitriev}}]{Song2006}%
  \BibitemOpen
  \bibfield  {author} {\bibinfo {author} {\bibfnamefont {D.~Y.}\ \bibnamefont
  {Song}}, \bibinfo {author} {\bibfnamefont {M.}~\bibnamefont {Basavaraj}},
  \bibinfo {author} {\bibfnamefont {S.~A.}\ \bibnamefont {Nikishin}}, \bibinfo
  {author} {\bibfnamefont {M.}~\bibnamefont {Holtz}}, \bibinfo {author}
  {\bibfnamefont {V.}~\bibnamefont {Soukhoveev}}, \bibinfo {author}
  {\bibfnamefont {A.}~\bibnamefont {Usikov}}, \ and\ \bibinfo {author}
  {\bibfnamefont {V.}~\bibnamefont {Dmitriev}},\ }\href {\doibase
  10.1063/1.2361159} {\bibfield  {journal} {\bibinfo  {journal} {Journal of
  Applied Physics}\ }\textbf {\bibinfo {volume} {100}},\ \bibinfo {pages}
  {113504} (\bibinfo {year} {2006})}\BibitemShut {NoStop}%
\bibitem [{\citenamefont {Callsen}\ \emph
  {et~al.}(2015{\natexlab{a}})\citenamefont {Callsen}, \citenamefont {Pahn},
  \citenamefont {Kalinowski}, \citenamefont {Kindel}, \citenamefont {Settke},
  \citenamefont {Brunnmeier}, \citenamefont {Nenstiel}, \citenamefont {Kure},
  \citenamefont {Nippert}, \citenamefont {Schliwa}, \citenamefont {Hoffmann},
  \citenamefont {Markurt}, \citenamefont {Schulz}, \citenamefont {Albrecht},
  \citenamefont {Kako}, \citenamefont {Arita},\ and\ \citenamefont
  {Arakawa}}]{Callsen2015d}%
  \BibitemOpen
  \bibfield  {author} {\bibinfo {author} {\bibfnamefont {G.}~\bibnamefont
  {Callsen}}, \bibinfo {author} {\bibfnamefont {G.}~\bibnamefont {Pahn}},
  \bibinfo {author} {\bibfnamefont {S.}~\bibnamefont {Kalinowski}}, \bibinfo
  {author} {\bibfnamefont {C.}~\bibnamefont {Kindel}}, \bibinfo {author}
  {\bibfnamefont {J.}~\bibnamefont {Settke}}, \bibinfo {author} {\bibfnamefont
  {J.}~\bibnamefont {Brunnmeier}}, \bibinfo {author} {\bibfnamefont
  {C.}~\bibnamefont {Nenstiel}}, \bibinfo {author} {\bibfnamefont
  {T.}~\bibnamefont {Kure}}, \bibinfo {author} {\bibfnamefont {F.}~\bibnamefont
  {Nippert}}, \bibinfo {author} {\bibfnamefont {A.}~\bibnamefont {Schliwa}},
  \bibinfo {author} {\bibfnamefont {A.}~\bibnamefont {Hoffmann}}, \bibinfo
  {author} {\bibfnamefont {T.}~\bibnamefont {Markurt}}, \bibinfo {author}
  {\bibfnamefont {T.}~\bibnamefont {Schulz}}, \bibinfo {author} {\bibfnamefont
  {M.}~\bibnamefont {Albrecht}}, \bibinfo {author} {\bibfnamefont
  {S.}~\bibnamefont {Kako}}, \bibinfo {author} {\bibfnamefont {M.}~\bibnamefont
  {Arita}}, \ and\ \bibinfo {author} {\bibfnamefont {Y.}~\bibnamefont
  {Arakawa}},\ }\href {\doibase 10.1103/PhysRevB.92.235439} {\bibfield
  {journal} {\bibinfo  {journal} {Physical Review B}\ }\textbf {\bibinfo
  {volume} {92}},\ \bibinfo {pages} {235439} (\bibinfo {year}
  {2015}{\natexlab{a}})}\BibitemShut {NoStop}%
\bibitem [{\citenamefont {Davydov}\ \emph {et~al.}(1999)\citenamefont
  {Davydov}, \citenamefont {Emtsev}, \citenamefont {Goncharuk}, \citenamefont
  {Smirnov}, \citenamefont {Petrikov}, \citenamefont {Mamutin}, \citenamefont
  {Vekshin}, \citenamefont {Ivanov}, \citenamefont {Smirnov},\ and\
  \citenamefont {Inushima}}]{Davydov1999}%
  \BibitemOpen
  \bibfield  {author} {\bibinfo {author} {\bibfnamefont {V.~Y.}\ \bibnamefont
  {Davydov}}, \bibinfo {author} {\bibfnamefont {V.~V.}\ \bibnamefont {Emtsev}},
  \bibinfo {author} {\bibfnamefont {I.~N.}\ \bibnamefont {Goncharuk}}, \bibinfo
  {author} {\bibfnamefont {A.~N.}\ \bibnamefont {Smirnov}}, \bibinfo {author}
  {\bibfnamefont {V.~D.}\ \bibnamefont {Petrikov}}, \bibinfo {author}
  {\bibfnamefont {V.~V.}\ \bibnamefont {Mamutin}}, \bibinfo {author}
  {\bibfnamefont {V.~A.}\ \bibnamefont {Vekshin}}, \bibinfo {author}
  {\bibfnamefont {S.~V.}\ \bibnamefont {Ivanov}}, \bibinfo {author}
  {\bibfnamefont {M.~B.}\ \bibnamefont {Smirnov}}, \ and\ \bibinfo {author}
  {\bibfnamefont {T.}~\bibnamefont {Inushima}},\ }\href
  {papers2://publication/uuid/04FCCD67-5005-40EB-A7CF-927476802887} {\bibfield
  {journal} {\bibinfo  {journal} {Applied Physics Letters}\ }\textbf {\bibinfo
  {volume} {75}},\ \bibinfo {pages} {3297} (\bibinfo {year}
  {1999})}\BibitemShut {NoStop}%
\bibitem [{\citenamefont {Nenstiel}\ \emph {et~al.}(2015)\citenamefont
  {Nenstiel}, \citenamefont {B{\"{u}}gler},\ and\ \citenamefont
  {Callsen}}]{Nenstiel2015a}%
  \BibitemOpen
  \bibfield  {author} {\bibinfo {author} {\bibfnamefont {C.}~\bibnamefont
  {Nenstiel}}, \bibinfo {author} {\bibfnamefont {M.}~\bibnamefont
  {B{\"{u}}gler}}, \ and\ \bibinfo {author} {\bibfnamefont {G.}~\bibnamefont
  {Callsen}},\ }\href
  {http://onlinelibrary.wiley.com/doi/10.1002/pssr.201510278/full
  file:///Users/gordito/Documents/Papers2/Articles/2015/Nenstiel/Unknown/2015
  Nenstiel.pdf papers2://publication/doi/10.1002/pssr.201510278} {\bibfield
  {journal} {\bibinfo  {journal} {Physica Status Solidi (RRL)}\ }\textbf
  {\bibinfo {volume} {9}},\ \bibinfo {pages} {716} (\bibinfo {year}
  {2015})}\BibitemShut {NoStop}%
\bibitem [{\citenamefont {Bezyazychnaya}\ \emph {et~al.}(2015)\citenamefont
  {Bezyazychnaya}, \citenamefont {Kabanau}, \citenamefont {Kabanov},
  \citenamefont {Lebiadok}, \citenamefont {Ryabtsev}, \citenamefont {Ryabtsev},
  \citenamefont {Zelenkovskii},\ and\ \citenamefont
  {Mehta}}]{Bezyazychnaya2015}%
  \BibitemOpen
  \bibfield  {author} {\bibinfo {author} {\bibfnamefont {T.~V.}\ \bibnamefont
  {Bezyazychnaya}}, \bibinfo {author} {\bibfnamefont {D.~M.}\ \bibnamefont
  {Kabanau}}, \bibinfo {author} {\bibfnamefont {V.~V.}\ \bibnamefont
  {Kabanov}}, \bibinfo {author} {\bibfnamefont {Y.~V.}\ \bibnamefont
  {Lebiadok}}, \bibinfo {author} {\bibfnamefont {A.~G.}\ \bibnamefont
  {Ryabtsev}}, \bibinfo {author} {\bibfnamefont {G.~I.}\ \bibnamefont
  {Ryabtsev}}, \bibinfo {author} {\bibfnamefont {V.~M.}\ \bibnamefont
  {Zelenkovskii}}, \ and\ \bibinfo {author} {\bibfnamefont {S.~K.}\
  \bibnamefont {Mehta}},\ }\href {\doibase 10.3952/physics.v55i1.3053}
  {\bibfield  {journal} {\bibinfo  {journal} {Lithuanian Journal of Physics}\
  }\textbf {\bibinfo {volume} {55}},\ \bibinfo {pages} {10} (\bibinfo {year}
  {2015})}\BibitemShut {NoStop}%
\bibitem [{\citenamefont {Liu}\ \emph {et~al.}(2016{\natexlab{b}})\citenamefont
  {Liu}, \citenamefont {Fu}, \citenamefont {Yi}, \citenamefont {Yuan},\ and\
  \citenamefont {Wang}}]{Liu2016}%
  \BibitemOpen
  \bibfield  {author} {\bibinfo {author} {\bibfnamefont {Z.}~\bibnamefont
  {Liu}}, \bibinfo {author} {\bibfnamefont {B.}~\bibnamefont {Fu}}, \bibinfo
  {author} {\bibfnamefont {X.}~\bibnamefont {Yi}}, \bibinfo {author}
  {\bibfnamefont {G.}~\bibnamefont {Yuan}}, \ and\ \bibinfo {author}
  {\bibfnamefont {J.}~\bibnamefont {Wang}},\ }\href {\doibase
  10.1039/C5RA24642C} {\bibfield  {journal} {\bibinfo  {journal} {RSC
  Advances}\ }\textbf {\bibinfo {volume} {6}},\ \bibinfo {pages} {5111}
  (\bibinfo {year} {2016}{\natexlab{b}})}\BibitemShut {NoStop}%
\bibitem [{\citenamefont {Reynolds}\ \emph {et~al.}(2000)\citenamefont
  {Reynolds}, \citenamefont {Look},\ and\ \citenamefont
  {Jogai}}]{Reynolds2000a}%
  \BibitemOpen
  \bibfield  {author} {\bibinfo {author} {\bibfnamefont {D.~C.}\ \bibnamefont
  {Reynolds}}, \bibinfo {author} {\bibfnamefont {D.~C.}\ \bibnamefont {Look}},
  \ and\ \bibinfo {author} {\bibfnamefont {B.}~\bibnamefont {Jogai}},\ }\href
  {papers2://publication/uuid/65E0DE3E-0ABB-4F3B-81FA-6421C1D89EFD} {\bibfield
  {journal} {\bibinfo  {journal} {Journal of Applied Physics}\ }\textbf
  {\bibinfo {volume} {88}},\ \bibinfo {pages} {5760} (\bibinfo {year}
  {2000})}\BibitemShut {NoStop}%
\bibitem [{\citenamefont {Kindel}(2010)}]{Kindel2010}%
  \BibitemOpen
  \bibfield  {author} {\bibinfo {author} {\bibfnamefont {C.~H.}\ \bibnamefont
  {Kindel}},\ }\emph {\bibinfo {title} {{Study on Optical Polarization in
  Hexagonal Gallium Nitride Quantum Dots}}},\ \href
  {papers2://publication/uuid/38700713-F1C0-4FA5-9D5A-B2A4CE61456A} {Ph.D.
  thesis},\ \bibinfo  {school} {University of Tokyo} (\bibinfo {year}
  {2010})\BibitemShut {NoStop}%
\bibitem [{\citenamefont {Kindel}\ \emph {et~al.}(2014)\citenamefont {Kindel},
  \citenamefont {Callsen}, \citenamefont {Kako}, \citenamefont {Kawano},
  \citenamefont {Oishi}, \citenamefont {H{\"{o}}nig}, \citenamefont {Schliwa},
  \citenamefont {Hoffmann},\ and\ \citenamefont {Arakawa}}]{Kindel2014}%
  \BibitemOpen
  \bibfield  {author} {\bibinfo {author} {\bibfnamefont {C.}~\bibnamefont
  {Kindel}}, \bibinfo {author} {\bibfnamefont {G.}~\bibnamefont {Callsen}},
  \bibinfo {author} {\bibfnamefont {S.}~\bibnamefont {Kako}}, \bibinfo {author}
  {\bibfnamefont {T.}~\bibnamefont {Kawano}}, \bibinfo {author} {\bibfnamefont
  {H.}~\bibnamefont {Oishi}}, \bibinfo {author} {\bibfnamefont
  {G.}~\bibnamefont {H{\"{o}}nig}}, \bibinfo {author} {\bibfnamefont
  {A.}~\bibnamefont {Schliwa}}, \bibinfo {author} {\bibfnamefont
  {A.}~\bibnamefont {Hoffmann}}, \ and\ \bibinfo {author} {\bibfnamefont
  {Y.}~\bibnamefont {Arakawa}},\ }\href
  {http://onlinelibrary.wiley.com/doi/10.1002/pssr.201409096/full
  file:///Users/gordito/Documents/Papers2/Articles/2014/Kindel/Phys Status
  Solidi - Rapid Research Letters/Phys Status Solidi - Rapid Research Letters
  2014 Kindel.pdf papers2://publication/doi/1} {\bibfield  {journal} {\bibinfo
  {journal} {Phys Status Solidi - Rapid Research Letters}\ }\textbf {\bibinfo
  {volume} {8}},\ \bibinfo {pages} {408} (\bibinfo {year} {2014})}\BibitemShut
  {NoStop}%
\bibitem [{\citenamefont {Holmes}\ \emph {et~al.}(2015)\citenamefont {Holmes},
  \citenamefont {Kako}, \citenamefont {Choi}, \citenamefont {Arita},\ and\
  \citenamefont {Arakawa}}]{Holmes2015}%
  \BibitemOpen
  \bibfield  {author} {\bibinfo {author} {\bibfnamefont {M.}~\bibnamefont
  {Holmes}}, \bibinfo {author} {\bibfnamefont {S.}~\bibnamefont {Kako}},
  \bibinfo {author} {\bibfnamefont {K.}~\bibnamefont {Choi}}, \bibinfo {author}
  {\bibfnamefont {M.}~\bibnamefont {Arita}}, \ and\ \bibinfo {author}
  {\bibfnamefont {Y.}~\bibnamefont {Arakawa}},\ }\href {\doibase
  10.1103/PhysRevB.92.115447} {\bibfield  {journal} {\bibinfo  {journal}
  {Physical Review B}\ }\textbf {\bibinfo {volume} {92}},\ \bibinfo {pages}
  {115447} (\bibinfo {year} {2015})}\BibitemShut {NoStop}%
\bibitem [{\citenamefont {Callsen}\ \emph
  {et~al.}(2015{\natexlab{b}})\citenamefont {Callsen}, \citenamefont {Pahn},
  \citenamefont {Kalinowski}, \citenamefont {Kindel}, \citenamefont {Settke},
  \citenamefont {Brunnmeier}, \citenamefont {Nenstiel}, \citenamefont {Kure},
  \citenamefont {Nippert}, \citenamefont {Schliwa}, \citenamefont {Hoffmann},
  \citenamefont {Markurt}, \citenamefont {Schulz}, \citenamefont {Albrecht},
  \citenamefont {Kako}, \citenamefont {Arita},\ and\ \citenamefont
  {Arakawa}}]{Callsen2015}%
  \BibitemOpen
  \bibfield  {author} {\bibinfo {author} {\bibfnamefont {G.}~\bibnamefont
  {Callsen}}, \bibinfo {author} {\bibfnamefont {G.}~\bibnamefont {Pahn}},
  \bibinfo {author} {\bibfnamefont {S.}~\bibnamefont {Kalinowski}}, \bibinfo
  {author} {\bibfnamefont {C.}~\bibnamefont {Kindel}}, \bibinfo {author}
  {\bibfnamefont {J.}~\bibnamefont {Settke}}, \bibinfo {author} {\bibfnamefont
  {J.}~\bibnamefont {Brunnmeier}}, \bibinfo {author} {\bibfnamefont
  {C.}~\bibnamefont {Nenstiel}}, \bibinfo {author} {\bibfnamefont
  {T.}~\bibnamefont {Kure}}, \bibinfo {author} {\bibfnamefont {F.}~\bibnamefont
  {Nippert}}, \bibinfo {author} {\bibfnamefont {A.}~\bibnamefont {Schliwa}},
  \bibinfo {author} {\bibfnamefont {A.}~\bibnamefont {Hoffmann}}, \bibinfo
  {author} {\bibfnamefont {T.}~\bibnamefont {Markurt}}, \bibinfo {author}
  {\bibfnamefont {T.}~\bibnamefont {Schulz}}, \bibinfo {author} {\bibfnamefont
  {M.}~\bibnamefont {Albrecht}}, \bibinfo {author} {\bibfnamefont
  {S.}~\bibnamefont {Kako}}, \bibinfo {author} {\bibfnamefont {M.}~\bibnamefont
  {Arita}}, \ and\ \bibinfo {author} {\bibfnamefont {Y.}~\bibnamefont
  {Arakawa}},\ }\href {\doibase 10.1103/PhysRevB.92.235439} {\bibfield
  {journal} {\bibinfo  {journal} {Physical Review B}\ }\textbf {\bibinfo
  {volume} {92}},\ \bibinfo {pages} {235439} (\bibinfo {year}
  {2015}{\natexlab{b}})}\BibitemShut {NoStop}%
\bibitem [{\citenamefont {Ostapenko}\ \emph {et~al.}(2012)\citenamefont
  {Ostapenko}, \citenamefont {H{\"{o}}nig}, \citenamefont {Rodt}, \citenamefont
  {Schliwa}, \citenamefont {Hoffmann}, \citenamefont {Bimberg}, \citenamefont
  {Dachner}, \citenamefont {Richter}, \citenamefont {Knorr}, \citenamefont
  {Kako},\ and\ \citenamefont {Arakawa}}]{Ostapenko2012}%
  \BibitemOpen
  \bibfield  {author} {\bibinfo {author} {\bibfnamefont {I.~A.}\ \bibnamefont
  {Ostapenko}}, \bibinfo {author} {\bibfnamefont {G.}~\bibnamefont
  {H{\"{o}}nig}}, \bibinfo {author} {\bibfnamefont {S.}~\bibnamefont {Rodt}},
  \bibinfo {author} {\bibfnamefont {A.}~\bibnamefont {Schliwa}}, \bibinfo
  {author} {\bibfnamefont {A.}~\bibnamefont {Hoffmann}}, \bibinfo {author}
  {\bibfnamefont {D.}~\bibnamefont {Bimberg}}, \bibinfo {author} {\bibfnamefont
  {M.~R.}\ \bibnamefont {Dachner}}, \bibinfo {author} {\bibfnamefont
  {M.}~\bibnamefont {Richter}}, \bibinfo {author} {\bibfnamefont
  {A.}~\bibnamefont {Knorr}}, \bibinfo {author} {\bibfnamefont
  {S.}~\bibnamefont {Kako}}, \ and\ \bibinfo {author} {\bibfnamefont
  {Y.}~\bibnamefont {Arakawa}},\ }\href {\doibase 10.1103/PhysRevB.85.081303}
  {\bibfield  {journal} {\bibinfo  {journal} {Physical Review B}\ }\textbf
  {\bibinfo {volume} {85}},\ \bibinfo {pages} {081303(R)} (\bibinfo {year}
  {2012})}\BibitemShut {NoStop}%
\bibitem [{\citenamefont {Meyer}\ \emph {et~al.}(2010)\citenamefont {Meyer},
  \citenamefont {Sann}, \citenamefont {Eisermann}, \citenamefont
  {Lautenschlaeger}, \citenamefont {Wagner}, \citenamefont {Kaiser},
  \citenamefont {Callsen}, \citenamefont {Reparaz},\ and\ \citenamefont
  {Hoffmann}}]{Meyer2010a}%
  \BibitemOpen
  \bibfield  {author} {\bibinfo {author} {\bibfnamefont {B.~K.}\ \bibnamefont
  {Meyer}}, \bibinfo {author} {\bibfnamefont {J.}~\bibnamefont {Sann}},
  \bibinfo {author} {\bibfnamefont {S.}~\bibnamefont {Eisermann}}, \bibinfo
  {author} {\bibfnamefont {S.}~\bibnamefont {Lautenschlaeger}}, \bibinfo
  {author} {\bibfnamefont {M.~R.}\ \bibnamefont {Wagner}}, \bibinfo {author}
  {\bibfnamefont {M.}~\bibnamefont {Kaiser}}, \bibinfo {author} {\bibfnamefont
  {G.}~\bibnamefont {Callsen}}, \bibinfo {author} {\bibfnamefont {J.~S.}\
  \bibnamefont {Reparaz}}, \ and\ \bibinfo {author} {\bibfnamefont
  {A.}~\bibnamefont {Hoffmann}},\ }\href
  {http://prb.aps.org/abstract/PRB/v82/i11/e115207
  file:///Users/gordito/Documents/Papers2/Articles/2010/Meyer/Phys Rev B/Phys
  Rev B 2010 Meyer.pdf papers2://publication/doi/10.1103/PhysRevB.82.115207}
  {\bibfield  {journal} {\bibinfo  {journal} {Physical Review B}\ }\textbf
  {\bibinfo {volume} {82}},\ \bibinfo {pages} {115207} (\bibinfo {year}
  {2010})}\BibitemShut {NoStop}%
\bibitem [{\citenamefont {Nozawa}\ and\ \citenamefont
  {Horikoshi}(1991)}]{Takano1991}%
  \BibitemOpen
  \bibfield  {author} {\bibinfo {author} {\bibfnamefont {K.}~\bibnamefont
  {Nozawa}}\ and\ \bibinfo {author} {\bibfnamefont {Y.}~\bibnamefont
  {Horikoshi}},\ }\href@noop {} {\bibfield  {journal} {\bibinfo  {journal}
  {Japanese Journal of Applied Physics}\ }\textbf {\bibinfo {volume} {30}},\
  \bibinfo {pages} {L668} (\bibinfo {year} {1991})}\BibitemShut {NoStop}%
\end{thebibliography}%


\begin{thebibliography}{14}%
\makeatletter
\providecommand \@ifxundefined [1]{%
 \@ifx{#1\undefined}
}%
\providecommand \@ifnum [1]{%
 \ifnum #1\expandafter \@firstoftwo
 \else \expandafter \@secondoftwo
 \fi
}%
\providecommand \@ifx [1]{%
 \ifx #1\expandafter \@firstoftwo
 \else \expandafter \@secondoftwo
 \fi
}%
\providecommand \natexlab [1]{#1}%
\providecommand \enquote  [1]{``#1''}%
\providecommand \bibnamefont  [1]{#1}%
\providecommand \bibfnamefont [1]{#1}%
\providecommand \citenamefont [1]{#1}%
\providecommand \href@noop [0]{\@secondoftwo}%
\providecommand \href [0]{\begingroup \@sanitize@url \@href}%
\providecommand \@href[1]{\@@startlink{#1}\@@href}%
\providecommand \@@href[1]{\endgroup#1\@@endlink}%
\providecommand \@sanitize@url [0]{\catcode `\\12\catcode `\$12\catcode
  `\&12\catcode `\#12\catcode `\^12\catcode `\_12\catcode `\%12\relax}%
\providecommand \@@startlink[1]{}%
\providecommand \@@endlink[0]{}%
\providecommand \url  [0]{\begingroup\@sanitize@url \@url }%
\providecommand \@url [1]{\endgroup\@href {#1}{\urlprefix }}%
\providecommand \urlprefix  [0]{URL }%
\providecommand \Eprint [0]{\href }%
\providecommand \doibase [0]{http://dx.doi.org/}%
\providecommand \selectlanguage [0]{\@gobble}%
\providecommand \bibinfo  [0]{\@secondoftwo}%
\providecommand \bibfield  [0]{\@secondoftwo}%
\providecommand \translation [1]{[#1]}%
\providecommand \BibitemOpen [0]{}%
\providecommand \bibitemStop [0]{}%
\providecommand \bibitemNoStop [0]{.\EOS\space}%
\providecommand \EOS [0]{\spacefactor3000\relax}%
\providecommand \BibitemShut  [1]{\csname bibitem#1\endcsname}%
\let\auto@bib@innerbib\@empty
\bibitem [{\citenamefont {Motoki}\ \emph {et~al.}(2001)\citenamefont {Motoki},
  \citenamefont {Kahisa}, \citenamefont {Atsumoto}, \citenamefont {Atsushima},\
  and\ \citenamefont {Imura}}]{Motoki2001}%
  \BibitemOpen
  \bibfield  {author} {\bibinfo {author} {\bibfnamefont {K.~M.}\ \bibnamefont
  {Motoki}}, \bibinfo {author} {\bibfnamefont {T.~O.}\ \bibnamefont {Kahisa}},
  \bibinfo {author} {\bibfnamefont {N.~M.}\ \bibnamefont {Atsumoto}}, \bibinfo
  {author} {\bibfnamefont {M.~M.}\ \bibnamefont {Atsushima}}, \ and\ \bibinfo
  {author} {\bibfnamefont {H.~K.}\ \bibnamefont {Imura}},\ }\href@noop {}
  {\bibfield  {journal} {\bibinfo  {journal} {Japanese Journal of Applied
  Physics}\ }\textbf {\bibinfo {volume} {40}},\ \bibinfo {pages} {L140}
  (\bibinfo {year} {2001})}\BibitemShut {NoStop}%
\bibitem [{\citenamefont {Rousseau}\ \emph {et~al.}(2018)\citenamefont
  {Rousseau}, \citenamefont {Callsen}, \citenamefont {Jacopin}, \citenamefont
  {Carlin}, \citenamefont {Butt{\'{e}}},\ and\ \citenamefont
  {Grandjean}}]{Rousseau2018}%
  \BibitemOpen
  \bibfield  {author} {\bibinfo {author} {\bibfnamefont {I.}~\bibnamefont
  {Rousseau}}, \bibinfo {author} {\bibfnamefont {G.}~\bibnamefont {Callsen}},
  \bibinfo {author} {\bibfnamefont {G.}~\bibnamefont {Jacopin}}, \bibinfo
  {author} {\bibfnamefont {J.-F.}\ \bibnamefont {Carlin}}, \bibinfo {author}
  {\bibfnamefont {R.}~\bibnamefont {Butt{\'{e}}}}, \ and\ \bibinfo {author}
  {\bibfnamefont {N.}~\bibnamefont {Grandjean}},\ }\href {\doibase
  10.1063/1.5022150} {\bibfield  {journal} {\bibinfo  {journal} {Journal of
  Applied Physics}\ }\textbf {\bibinfo {volume} {123}},\ \bibinfo {pages}
  {113103} (\bibinfo {year} {2018})}\BibitemShut {NoStop}%
\bibitem [{\citenamefont {Kako}(2007)}]{Kako2007}%
  \BibitemOpen
  \bibfield  {author} {\bibinfo {author} {\bibfnamefont {S.}~\bibnamefont
  {Kako}},\ }\emph {\bibinfo {title} {{Optical properties of gallium nitride
  self-assembled quantum dots and application to generation of non-classical
  light}}},\ \href
  {papers2://publication/uuid/62F57316-1A23-48E8-8CF8-0FCB2F1130BA} {Ph.D.
  thesis},\ \bibinfo  {school} {University of Tokyo} (\bibinfo {year}
  {2007})\BibitemShut {NoStop}%
\bibitem [{\citenamefont {Mott}(1987)}]{Mott1987}%
  \BibitemOpen
  \bibfield  {author} {\bibinfo {author} {\bibfnamefont {N.}~\bibnamefont
  {Mott}},\ }\href@noop {} {\bibfield  {journal} {\bibinfo  {journal} {Journal
  of Physics C}\ }\textbf {\bibinfo {volume} {20}},\ \bibinfo {pages} {3075}
  (\bibinfo {year} {1987})}\BibitemShut {NoStop}%
\bibitem [{\citenamefont {Rodina}\ \emph {et~al.}(2001)\citenamefont {Rodina},
  \citenamefont {Dietrich}, \citenamefont {G{\"{o}}ldner}, \citenamefont
  {Eckey}, \citenamefont {Hoffmann}, \citenamefont {Efros}, \citenamefont
  {Rosen},\ and\ \citenamefont {Meyer}}]{Rodina2001}%
  \BibitemOpen
  \bibfield  {author} {\bibinfo {author} {\bibfnamefont {A.}~\bibnamefont
  {Rodina}}, \bibinfo {author} {\bibfnamefont {M.}~\bibnamefont {Dietrich}},
  \bibinfo {author} {\bibfnamefont {A.}~\bibnamefont {G{\"{o}}ldner}}, \bibinfo
  {author} {\bibfnamefont {L.}~\bibnamefont {Eckey}}, \bibinfo {author}
  {\bibfnamefont {A.}~\bibnamefont {Hoffmann}}, \bibinfo {author}
  {\bibfnamefont {A.}~\bibnamefont {Efros}}, \bibinfo {author} {\bibfnamefont
  {M.}~\bibnamefont {Rosen}}, \ and\ \bibinfo {author} {\bibfnamefont
  {B.}~\bibnamefont {Meyer}},\ }\href {\doibase 10.1103/PhysRevB.64.115204}
  {\bibfield  {journal} {\bibinfo  {journal} {Physical Review B}\ }\textbf
  {\bibinfo {volume} {64}},\ \bibinfo {pages} {115204} (\bibinfo {year}
  {2001})}\BibitemShut {NoStop}%
\bibitem [{\citenamefont {Bimberg}\ \emph {et~al.}(1971)\citenamefont
  {Bimberg}, \citenamefont {Sondergeld},\ and\ \citenamefont
  {Grobe}}]{Bimberg1971}%
  \BibitemOpen
  \bibfield  {author} {\bibinfo {author} {\bibfnamefont {D.}~\bibnamefont
  {Bimberg}}, \bibinfo {author} {\bibfnamefont {M.}~\bibnamefont {Sondergeld}},
  \ and\ \bibinfo {author} {\bibfnamefont {E.}~\bibnamefont {Grobe}},\ }\href
  {http://prb.aps.org/abstract/PRB/v4/i10/p3451{\_}1
  papers2://publication/uuid/4CF4FE39-4DBD-4028-A9A3-6C2B4278C4FD} {\bibfield
  {journal} {\bibinfo  {journal} {Physical Review B}\ }\textbf {\bibinfo
  {volume} {4}},\ \bibinfo {pages} {3451} (\bibinfo {year} {1971})}\BibitemShut
  {NoStop}%
\bibitem [{\citenamefont {Callsen}\ \emph {et~al.}(2011)\citenamefont
  {Callsen}, \citenamefont {Reparaz}, \citenamefont {Wagner}, \citenamefont
  {Kirste}, \citenamefont {Nenstiel}, \citenamefont {Hoffmann},\ and\
  \citenamefont {Phillips}}]{Callsen2011}%
  \BibitemOpen
  \bibfield  {author} {\bibinfo {author} {\bibfnamefont {G.}~\bibnamefont
  {Callsen}}, \bibinfo {author} {\bibfnamefont {J.~S.}\ \bibnamefont
  {Reparaz}}, \bibinfo {author} {\bibfnamefont {M.~R.}\ \bibnamefont {Wagner}},
  \bibinfo {author} {\bibfnamefont {R.}~\bibnamefont {Kirste}}, \bibinfo
  {author} {\bibfnamefont {C.}~\bibnamefont {Nenstiel}}, \bibinfo {author}
  {\bibfnamefont {A.}~\bibnamefont {Hoffmann}}, \ and\ \bibinfo {author}
  {\bibfnamefont {M.~R.}\ \bibnamefont {Phillips}},\ }\href {\doibase
  10.1063/1.3554434} {\bibfield  {journal} {\bibinfo  {journal} {Applied
  Physics Letters}\ }\textbf {\bibinfo {volume} {98}},\ \bibinfo {pages}
  {061906} (\bibinfo {year} {2011})}\BibitemShut {NoStop}%
\bibitem [{\citenamefont {Morko{\c{c}}}(2009)}]{Morkoc2009}%
  \BibitemOpen
  \bibfield  {author} {\bibinfo {author} {\bibfnamefont {H.}~\bibnamefont
  {Morko{\c{c}}}},\ }\href
  {http://books.google.com/books?hl=en{\&}lr={\&}id=00TL1DmoA0QC{\&}oi=fnd{\&}pg=PR5{\&}dq=Handbook+of+Nitride+Semiconductors+and+Devices{\&}ots=YPHWwCzocx{\&}sig=Foxz8pKTwIeDVUCgQsPaNPYg0qA
  papers2://publication/uuid/335DFA91-CC0F-45AB-927F-E3B13BAF794B} {\emph
  {\bibinfo {title} {{Handbook of Nitride Semiconductors and Devices}}}},\
  \bibinfo {edition} {1st}\ ed.\ (\bibinfo  {publisher} {Wiley},\ \bibinfo
  {address} {Weinheim},\ \bibinfo {year} {2009})\BibitemShut {NoStop}%
\bibitem [{\citenamefont {Smeeton}\ \emph {et~al.}(2006)\citenamefont
  {Smeeton}, \citenamefont {Humphreys}, \citenamefont {Barnard},\ and\
  \citenamefont {Kappers}}]{Smeeton2006}%
  \BibitemOpen
  \bibfield  {author} {\bibinfo {author} {\bibfnamefont {T.~M.}\ \bibnamefont
  {Smeeton}}, \bibinfo {author} {\bibfnamefont {C.~J.}\ \bibnamefont
  {Humphreys}}, \bibinfo {author} {\bibfnamefont {J.~S.}\ \bibnamefont
  {Barnard}}, \ and\ \bibinfo {author} {\bibfnamefont {M.~J.}\ \bibnamefont
  {Kappers}},\ }\href {\doibase 10.1007/s10853-006-7876-x} {\bibfield
  {journal} {\bibinfo  {journal} {Journal of Materials Science}\ }\textbf
  {\bibinfo {volume} {41}},\ \bibinfo {pages} {2729} (\bibinfo {year}
  {2006})}\BibitemShut {NoStop}%
\bibitem [{\citenamefont {Cerezo}\ \emph {et~al.}(2007)\citenamefont {Cerezo},
  \citenamefont {Clifton}, \citenamefont {Galtrey}, \citenamefont {Humphreys},
  \citenamefont {Kelly}, \citenamefont {Larson}, \citenamefont {Lozano-Perez},
  \citenamefont {Marquis}, \citenamefont {Oliver}, \citenamefont {Sha},
  \citenamefont {Thompson}, \citenamefont {Zandbergen},\ and\ \citenamefont
  {Alvis}}]{Cerezo2007}%
  \BibitemOpen
  \bibfield  {author} {\bibinfo {author} {\bibfnamefont {A.}~\bibnamefont
  {Cerezo}}, \bibinfo {author} {\bibfnamefont {P.~H.}\ \bibnamefont {Clifton}},
  \bibinfo {author} {\bibfnamefont {M.~J.}\ \bibnamefont {Galtrey}}, \bibinfo
  {author} {\bibfnamefont {C.~J.}\ \bibnamefont {Humphreys}}, \bibinfo {author}
  {\bibfnamefont {T.~F.}\ \bibnamefont {Kelly}}, \bibinfo {author}
  {\bibfnamefont {D.~J.}\ \bibnamefont {Larson}}, \bibinfo {author}
  {\bibfnamefont {S.}~\bibnamefont {Lozano-Perez}}, \bibinfo {author}
  {\bibfnamefont {E.~A.}\ \bibnamefont {Marquis}}, \bibinfo {author}
  {\bibfnamefont {R.~A.}\ \bibnamefont {Oliver}}, \bibinfo {author}
  {\bibfnamefont {G.}~\bibnamefont {Sha}}, \bibinfo {author} {\bibfnamefont
  {K.}~\bibnamefont {Thompson}}, \bibinfo {author} {\bibfnamefont
  {M.}~\bibnamefont {Zandbergen}}, \ and\ \bibinfo {author} {\bibfnamefont
  {R.~L.}\ \bibnamefont {Alvis}},\ }\href {\doibase
  10.1016/S1369-7021(07)70306-1} {\bibfield  {journal} {\bibinfo  {journal}
  {Materials Today}\ }\textbf {\bibinfo {volume} {10}},\ \bibinfo {pages} {36}
  (\bibinfo {year} {2007})}\BibitemShut {NoStop}%
\bibitem [{\citenamefont {Narukawa}\ \emph {et~al.}(1997)\citenamefont
  {Narukawa}, \citenamefont {Kawakami}, \citenamefont {Funato}, \citenamefont
  {Fujita}, \citenamefont {Fujita},\ and\ \citenamefont
  {Nakamura}}]{Narukawa1997}%
  \BibitemOpen
  \bibfield  {author} {\bibinfo {author} {\bibfnamefont {Y.}~\bibnamefont
  {Narukawa}}, \bibinfo {author} {\bibfnamefont {Y.}~\bibnamefont {Kawakami}},
  \bibinfo {author} {\bibfnamefont {M.}~\bibnamefont {Funato}}, \bibinfo
  {author} {\bibfnamefont {S.}~\bibnamefont {Fujita}}, \bibinfo {author}
  {\bibfnamefont {S.}~\bibnamefont {Fujita}}, \ and\ \bibinfo {author}
  {\bibfnamefont {S.}~\bibnamefont {Nakamura}},\ }\href {\doibase
  10.1063/1.118455} {\bibfield  {journal} {\bibinfo  {journal} {Applied Physics
  Letters}\ }\textbf {\bibinfo {volume} {70}},\ \bibinfo {pages} {981}
  (\bibinfo {year} {1997})}\BibitemShut {NoStop}%
\bibitem [{\citenamefont {Galtrey}\ \emph {et~al.}(2007)\citenamefont
  {Galtrey}, \citenamefont {Oliver}, \citenamefont {Kappers}, \citenamefont
  {Humphreys}, \citenamefont {Stokes}, \citenamefont {Clifton},\ and\
  \citenamefont {Cerezo}}]{Galtrey2007}%
  \BibitemOpen
  \bibfield  {author} {\bibinfo {author} {\bibfnamefont {M.~J.}\ \bibnamefont
  {Galtrey}}, \bibinfo {author} {\bibfnamefont {R.~A.}\ \bibnamefont {Oliver}},
  \bibinfo {author} {\bibfnamefont {M.~J.}\ \bibnamefont {Kappers}}, \bibinfo
  {author} {\bibfnamefont {C.~J.}\ \bibnamefont {Humphreys}}, \bibinfo {author}
  {\bibfnamefont {D.~J.}\ \bibnamefont {Stokes}}, \bibinfo {author}
  {\bibfnamefont {P.~H.}\ \bibnamefont {Clifton}}, \ and\ \bibinfo {author}
  {\bibfnamefont {A.}~\bibnamefont {Cerezo}},\ }\href {\doibase
  10.1063/1.2431573} {\bibfield  {journal} {\bibinfo  {journal} {Applied
  Physics Letters}\ }\textbf {\bibinfo {volume} {90}},\ \bibinfo {pages}
  {061903} (\bibinfo {year} {2007})}\BibitemShut {NoStop}%
\bibitem [{\citenamefont {Wu}\ \emph {et~al.}(2012)\citenamefont {Wu},
  \citenamefont {Shivaraman}, \citenamefont {Wang},\ and\ \citenamefont
  {Speck}}]{Wu2012}%
  \BibitemOpen
  \bibfield  {author} {\bibinfo {author} {\bibfnamefont {Y.-R.}\ \bibnamefont
  {Wu}}, \bibinfo {author} {\bibfnamefont {R.}~\bibnamefont {Shivaraman}},
  \bibinfo {author} {\bibfnamefont {K.-C.}\ \bibnamefont {Wang}}, \ and\
  \bibinfo {author} {\bibfnamefont {J.~S.}\ \bibnamefont {Speck}},\ }\href
  {\doibase 10.1063/1.4747532} {\bibfield  {journal} {\bibinfo  {journal}
  {Applied Physics Letters}\ }\textbf {\bibinfo {volume} {101}},\ \bibinfo
  {pages} {083505} (\bibinfo {year} {2012})}\BibitemShut {NoStop}%
\bibitem [{\citenamefont {Schulz}\ \emph {et~al.}(2012)\citenamefont {Schulz},
  \citenamefont {Remmele}, \citenamefont {Markurt}, \citenamefont {Korytov},\
  and\ \citenamefont {Albrecht}}]{Schulz2012}%
  \BibitemOpen
  \bibfield  {author} {\bibinfo {author} {\bibfnamefont {T.}~\bibnamefont
  {Schulz}}, \bibinfo {author} {\bibfnamefont {T.}~\bibnamefont {Remmele}},
  \bibinfo {author} {\bibfnamefont {T.}~\bibnamefont {Markurt}}, \bibinfo
  {author} {\bibfnamefont {M.}~\bibnamefont {Korytov}}, \ and\ \bibinfo
  {author} {\bibfnamefont {M.}~\bibnamefont {Albrecht}},\ }\href {\doibase
  10.1063/1.4742015} {\bibfield  {journal} {\bibinfo  {journal} {Journal of
  Applied Physics}\ }\textbf {\bibinfo {volume} {112}},\ \bibinfo {pages}
  {033106} (\bibinfo {year} {2012})}\BibitemShut {NoStop}%
\end{thebibliography}%
\end{document}


\title{Supplementary Information \\ - \\ Probing alloy formation using different excitonic species: \\ The particular case of InGaN}
\author{G. Callsen}
\email{gordon.callsen@epfl.ch}
\author{R. Butt\'{e}}
\author{N. Grandjean}
\affiliation{Institute of Physics, \'{E}cole Polytechnique F\'{e}d\'{e}rale de Lausanne (EPFL), CH-1015 Lausanne, Switzerland}
\date{\today}
%
%
%
%

%
%
%
%
\pacs{}
\maketitle

\section{Experimental details} \label{Experimental}

The investigated In$_x$Ga$_{1-x}$N (0\,$\leq$\,$x$\,$\leq$\,2.4\%) samples were grown by metalorganic vapor-phase epitaxy on $c$-plane, random core, freestanding GaN substrates \cite{Motoki2001} ($n$-type) obtained from a hydride vapor phase epitaxy based overgrowth technique yielding a dislocation density of $\approx10^{6}\,\text{cm}^{-2}$. First, a 360-nm-thick GaN buffer is grown at a temperature of 1000\,$^\circ$C with tri-methyl-gallium and H$_2$ used as carrier gas. Second, a 500-nm-thick Al$_{0.06}$Ga$_{0.94}$N layer is deposited at 1000\,$^\circ$C with tri-methyl-aluminum, tri-methyl-gallium, and H$_2$ used as carrier gas. Subsequently, N$_2$ is used as carrier gas in order to deposit the final epilayer of InGaN with a thickness of 100\,nm based on tri-methyl-gallium and tri-methyl-indium at a temperature of around 770\,$^\circ$C. The use of an Al$_{0.06}$Ga$_{0.94}$N interlayer allows to inhibit carrier migration and to study the top In$_x$Ga$_{1-x}$N (0\,$\leq$\,$x$\,$\leq$\,2.4\%) layer optically without any significant luminescence contribution from the underlying GaN buffer and substrate. 

In order to control the indium incorporation in between $x=0-2.4\%$ we grew a sequence of reference InGaN epilayers at a constant flow rate of the tri-methyl-indium source, while the growth temperature ($T_g$) was varied around 770\,$^\circ$C in sufficiently small steps derived from the method described in the following. We determined the resulting indium concentration down to $x\approx 1\%$ by high-resolution x-ray diffraction (HRXRD) measurements that we partially confirmed by a secondary ion mass spectroscopy (SIMS) analysis. As a result, we obtained $x(T_{g})$, which we linearly extrapolated for $x \rightarrow 0$ in order to control the indium incorporation for $x<1\%$ by just varying $T_g$. Again, for $x=0.05\%$ we cross-checked the feasibility of our approach by SIMS. From our PL spectra we can extract a linear dependence for the peak position of $\mathrm{X_A}$ as shown in the inset of Fig.\,2(a) for $x=0-2.4\%$. Again, this observation supports the feasibility of our linear approximation for $x(T_{g})$ for $x \rightarrow 0$. Finally, photoluminescence (PL) proves the most sensitive and versatile technique in order to study the alloy content at the very onset of alloy formation. Further SIMS measurements confirmed that the oxygen and silicon concentrations in our samples are below the detection limit ($\leq 2 \times 10^{16}\,\text{cm}^{-3}$), while the carbon concentration amounts to $\approx 1 \times 10^{16}\,\text{cm}^{-3}$.

All macro-PL spectra were measured with the samples situated in a helium-flow or a closed-cycle cryostat. A cw HeCd laser emitting at 325\,nm (Kimmon Koha Co., Ltd.) was used as excitation source. For the macro-PL measurements the excitation light was guided towards the sample under normal incidence ($\textbf{k}\,\parallel\,\textbf{c}$), while being focused by a conventional lens with a focal length of 30\,cm (excitation spot size diameter: $d_{exc}\approx 100\,\mu\text{m}$). The samples' luminescence was also collected with the $\textbf{k}\,\parallel\,\textbf{c}$ geometry using a lens with a numerical aperture (NA) of $\approx$\,0.2, before being imaged onto a Horiba Jobin Yvon iHR320 single monochromator (32\,cm focal length) equipped with a 2400\,l/mm grating (400\,nm blaze wavelength), which we employed in second order to achieve an optical resolution better than 300\,$\mu$eV in the energy regime of interest. Finally, the dispersed light was monitored with a UV-enhanced, charge-coupled-device (CCD) array from Horiba (128$\times$1024 pixels). Each emission linewidth value was determined by recording PL spectra from up to ten different positions on the sample. The error bars shown in Fig.\,3(a) of the main text originate from such multiple measurements on each sample.

The micro-photoluminescence ($\mu$-PL) measurements were performed with a fully customized setup optimized for the ultraviolet (UV) spectral range. All details of the setup can be found in the supplementary material of Ref.\,\cite{Rousseau2018}. The samples were mounted in a closed-cycle helium cryostat (Cryostation C2 from Montana Instruments, Inc.), while being excited by a cw HeCd laser emitting at 325\,nm. The diameter of the laser spot amounted to $d_{exc}\approx\,1 \mu \text{m}$ as the laser beam was expanded in order to fill around 80\% of the objective back aperture (20$\times$, NA\,=\,0.4). Detection of the luminescence was based on a nitrogen-cooled, UV-enhanced charge-coupled device (Symphony II from Horiba Jobin-Yvon), while the light was dispersed by a FHR 640 single monochromator (Horiba Jobin-Yvon) with a focal length of 64\,cm. We used a 1800\,l/mm grating (blaze wavelength of 400\,nm) in second order and slit settings of 10-25\,$\mu$m to facilitate a spectral resolution of better than 150\,$\mu$eV in the energy range of interest. Suitable optical filters were applied in order to lower the straylight level in the single monochromator (required to enable integration times up to 30\,min for selected $\mu$-PL spectra comprising a high dynamic range). All spectra  were calibrated based on a neon spectral calibration lamp. Subsequently, wavelength was converted to energy without applying a correction for the refractive index of air in order to make the present results more comparable to the commonly available literature treating bound excitons in GaN. Application of such a vacuum correction would shift in the following all reported absolute peak energies by around 900\,$\mu$eV to lower energies.

Apertures were defined in an opaque aluminum film by electron-beam- (diameters down to 200\,nm) and photo-lithography (diameters down to 1\,$\mu$m), while the metal film itself was always deposited using electron-beam deposition. During the deposition of the film the distance between the sample and the metal source heated by an electron-beam was about 1\,m in order to exclude any electron-beam-induced sample damage \cite{Kako2007}. Finally, the apertures were either etched into the metal film or generated by a metal lift-off procedure (positive vs. negative resists). For all our processing attempts we observed closely matching $\mu$-PL results.

\section{Temperature-dependent photoluminescence}
\label{temperature}
\begin{figure}[]
\includegraphics[scale=0.85]{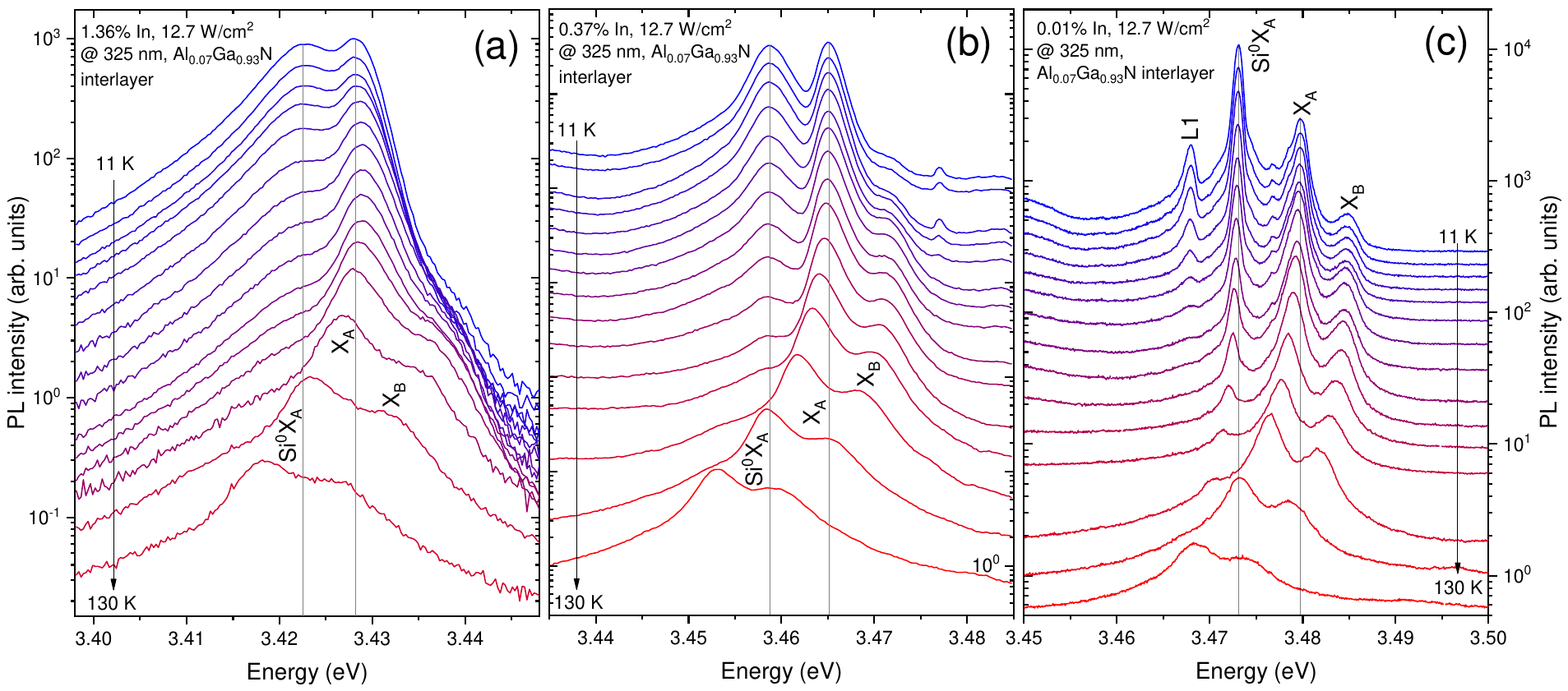} 
\caption{(color online) Onset of the s-like shift of the $\mathrm{X_{A}}$ transition in In$_x$Ga$_{1-x}$N epilayers for indium contents $x$ up to 1.36\%. (c) At $x=0.01\%$ all bound (L1 and $\mathrm{Si^{0}X_{A}}$) and free excitonic emission lines ($\mathrm{X_{A}}$ and $\mathrm{X_{B}}$) continuously shift towards lower energies with rising temperature following the evolution of the bandgap. (b) At $x=0.37\%$ this redshift diminishes for $\mathrm{X_{A}}$ due to the onset of exciton localization. Finally, at $x=1.36\%$ the typical, s-like shift of the $\mathrm{X_{A}}$ transition becomes noticeable, while such a behavior is not present for $\mathrm{Si^{0}X_{A}}$. All peak positions are summarized in Fig.\,\ref{temperature_shifts}.}
\label{temperature_series}
\end{figure}

Figure\,\ref{temperature_series} introduces exemplary temperature-dependent PL data, which we recorded for our entire In$_x$Ga$_{1-x}$N sample series (0\,$\leq$\,$x$\,$\leq$\,2.4\%). As introduced in the main text, a set of impurity-bound (L1 and  $\mathrm{Si^{0}X_{A}}$) and "free" excitons ($\mathrm{X_{A}}$ and $\mathrm{X_{B}}$) can be observed. Here, the terminology of free excitons needs to be treated with care as those particles get increasingly bound in the alloy at indium-related assemblies with rising $x$. Figure\,3(b) of the main text illustrates this continuous transition from free to bound excitons (two-particle complexes), which must be clearly distinguished from impurity-bound complexes such as $\mathrm{Si^{0}X_{A}}$ (three-particle complex). At the onset of our alloy series ($x=0.01\%$), all transitions follow the evolution of the bandgap with rising temperature as shown in Fig.\,\ref{temperature_series}(c). However, as soon as $x$ increases to 0.37\% [Fig.\,\ref{temperature_series}(b)] the redshift, e.g., for $\mathrm{X_A}$ is suppressed at the very onset of the temperature series and turns into a subtle blueshift reaching a maximum value of $\approx0.8\,\text{meV}$ at a temperature of $\approx 35\,\text{K}$ as depicted by the spectrum of Fig.\,\ref{temperature_series}(c) for $x=1.36\%$. In contrast, the $\mathrm{Si^{0}X_{A}}$ emission only shows a continuous redshift with rising temperature for all $x$ values we analyzed.

\begin{figure}[]
\includegraphics[width=10cm]{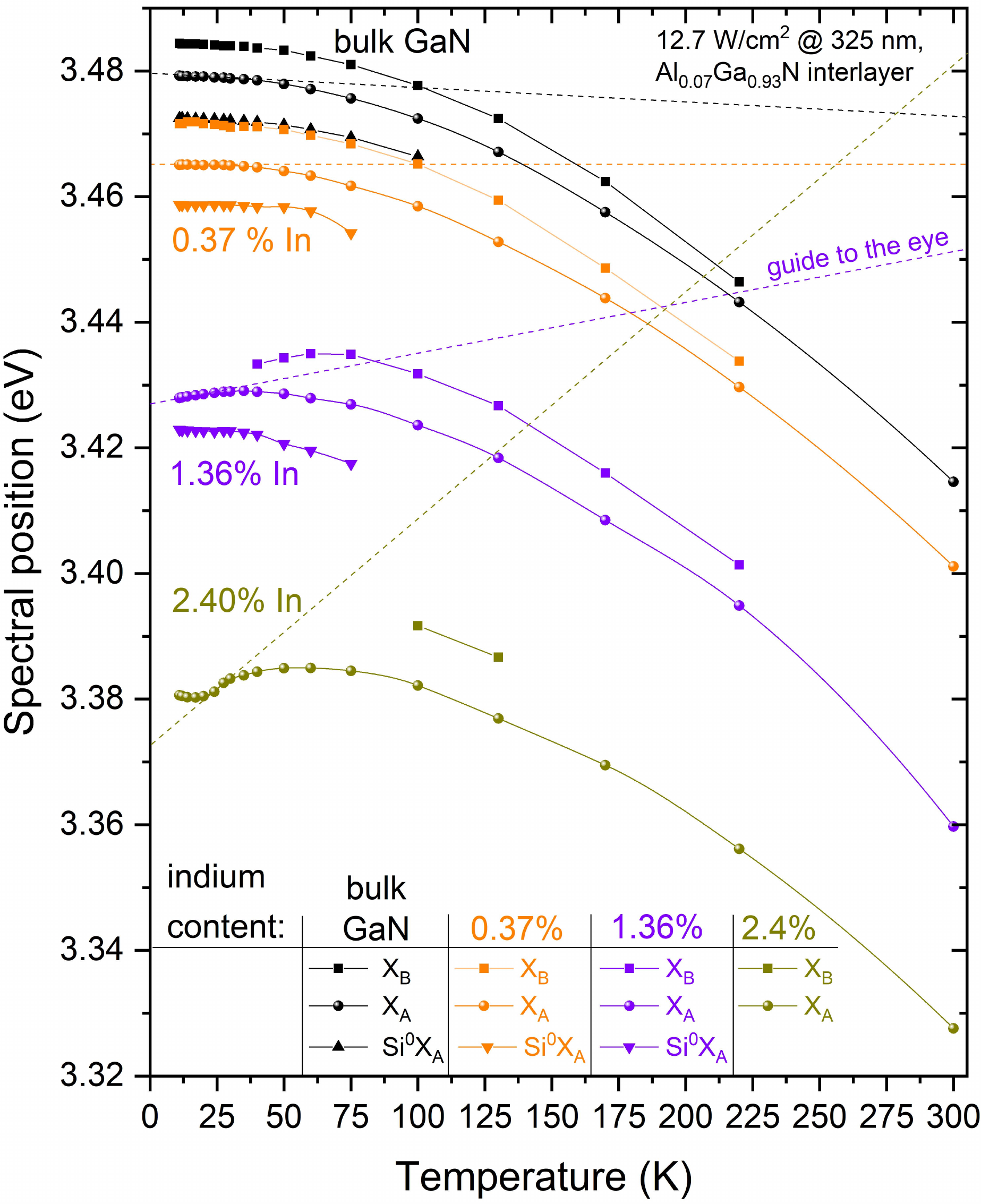} 
\caption{(color online) Summary of the emission lines' shifts for $\mathrm{X_{A}}$, $\mathrm{X_{B}}$, and $\mathrm{Si^{0}X_{A}}$ partially extracted from Fig.\,\ref{temperature_series}. At an indium concentration of $x$\,=\,0.37\% one can observe that the common redshift of the $\mathrm{X_{A}}$ transitions with rising temperature is diminished. Here, the dashed lines are guides to the eye in order to illustrate this aspect. At $x=1.36\%$ a blueshift of $\approx0.8\,\text{meV}$ can be observed for $\mathrm{X_A}$ due to exciton re-distribution into deeper potential traps. With rising indium content (see $x=2.4\%$) more and also energetically deeper, indium-related potential traps are formed, leading to a more pronounced s-like shift of the main emission line $\mathrm{X_{A}}$.}
\label{temperature_shifts}
\end{figure}

Figure\,\ref{temperature_shifts} summarizes the temperature-dependent spectral positions for $\mathrm{Si^{0}X_{A}}$, $\mathrm{X_{A}}$, and $\mathrm{X_{B}}$ for 0\,$\leq$\,$x$\,$\leq$\,2.4\%. At $x=2.4\%$ it is not possible to track the position of $\mathrm{Si^{0}X_{A}}$ vs. temperature anymore as already at a temperature of 12\,K the spectral separation in between $\mathrm{X_{A}}$ and $\mathrm{Si^{0}X_{A}}$ is on the order of the corresponding full width at half maximum (FWHM) values, cf. Fig.\,2(a) of the main text. Clearly, for $x=0.37\%$ the redshift of $\mathrm{X_{A}}$ with rising temperature is delayed in comparison to pure GaN (cf. Fig.\,\ref{temperature_shifts}). However, an indium concentration of $\sim$1.36\% is required in order to observe the onset of the typical s-like shift of $\mathrm{X_{A}}$. Figure\,\ref{temperature_shifts} includes some guides to the eye in order to illustrate such emission line shifts at the beginning of the temperature series. For $x=1.36\%$ we observe a maximum blueshift of just $\approx0.8\,\text{meV}$ at 35\,K. As shown in Fig.\,\ref{temperature_shifts} the absolute value of this blueshift increases to $\approx4.7\,\text{meV}$ for $x=2.4\%$ and reaches its maximum at an accordingly higher temperature of $\approx 60\,\text{K}$, which will be explained in the following. Naturally, with increasing $x$ the average size of indium-related assemblies increases, causing more pronounced fluctuations in the potential landscape of the alloy. It is exactly the rising number and also the size of such indium assemblies that impact the associated exciton localization energy ($E_{loc}$) and facilitates the transition from a free to an increasingly bound $\mathrm{X_{A}}$ complex as described in the context of Fig.\,3(b) in the main text. Upon rising temperature the $E_{loc}$ values of such bound $\mathrm{X_{A}}$ complexes can be overcome, leading to their thermal redistribution in the potential landscape. As a result, higher energy states (more shallowly bound $\mathrm{X_{A}}$) become accessible with rising temperature, leading to the clear, s-like shift of $\mathrm{X_{A}}$ at $x=2.4\%$. At the onset of the corresponding temperature series a minor redshift is observed due to bandgap shrinkage that still dominates the thermal redistribution of $\mathrm{X_{A}}$ in the potential landscape. However, a blueshift is observed as soon as this thermal redistribution of $\mathrm{X_{A}}$ takes over the bandgap shrinkage. This evolution is followed by a pronounced redshift at elevated temperatures that is accompanied by the conversion of bound excitons into free ones, a process which is often referred to in the context of the mobility edge \cite{Mott1987}. Figure\,\ref{temperature_shifts} demonstrates that the blueshift for the $\mathrm{X_{B}}$ complex is delayed with respect to $\mathrm{X_{A}}$ for $x=1.36\%$. We ascribe this difference to the change in hole mass between the A- and B-valence subbands in GaN \cite{Rodina2001} and ultimately InGaN.

Interestingly, the temperature-dependent shift of $\mathrm{Si^{0}X_{A}}$ does not exhibit any blueshift nor delayed redshift as depicted in Fig.\,\ref{temperature_shifts} for up to $x=1.36\%$. The excitons bound to $\mathrm{Si^{0}}$ are distributed in the potential landscape of the alloy. However, no thermal redistribution is noticeable as the binding energy $E_{bind}$ (i.e., the splitting between $\mathrm{Si^{0}X_{A}}$ and $\mathrm{X_{A}}$ measured at 12\,K) of the exciton is always larger than the thermal energy $E_{bind}>E_{th}$, otherwise dissociation of the bound exciton occurs in a multi-step process \cite{Bimberg1971}. Naturally, the two competing statistical processes of exciton capture by $\mathrm{Si^{0}}$ and the subsequent thermal release of the exciton contribute to the particular lineshape of the $\mathrm{Si^{0}X_{A}}$ emission band with rising $x$ and temperature. $E_{bind}$ of $\mathrm{Si^{0}X_{A}}$ varies from $6.8 \pm 0.1\,\text{meV}$, over $6.3 \pm 0.1\,\text{meV}$, to $5.0 \pm 0.1\,\text{meV}$ for $x=0$, 0.37\%, and 1.36\%, respectively, pointing towards a slight reduction in the effective donor binding energy in the In$_x$Ga$_{1-x}$N alloy. The maximum total redshift of $\mathrm{Si^{0}X_{A}}$ that occurs in between 11\,K and 75\,K can be compared for these three indium contents. Here, we observe an increasing redshift of $\mathrm{Si^{0}X_{A}}$ with respect to $\mathrm{X_{A}}$ that varies from $3.1 \pm 0.1\,\text{meV}$ for $x=0$, to $4.5 \pm 0.1\,\text{meV}$ for $x=0.37\%$, and $5.4 \pm 0.1\,\text{meV}$ for $x=1.36\%$.

Hence, with rising $x$ an increasing number of deeply bound $\mathrm{Si^{0}X_{A}-In^{\textit{n}}}$ complexes is formed comprising a rising number of indium atoms $n$ in agreement with our discussion given for Fig.\,4 of the main text. In addition, Fig.\,5 therein introduced the $\mu$-PL signature of such complexes, which are mainly distributed over an energy range given by the FWHM of the $\mathrm{Si^{0}X_{A}}$ emission band (see the PL measurements from Fig.\,2 in the main text). Our $\mu$-PL study from Sec.\,IIID of the main text finally suggests two different types of transitions that contribute to the overall $\mathrm{Si^{0}X_{A}}$ emission and the associated particular linewidth evolution over temperature that is impacted by the indium content.

\section{Temperature-dependent linewidth broadening} 
\label{TLWB}
\begin{figure}[]
\includegraphics[width=10cm]{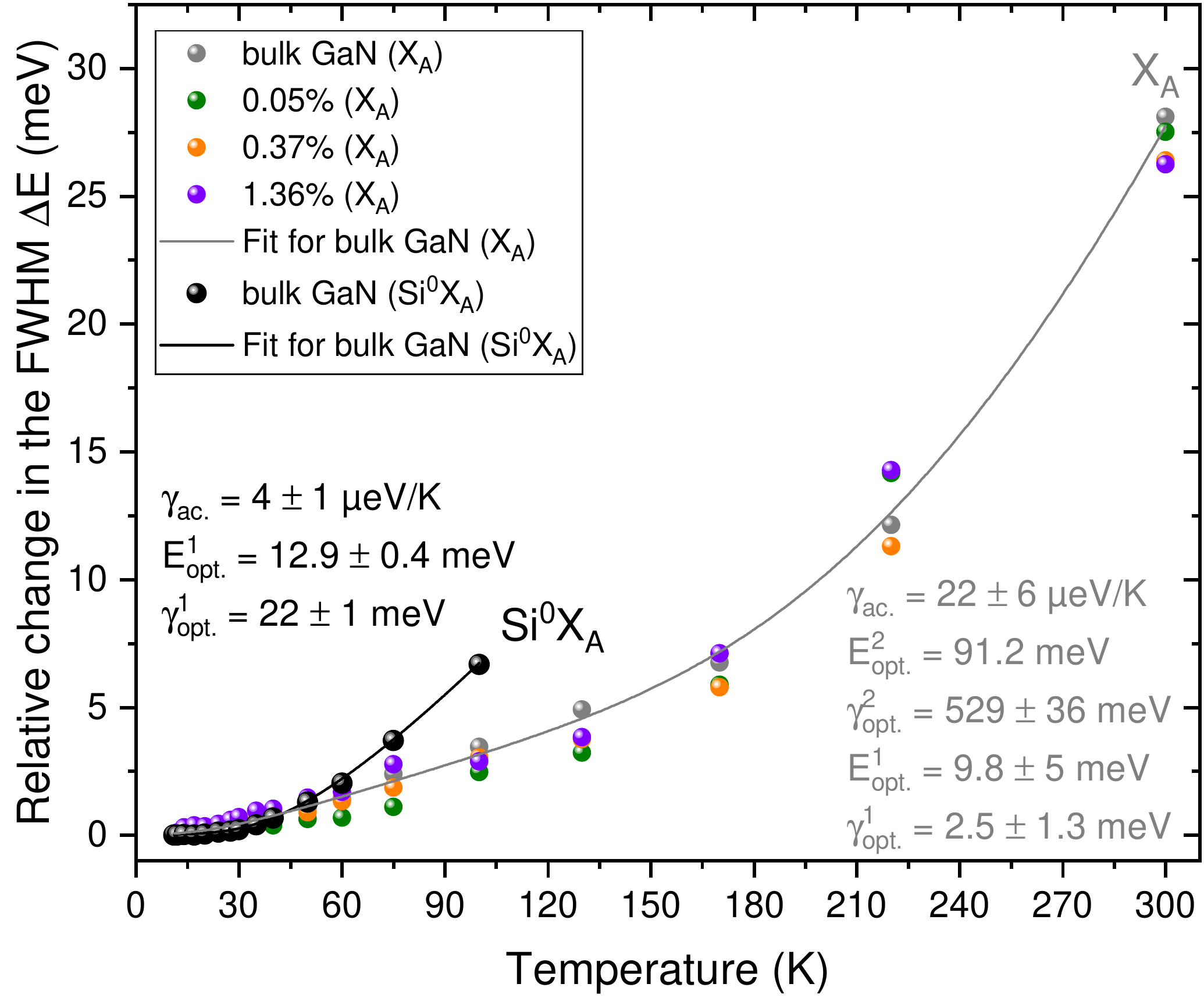}
\caption{(color online) The temperature-dependent evolution of the FWHM ($\Delta E$) of the $\mathrm{X_{A}}$ transition does not show a pronounced dependence on the alloy content for the analyzed indium content interval 0\,$\leq$\,$x$\,$\leq$\,1.36\%. In contrast, the temperature-dependent broadening of $\mathrm{Si^{0}X_{A}}$  is heavily influenced by $x$ as shown in Fig.\,4 of the main text. All symbols and values are explained in the main text of the manuscript along with the fits (solid lines) that are based on Eq.\,3 therein.}
\label{temperature_FWHM}
\end{figure}

Basic features of the temperature-dependent linewidth broadening for both, the $\mathrm{Si^{0}X_{A}}$ and the $\mathrm{X_{A}}$ complexes, are shown in Fig.\,\ref{temperature_FWHM}. For $\mathrm{X_{A}}$ no clear alloy-dependent trend can be identified - the temperature-dependent FWHM values ($\Delta E$) overlap within the experimental error. The best fit to the data for the $\mathrm{X_{A}}$ complex based on Eq.\,3 of the main text is obtained for $\gamma_{ac} = 22 \pm 6 \, \mu \text{eV/K}$ and two sets of $\gamma_{opt}^{i}$ and $E_{opt}^{i}$ values as denoted in Fig.\,\ref{temperature_FWHM}. At room temperature the broadening of $\mathrm{X_{A}}$ is mainly affected by longitudinal-optical (LO) phonons in GaN (energy fixed at $E^{2}_{opt}=91.2\,\text{meV}$ \cite{Callsen2011} for the fitting procedure), while with decreasing temperature, first optical phonons with an average energy of $E_{opt}^{1}=9.8 \pm 5\,\text{meV}$ account for the broadening, followed by acoustic phonons. Here, $\gamma_{ac}=22 \pm 6\,\mu\text{eV/K}$ is close to reported literature values of $\approx 15-16\,\mu\text{eV/K}$ \cite{Morkoc2009} for $\mathrm{X_{A}}$ and the large error for the $E_{opt}^{1}$ phonon energy originates from the over-parametrized fitting function, despite the fact that a fixed phonon energy of 91.2\,meV was used for $E_{opt}^{2}$. Nevertheless, the value of $E_{opt}^{1}$ is lower than the energy of the $E_{2}^{low}$ phonon mode in GaN of 17.85\,meV \cite{Callsen2011}, as the momentum distribution of $\mathrm{X_{A}}$ allows accessing a larger fraction of the first Brillouin zone as shown in the inset of Fig.\,4(a) of the main text.

%
%

Figure\,\ref{temperature_FWHM} also includes the temperature-induced broadening of $\mathrm{Si^{0}X_{A}}$ at $x=0$ as introduced in Fig.\,4(a) of the main text. Here, the broadening is more pronounced with respect to the case of $\mathrm{X_{A}}$ as the coupling constant $\gamma_{opt}^1$ related to $E_{opt}^{1}$ is larger in comparison to $\mathrm{X_{A}}$ due to the strong localization of impurity bound excitons. 


%
%
%
\section{Micro-photoluminescence spectroscopy} 
\label{muPL}
%
%
%

%
%

Figure\,5(a) of the main text shows two $\mu$-PL spectra on a relative energy scale recorded for $x=0.37\%$ and 1.36\% on the bare, unprocessed sample surface (laser spot diameter $d_{exc}\approx 1\,\mu \text{m}$). In both cases the $\mathrm{X_{A}}$ emission band is unstructured, while the onset of an ensemble of sharp emission lines can be observed around the relative energy of $\mathrm{Si^{0}X_{A}}$. Generally, the detection of these sharp emission lines at $x=0.37\%$ proves very sensitive to any rise in excitation density or temperature. Naturally, one expects that an effective reduction of $d_{exc}$ should lead to clearly resolved emission lines. Hence, we fabricated apertures with diameters down to 200\,nm into an opaque aluminum layer deposited on the sample surface by different processing techniques (see SI Sec.\,\ref{Experimental}). Nevertheless, the observation shown in Fig.\,5(a) of the main text is crucial as it proves that such sharp emission lines are not induced by the processing steps. Generally, any such steps (e.g., 100\,kV e-beam exposure, metal deposition, lift-off, etching, etc.) could damage the InGaN material leading to exciton localization and sharp emission lines. For instance, it was found that InGaN is very sensitive to high-voltage electron irradiation causing a migration of indium atoms \cite{Smeeton2006,Cerezo2007}. Hence, in the early days of InGaN research the material was treated like a quantum dot solid comprising indium-rich islands \cite{Narukawa1997}, while more recent measurements based on atom probe tomography \cite{Galtrey2007,Wu2012} and scanning transmission electron microscopy \cite{Schulz2012} have confirmed a random distribution of indium atoms in state-of-the-art samples. In addition, it is reassuring that the emission around $\mathrm{X_{A}}$ shows no clear substructure of sharp emission lines in Fig.\,5(b) of the main text as the density of $\mathrm{X_{A}-In^{\textit{n}}}$ complexes is much higher when being compared to their $\mathrm{Si^{0}X_{A}-In^{\textit{n}}}$ counterparts (cf. the discussion about the inset of Fig.\,5(b) in the main text).

\bibliography{Alloying1}